\documentclass[journal, 10pt, onecolumn, letterpaper, oneside, pdftex]{IEEEtran}

\IEEEoverridecommandlockouts

\usepackage{ifpdf}

\usepackage[noadjust]{cite}

\ifCLASSINFOpdf
  \usepackage[pdftex]{graphicx}
\else
  \usepackage[dvips]{graphicx}
\fi

\usepackage[cmex10]{amsmath}
\usepackage{array}

\ifCLASSOPTIONcompsoc
  \usepackage[caption=false,font=normalsize,labelfont=sf,textfont=sf]{subfig}
\else
  \usepackage[caption=false,font=footnotesize]{subfig}
\fi

\usepackage{fixltx2e}

\usepackage{url}

\usepackage[utf8]{inputenc} 
\usepackage[T1]{fontenc}
\usepackage{ifthen}

\usepackage{mathtools}
\usepackage{amssymb}
\usepackage{relsize}
\usepackage{bbm}
\usepackage{xfrac}
\usepackage{amsthm}
\theoremstyle{plain}
\newtheorem{definition}{Definition}
\newtheorem{lemma}{Lemma}

\newtheorem{theorem}{Theorem}
\newtheorem{corollary}{Corollary}
\newtheorem{remark}{Remark}

\newtheorem{example}{Example}
\newtheorem{proposition}{Proposition}
\usepackage[varg, cmbraces]{newtxmath}

\usepackage{accents}

\usepackage{xcolor}
\definecolor{burgundy}{rgb}{0.545098,0,0}
\definecolor{navyblue}{rgb}{0.0, 0.0, 0.5}
\definecolor{leafgreen}{rgb}{0.290196, 0.470588, 0.0}
\definecolor{bluegreen}{rgb}{0, 0.470588, 0.415686}
\definecolor{zuhl}{rgb}{0.1875, 0.26171875, 0.46484375}
\definecolor{orange}{rgb}{1, 0.6470588235, 0}
\definecolor{red}{rgb}{1, 0, 0}

\usepackage{color}
\usepackage{overpic}
\usepackage{pict2e}
\usepackage{booktabs}

\usepackage{tikz}
\usetikzlibrary{arrows}
\usetikzlibrary{positioning}
\usetikzlibrary{patterns}
\usetikzlibrary{calc}
\tikzset{
    partial ellipse/.style args={#1:#2:#3}{
        insert path={+ (#1:#3) arc (#1:#2:#3)}
    }
}
\usepackage{pgfplots}

\newcommand{\bvec}[1]{\boldsymbol{#1}}

\newcommand{\supp}{\operatorname{supp}}

\newcommand{\figref}[1]{Fig.~\ref{#1}}

\newcommand{\lemref}[1]{Lemma~\ref{#1}}
\newcommand{\propref}[1]{Proposition~\ref{#1}}
\newcommand{\thref}[1]{Theorem~\ref{#1}}
\newcommand{\defref}[1]{Definition~\ref{#1}}
\newcommand{\corref}[1]{Corollary~\ref{#1}}
\newcommand{\sectref}[1]{Section~\ref{#1}}
\newcommand{\remref}[1]{Remark~\ref{#1}}
\newcommand{\exref}[1]{Example~\ref{#1}}
\newcommand{\appref}[1]{Appendix~\ref{#1}}

\usepackage{soul}
\setstcolor{red}

\allowdisplaybreaks[3]

\usepackage{array}
\usepackage[linesnumbered, algoruled, boxed, lined, vlined]{algorithm2e}
\SetKwRepeat{Do}{do}{while}

\usepackage{hyperref}
\hypersetup{
    colorlinks,
    linkcolor = {red!60!black},
    citecolor = {blue!60!black},
    urlcolor = {blue!50!black}
}
\usepackage{microtype}

\begin{document}

\title{Generalizations of Fano's Inequality for Conditional Information Measures via Majorization Theory}

\IEEEoverridecommandlockouts

\author{%
\IEEEauthorblockN{%
Yuta~Sakai}%
\thanks{This research is supported in part by JSPS KAKENHI Grant Number 17J11247.}
\thanks{Y.~Sakai is with the Department of Electrical and Computer Engineering, National University of Singapore, Singapore, Email: \url{eleyuta@nus.edu.sg}, \url{yuta.sakai@m.ieice.org}, \url{yuta.sakai@ieee.org}}%
}%

\maketitle

\begin{abstract}
Fano's inequality is one of the most elementary, ubiquitous, and important tools in information theory.
Using majorization theory, Fano’s inequality is generalized to a broad class of information measures, which contains those of Shannon and R\'{e}nyi.
When specialized to these measures, it recovers and generalizes the classical inequalities.
Key to the derivation is the construction of an appropriate conditional distribution inducing a desired marginal distribution on a countably infinite alphabet.
The construction is based on the infinite-dimensional version of Birkhoff's theorem proven by R\'{e}v\'{e}sz [\emph{Acta\ Math.\ Hungar.}\ \textbf{1962}, \emph{3}, 188--198], and the constraint of maintaining a desired marginal distribution is similar to coupling in probability theory.
Using our Fano-type inequalities for Shannon's and R\'{e}nyi's information measures, we also investigate the asymptotic behavior of the sequence of Shannon's and R\'{e}nyi's equivocations when the error probabilities vanish.
This asymptotic behavior provides a novel characterization of the asymptotic equipartition property (AEP) via Fano's inequality.
\end{abstract}

\begin{IEEEkeywords}
Fano's inequality;
countably infinite alphabet;
list decoding;
general class of conditional information measures;
conditional R\'{e}nyi entropies;
$\alpha$-mutual information;
majorization theory;
the infinite-dimensional version of Birkhoff's theorem;
the Birkhoff--von~Neumann decomposition;
asymptotic equipartition property (AEP)
\end{IEEEkeywords}

\IEEEpeerreviewmaketitle

\section{Introduction}

Inequalities relating probabilities to various information measures are fundamental tools for proving various coding theorems in information theory.
Fano's inequality \cite{fano_1952} is one such paradigmatic example of an information-theoretic inequality;
it elucidates the interplay between the conditional Shannon entropy $H(X \mid Y)$ and the error probability $\mathbb{P}\{ X \neq Y \}$.
Denoting by $h_{2} : u \mapsto - u \log u - (1-u) \log (1-u)$ the binary entropy function on $[0, 1]$ with the conventional hypothesis that $h_{2}( 0 ) = h_{2}( 1 ) = 0$, Fano's inequality can be written as
\begin{align}
\max_{(X, Y) : \mathbb{P}\{ X \neq Y \} \le \varepsilon} H(X \mid Y)
=
h_{2}( \varepsilon )
+
\varepsilon \log(M - 1) 
\label{eq:Fano}
\end{align}
for every $0 \le \varepsilon \le 1 - 1/M$, where $\log$ stands for the natural logarithm, and the maximization is taken over the jointly distributed pairs of $\{ 1, \dots, M \}$-valued random variables (r.v.'s) $X$ and $Y$ satisfying $\mathbb{P}\{ X \neq Y \} \le \varepsilon$.
An important consequence of Fano's inequality is that if the error probabilities vanish, so do the normalized equivocations.
In other words,
\begin{align}
\lim_{n \to \infty} \mathbb{P}\{ X^{n} \neq Y^{n} \} = 0
\quad \Longrightarrow \quad
\lim_{n \to \infty} \frac{ 1 }{ n } H(X^{n} \mid Y^{n}) = 0 ,
\label{eq:classical_Pe-vs-N}
\end{align}
where both $X^{n} = (X_{1}, \dots, X_{n})$ and $Y^{n} = (Y_{1}, \dots, Y_{n})$ are random vectors in which each component is a $\{ 1, \dots, M \}$-valued r.v.
This is the key in proving weak converse results in various communication models (cf.\ \cite{cover_thomas_ElementsofInformationTheory, ElGamal_Kim_2011, yeung_2008}).
Moreover, Fano's inequality also shows that
\begin{align}
\lim_{n \to \infty} \mathbb{P}\{ X_{n} \neq Y_{n} \} = 0
\quad \Longrightarrow \quad
\lim_{n \to \infty} H(X_{n} \mid Y_{n}) = 0 ,
\label{eq:classical_Pe-vs-U}
\end{align}
where $X_{n}$ and $Y_{n}$ are $\{ 1, \dots, M \}$-valued r.v.'s for each $n \ge 1$.
This implication is used, for example, to prove that various Shannon's information measures are continuous in the error metric $\mathbb{P}\{ X_{n} \neq Y_{n} \}$ or the variational distance (cf.\ \cite{zhang_2007, csiszar_korner_2011, sason_2013}).

\subsection{Main Contributions}

In this study, we consider general maximization problems that can be specialized to the left-hand side of \eqref{eq:Fano}; we generalize Fano's inequality in the following four ways:
\begin{itemize}
\item[(i)]
the alphabet $\mathcal{X}$ of a discrete r.v.\ $X$ to be estimated is countably infinite,
\item[(ii)] the marginal distribution $P_{X}$ of $X$ is fixed,
\item[(iii)] the inequality is established on a general class of conditional information measures, and
\item[(iv)] the decoding rule is a list decoding scheme in contrast to a unique decoding scheme.
\end{itemize}
Specifically, given an $\mathcal{X}$-valued r.v.\ $X$ with a countably infinite alphabet $\mathcal{X}$ and a $\mathcal{Y}$-valued r.v.\ $Y$ with an abstract alphabet $\mathcal{Y}$, this study considers a generalized conditional information measure defined by
\begin{align}
\mathsf{H}_{\phi}(X \mid Y)
\coloneqq
\mathbb{E}[ \phi( P_{X|Y} ) ] ,
\label{def:A}
\end{align}
where $P_{X|Y}( x )$ stands for a version of the conditional probability $\mathbb{P}\{ X = x \mid Y \}$ for each $x \in \mathcal{X}$, and $\mathbb{E}[ Z ]$ stands for the expectation of the real-valued r.v.\ $Z$.
Here, this function $\phi : \mathcal{P}(\mathcal{X}) \to [0, \infty]$ defined on the set $\mathcal{P}(\mathcal{X})$ of discrete probability distributions on $\mathcal{X}$ plays the role of an \emph{information measure} of a discrete probability distribution.
When $\mathcal{Y}$ is a countable alphabet, the right-hand side of \eqref{def:A} can be written as
\begin{align}
\mathsf{H}_{\phi}(X \mid Y)
=
\sum_{\substack{ y \in \mathcal{Y} : \\ P_{Y}( y ) > 0 }} P_{Y}( y ) \, \phi( P_{X|Y=y} ) ,
\end{align}
where $P_{Y} = \mathbb{P} \circ Y^{-1}$ denotes the probability law of $Y$, and $P_{X|Y=y}( x ) \coloneqq \mathbb{P}\{ X = x \mid Y = y \}$ denotes the conditional probability for each $(x, y) \in \mathcal{X} \times \mathcal{Y}$.
In this study, we impose some postulates on $\phi$ for technical reasons.
Choosing $\phi$ appropriately, we can specialize $\mathsf{H}_{\phi}(X \mid Y)$ to the conditional Shannon entropy $H(X \mid Y)$, Arimoto's and Hayashi's conditional R\'{e}nyi entropies \cite{arimoto_1977, hayashi_2011}, and so on.
For example, if $\phi$ is given as
\begin{align}
\phi( P )
=
\sum_{x \in \mathcal{X}} P( x ) \log \frac{ 1 }{ P( x ) } ,
\end{align}
then $\mathsf{H}_{\phi}(X \mid Y)$ coincides with the conditional Shannon entropy $H(X \mid Y)$.
Denoting by $P_{\mathrm{e}}^{(L)}(X \mid Y)$ the minimum average probability of list decoding error with a list size $L$, the principal maximization problem considered in this study can be written as
\begin{align}
\mathbb{H}_{\phi}(Q, L, \varepsilon, \mathcal{Y})
\coloneqq
\sup_{(X, Y) : P_{\mathrm{e}}^{(L)}(X \mid Y) \le \varepsilon, P_{X} = Q} \mathsf{H}_{\phi}(X \mid Y) ,
\label{def:main_object}
\end{align}
where the supremum is taken over the pairs $(X, Y)$ satisfying $P_{\mathrm{e}}^{(L)}(X \mid Y) \le \varepsilon$ and fixing the $\mathcal{X}$-marginal $P_{X}$ to a given distribution $Q$.
The feasible region of systems $(Q, L, \varepsilon, \mathcal{Y})$ will be characterized in this paper to ensure that $\mathbb{H}_{\phi}(Q, L, \varepsilon, \mathcal{Y})$ is well-defined.
Under some mild conditions on a given system $(Q, L, \varepsilon, \mathcal{Y})$, especially on the cardinality of $\mathcal{Y}$, we derive explicit formulas of $\mathbb{H}_{\phi}(Q, L, \varepsilon, \mathcal{Y})$; otherwise, we establish tight upper bounds on $\mathbb{H}_{\phi}(Q, L, \varepsilon, \mathcal{Y})$.
As $\mathbb{H}_{\phi}(Q, L, \varepsilon, \mathcal{Y})$ can be thought of as a generalization of the maximization problem stated in \eqref{eq:Fano}, we call these results \emph{Fano-type inequalities} in this paper.
These Fano-type inequalities are formulated by the considered information measures $\phi( P_{\text{type-}\ast} )$ of certain (extremal) probability distributions $P_{\text{type-}\ast}$ depending only on the system $(Q, L, \varepsilon, \mathcal{Y})$.

In this study, we provide Fano-type inequalities via majorization theory \cite{marshall_olkin_arnold_majorization}.
A proof outline to obtain our Fano-type inequalities is as follows.
\begin{enumerate}
\item
First, we show that a generalized conditional information measure $\mathsf{H}_{\phi}(X \mid Y)$ can be bounded from above by $\mathsf{H}_{\phi}(U \mid V)$ with a certain pair $(U, V)$ in which the conditional distribution $P_{U|V}$ of $U$ given $V$ can be thought of as a so-called uniformly dispersive channel \cite{fano_1961, massey_1996} (see also Section~II-A of \cite{sakai_iwata_2018}).
We prove this fact via Jensen's inequality (cf.\ Proposition~A-2 of \cite{shirokov_2010}) and the symmetry of the considered information measures $\phi$.
Moreover, we establish a novel characterization of uniformly dispersive channels via a certain majorization relation; we show that the output distribution of a uniformly dispersive channel is majorized by its transition probability distribution for any fixed input symbol.
This majorization relation is used to obtain a sharp upper bound via the Schur-concavity property of the considered information measures~$\phi$.
\item
Second, we ensure the existence of a joint distribution $P_{X, Y}$ of $(X, Y)$ which satisfies all constraints in our maximization problems $\mathbb{H}_{\phi}(Q, L, \varepsilon, \mathcal{Y})$ stated in \eqref{def:main_object} and the conditional distribution $P_{X|Y}$ is uniformly dispersive.
Here, a main technical difficulty is to maintain a marginal distribution $P_{X}$ of $X$ over a countably infinite alphabet $\mathcal{X}$; see (ii) above.
Using a majorization relation for a uniformly dispersive channel, we express a desired marginal distribution $P_{X}$ by the multiplication of a doubly stochastic matrix and a uniformly dispersive $P_{X|Y}$.
This characterization of the majorization relation via a doubly stochastic matrix was proven by Hardy--Littleweed--P\'{o}lya \cite{hardy_littlewood_polya_1928} in the finite-dimensional case, and by Markus \cite{markus_1964} in the infinite-dimensional case.
From this doubly stochastic matrix, we construct a marginal distribution $P_{Y}$ of $Y$ so that the joint distribution $P_{X, Y} = P_{X|Y} P_{Y}$ has the desired marginal distribution $P_{X}$.
The construction of $P_{Y}$ is based on the infinite-dimensional version of Birkhoff's theorem, which was posed by Birkhoff \cite{birkhoff_1967} and was proven by R\'{e}v\'{e}sz \cite{revesz_1962} via Kolmogorov's extension theorem.
Although the finite-dimensional version of Birkhoff's theorem \cite{birkhoff_1946} (also known as the Birkhoff--von~Neumann decomposition) is well-known, the application of the infinite-dimensional version of Birkhoff's theorem in information theory appears to be novel;
its application aids in dealing with communication systems over countably infinite alphabets.
\item
Third, we introduce an extremal distribution $P_{\text{type-}\ast}$ on a countably infinite alphabet $\mathcal{X}$.
Showing that $P_{\text{type-}\ast}$ is the infimum of a certain class of discrete probability distributions with respect to the majorization relation, our maximization problems can be bounded from above by the considered information measure $\phi(P_{\text{type-}\ast})$.
Namely, our Fano-type inequality is expressed by a certain information measure of the extremal distribution.
When the cardinality of the alphabet of $Y$ is large enough, by constructing a joint distribution $P_{X, Y}$ achieving equality in our generalized Fano-type inequality, we say that the inequality is sharp.
\end{enumerate}

When the alphabet of $Y$ is finite, we further tighten our Fano-type inequality.
To do so, we prove a reduction lemma for the principal maximization problem from an infinite- to a finite-dimensional feasible region.
Therefore, when the alphabet of $Y$ is finite, we do not have to employ technical tools in infinite-dimensional majorization theory, e.g., the infinite-dimensional version of Birkhoff's theorem.
This reduction lemma is useful not only to tighten our Fano-type inequality but also to characterize a sufficient condition of the considered information measure $\phi$ in which $\mathbb{H}_{\phi} (Q, L, \varepsilon, \mathcal{Y})$ is finite if and only if $\phi(Q)$ is also finite.
In fact, Shannon's and R\'{e}nyi's information measures meet this sufficient~condition.

We show that our Fano-type inequalities can be specialized to some known generalizations of Fano's inequality \cite{erokhin_1958, ho_verdu_2010, sakai_iwata_isit2017, sason_verdu_2017} on Shannon's and R\'{e}nyi's information measures.
Therefore, one of our technical contributions is a unified proof of Fano's inequality for conditional information measures via majorization theory.
Generalizations of Erokhin's function \cite{erokhin_1958} from the ordinary mutual information to Sibson's and Arimoto's $\alpha$-mutual information \cite{sibson_1969, arimoto_1977} are also discussed.

Via our generalized Fano-type inequalities, we investigate sufficient conditions on a general source $\mathbf{X} = \{ X_{n} = (Z_{1}^{(n)}, \dots, Z_{n}^{(n)}) \}_{n = 1}^{\infty}$ in which vanishing error probabilities implies vanishing equivocations (cf.\ \eqref{eq:classical_Pe-vs-N} and \eqref{eq:classical_Pe-vs-U}).
We show that the asymptotic equipartition property (AEP) as defined by Verd\'{u}--Han~\cite{verdu_han_1997} is indeed such a sufficient condition.
In other words, if a general source $\mathbf{X} = \{ X_{n} \}_{n = 1}^{\infty}$ satisfies the AEP and $H( X_{n} ) = \Omega( 1 )$ as $n \to \infty$, then we prove that
\begin{align}
\lim_{n \to \infty} P_{\mathrm{e}}^{(L_{n})}(X_{n} \mid Y_{n})
=
\lim_{n \to \infty} \frac{ \log L_{n} }{ H( X_{n} ) }
=
0
\quad \Longrightarrow \quad
\lim_{n \to \infty} \frac{ H(X_{n} \mid Y_{n}) }{ H( X_{n} ) }
=
0 ,
\end{align}
where $\{ L_{n} \}_{n = 1}^{\infty}$ is an arbitrary sequence of list sizes.
This is a generalization of \eqref{eq:classical_Pe-vs-N} and \eqref{eq:classical_Pe-vs-U} and, to the best of the author's knowledge, a novel connection between the AEP and Fano's inequality.
We prove this connection by using the splitting technique of a probability distribution;
this technique was used to derive limit theorems of Markov processes by Nummelin \cite{nummelin_1978} and Athreya--Ney \cite{athreya_ney_1978}.
Note that there are also many applications of the splitting technique in information theory (cf.\ \cite{ho_verdu_2010, kumar_li_elgamal_isit2014, vellambi_kliewer_2016, vellambi_kliewer_isit2018, yu_tan_2018_common, yu_tan_2018_channel}).
In addition, we extend Ho--Verd\'{u}'s sufficient conditions (See Section~V of \cite{ho_verdu_2010}) and Sason--Verd\'{u}'s sufficient conditions (see Theorem~4 of \cite{sason_verdu_2017}) on a general source $\mathbf{X} = \{ X_{n} \}_{n = 1}^{\infty}$ in which equivocations vanish if the error probabilities vanish.

\subsection{Related Works}

\subsubsection{Information Theoretic Tools on Countably Infinite Alphabets}
\label{sect:countably-infinite}

As the right-hand side of \eqref{eq:Fano} diverges as $M$ goes to infinity whenever $\varepsilon > 0$ is fixed, the classical Fano inequality is applicable only if $X$ is supported on a finite alphabet (see also Chapter~1 of \cite{han_InformationSpectrum}).
In fact, if both $X_{n}$ and $Y_{n}$ are supported on the same countably infinite alphabet for each $n \ge 1$, one can construct a somewhat pathological example so that $\mathbb{P}\{ X_{n} \neq Y_{n} \} = \mathrm{o}( 1 )$ as $n \to \infty$ but $H(X_{n} \mid Y_{n}) = \infty$ for every $n \ge 1$ (cf.\ Example~2.49 of \cite{yeung_2008}).

Usually, it is not straightforward to generalize information theoretic tools for systems defined on a finite alphabet to systems defined on a countably infinite alphabet.
Ho--Yeung \cite{ho_yeung_2009} showed that Shannon's information measures defined on countably infinite alphabets are \emph{not} continuous with respect to the following distances; the $\chi^{2}$-divergence, the relative entropy, and the variational distance.
Continuity issues of R\'{e}nyi's information measures defined on countably infinite alphabets were explored by Kova\v{c}evi\'{c}--Stanojevi\'{c}--\v{S}enk \cite{kovacevic_stanojevic_senk_2013}.
In addition, although weak typicality (cf.\ Chapter~3 of \cite{cover_thomas_ElementsofInformationTheory}) that is also known as the entropy-typical sequences (cf.\ Problem~2.5 of \cite{csiszar_korner_2011}) is a convenient tool in proving achievability theorems for sources and channels with defined on  countably infinite (or even uncountable) alphabets, strong typicality \cite{csiszar_korner_2011} is only amenable in situations with finite alphabets.
To ameliorate this issue, Ho--Yeung \cite{ho_yeung_2010} proposed a notion known as unified typicality that ensures that the desirable properties of weak and strong typicality are retained  when one is working with countably infinite~alphabets.

Recently, Madiman--Wang--Woo \cite{madiman_wang_woo_2019} investigated relations between majorization and the strong Sperner property \cite{sperner_1928} of posets together with applications to the R\'{e}nyi entropy power inequality for sums of independent and integer-valued r.v.'s, i.e., supported on countably infinite alphabets.

To the best of the author's knowledge, a generalization of Fano's inequality to the case when $X$ is supported on a countably infinite alphabet was initiated by Erokhin \cite{erokhin_1958}.
Given a discrete probability distribution $Q$ on a countably infinite alphabet $\mathcal{X} = \{ 1, 2, \dots \}$, Erokhin established in Equation~(11) of \cite{erokhin_1958} an explicit formula of the function:
\begin{align}
\mathbb{I}(Q, \varepsilon)
\coloneqq
\min_{(X, Y) : \mathbb{P}\{ X \neq Y \} \le \varepsilon, P_{X} = Q} I(X \wedge Y) ,
\label{eq:Erokhin}
\end{align}
where the minimization is taken over the pairs of $\mathcal{X}$-valued r.v.'s $X$ and $Y$ satisfying $\mathbb{P}\{ X \neq Y \} \le \varepsilon$ and $\mathbb{P}\{ X = x \} = Q( x )$ for each $x \in \mathcal{X}$, and $I(X \wedge Y)$ stands for the mutual information between $X$ and $Y$.
Note that Erokhin's function $\mathbb{I}(Q, \varepsilon)$ is the rate-distortion function with Hamming distortion measures (cf.\ \cite{berger_1971, ahlswede_1990}).
As the well-known identity $I(X \wedge Y) = H(X) - H(X \mid Y)$ implies that
\begin{align}
\mathbb{I}(Q, \varepsilon)
=
H(X) - \max_{(X, Y) : \mathbb{P}\{ X \neq Y \} \le \varepsilon, P_{X} = Q} H(X \mid Y) ,
\label{eq:Erokhin_type2}
\end{align}
Erokhin's function $\mathbb{I}(Q, \varepsilon)$ can be naturally thought of as a generalization of the classical Fano inequality stated in \eqref{eq:Fano}, where $H(X)$ stands for the Shannon entropy of $X$, and the probability distribution of $X$ is given by $\mathbb{P}\{ X = x \} = Q( x )$ for each $x \in \mathcal{X}$.
Kostina--Polyanskiy--Verd\'{u} \cite{kostina_polyanskiy_verdu_2015} derived a second-order asymptotic expansion of $\mathbb{I}(Q^{n}, \varepsilon)$ as $n \to \infty$, where $Q^{n}$ stands for the $n$-fold product of $Q$.
Their asymptotic expansion is closely related to the second-order asymptotics of the variable-length compression allowing errors; see (\cite{kostina_polyanskiy_verdu_2015}, Theorem~4).

Ho--Verd\'{u} \cite{ho_verdu_2010} gave an explicit formula of the maximization in the right-hand side of \eqref{eq:Erokhin_type2}; they proved it via the additivity of Shannon's information measures.
Note that Ho--Verd\'{u}'s formula (cf.\ Theorem~1 of \cite{ho_verdu_2010}) coincides with Erokhin's formula (cf.\ Equation~(11) of \cite{erokhin_1958}) via the identity stated in \eqref{eq:Erokhin_type2}.
In Theorems~2 and~4 of \cite{ho_verdu_2010}, Ho--Verd\'{u} also tightened the maximization in the right-hand side of \eqref{eq:Erokhin_type2} when $Y$ is supported on a proper subalphabet of $\mathcal{X}$.
Moreover, they provided in Section~V of \cite{ho_verdu_2010} some sufficient conditions on a general source in which vanishing error probabilities (i.e., $\mathbb{P}\{ X_{n} \neq Y_{n} \} = \mathrm{o}( 1 )$) implies vanishing unnormalized or normalized equivocations (i.e., $H(X_{n} \mid Y_{n}) = \mathrm{o}( 1 )$ or $H(X_{n} \mid Y_{n}) = \mathrm{o}( n )$).

\subsubsection{Fano's Inequality with List Decoding}

Fano's inequality with list decoding was initiated by Ahlswede--G\'{a}cs--K\"{o}rner \cite{ahlswede_gacs_korner_1976}.
By a minor extension of the usual proof (see, e.g., Lemma~3.8 of \cite{csiszar_korner_2011}), one can see that
\begin{align}
\max_{(X, Y) : P_{\mathrm{e}}^{(L)}(X \mid Y) \le \varepsilon} H(X \mid Y)
=
h_{2}( \varepsilon ) + (1 - \varepsilon) \log L + \varepsilon \log (M - L)
\label{eq:Fano_list}
\end{align}
for every integers $1 \le L < M$ and every real number $0 \le \varepsilon \le 1 - L/M$, where the maximization is taken over the pairs of a $\{ 1, \dots, M \}$-valued r.v.\ $X$ and a $\mathcal{Y}$-valued r.v.\ $Y$ satisfying $P_{\mathrm{e}}^{(L)}(X \mid Y) \le \varepsilon$.
Note that the right-hand side of \eqref{eq:Fano_list} coincides with the Shannon entropy of the \emph{extremal distribution of type-$0$} defined by
\begin{align}
P_{\operatorname{type-0}}( x )
=
P_{\operatorname{type-0}}^{(M, L, \varepsilon)}( x )
\coloneqq
\begin{dcases}
\frac{ 1 - \varepsilon }{ L }
& \mathrm{if} \ 1 \le x \le L ,
\\
\frac{ \varepsilon }{ M - L }
& \mathrm{if} \ L < x \le M ,
\\
0
& \mathrm{if} \ M < x < \infty
\end{dcases}
\label{def:type0}
\end{align}
for each integer $x \ge 1$.
A graphical representation of this extremal distribution is plotted in \figref{fig:type0_form}.

\begin{figure}[!t]
\centering
\begin{tikzpicture}
\draw [very thick, -latex] (0, -1) -- (0, 5);
\draw [very thick] (-1, 0) -- (9, 0);
\draw [thick] (0, 0) rectangle (1, 4);
\draw [thick] (1, 0) rectangle (2, 4);
\draw [thick] (2, 0) rectangle (3, 4);
\draw [thick] (3, 0) rectangle (4, 2.5);
\draw [thick] (4, 0) rectangle (5, 2.5);
\draw [thick] (5, 0) rectangle (6, 2.5);
\draw [thick] (6, 0) rectangle (7, 2.5);
\draw [thick] (7, 0) rectangle (8, 2.5);
\foreach \i in {1,...,8}{\draw (\i - 0.5, 0) node [below = 0.75em] {\Large $\i$};}
\draw [very thick] (-0.1, 4) node [left] {$\dfrac{ 1 - \varepsilon }{ L }$} -- (0.1, 4);
\draw [very thick] (-0.1, 2.5) node [left] {$\dfrac{ \varepsilon }{ M - L }$} -- (0.1, 2.5);
\draw [thick, loosely dotted] (0, 2.5) -- (3, 2.5);
\end{tikzpicture}
\caption{Each bar represents a probability mass of the extremal distribution of type-0 defined in \eqref{def:type0}, where $M = 8$ and $L = 3$.}
\label{fig:type0_form}
\end{figure}

Combining \eqref{eq:Fano_list} and the blowing-up technique (cf.\ Chapter~5 of \cite{csiszar_korner_2011} or Section~3.6.2 of \cite{raginsky_sason_2014}), Ahlswede--G\'{a}cs--K\"{o}rner \cite{ahlswede_gacs_korner_1976} proved the strong converse property (in Wolfowitz's sense \cite{wolfowitz_1978}) of degraded broadcast channels under the maximum error probability criterion.
Extending the proof technique in \cite{ahlswede_gacs_korner_1976} together with the wringing technique, Dueck \cite{dueck_1981} proved the strong converse property of multiple-access channels under the average error probability criterion.
As these proofs rely on a combinatorial lemma (cf.\ Lemma~5.1 of \cite{csiszar_korner_2011}), they work only when the channel output alphabet is finite; but see recent work by Fong--Tan \cite{FongTanMAC, FongTanBC} in which such techniques have been extended to Gaussian channels.
On the other hand, Kim--Sutivong--Cover \cite{kim_sutivong_cover_2008} investigated a trade-off between the channel coding rate and the state uncertainty reduction of a channel with state information available only at the sender, and derived its trade-off region in the weak converse regime by employing \eqref{eq:Fano_list}.

\subsubsection{Fano's Inequality for R\'{e}nyi's Information Measures}

So far, many researchers have considered various directions for generalizing Fano's inequality. An interesting study involves \emph{reversing} the usual Fano inequality.
In this regard, lower bounds on $H(X \mid Y)$ subject to $\mathbb{P}\{ X \neq Y \} = \varepsilon$ were independently established by Kovalevsky \cite{kovalevsky_1968}, Chu--Cheuh \cite{chu_chueh_1966}, and Tebbe--Dwyer \cite{tebbe_dwyer_1968} (see also Feder--Merhav's study~\cite{feder_merhav_1994}).
Prasad \cite{prasad_2015} provided several refinements of the reverse/forward Fano inequalities for Shannon's information measures.

In \cite{ben-bassat_raviv_1978}, Ben-Bassat--Raviv explored several inequalities between the (unconditional) R\'{e}nyi entropy and the error probability.
Generalizations of Fano's inequality from the conditional Shannon entropy $H(X \mid Y)$ to Arimoto's conditional R\'{e}nyi entropy $H_{\alpha}^{\mathrm{Arimoto}}(X \mid Y)$ introduced in \cite{arimoto_1977} were recently and independently investigated by Sakai--Iwata \cite{sakai_iwata_isit2017} and Sason--Verd\'{u} \cite{sason_verdu_2017}.
Specifically, Sakai--Iwata~\cite{sakai_iwata_isit2017} provided sharp upper/lower bounds on $H_{\alpha}^{\mathrm{Arimoto}}(X \mid Y)$ for fixed $H_{\beta}^{\mathrm{Arimoto}}(X \mid Y)$ with two distinct orders $\alpha \neq \beta$.
In other words, they gave explicit formulas of the following minimization and maximization,
\begin{align}
f_{\min}(\alpha, \beta, \gamma)
& \coloneqq
\min_{(X, Y) : H_{\beta}^{\mathrm{Arimoto}}(X \mid Y) = \gamma} H_{\alpha}^{\mathrm{Arimoto}}(X \mid Y) ,
\label{eq:AR_min} \\
f_{\max}(\alpha, \beta, \gamma)
& \coloneqq
\max_{(X, Y) : H_{\beta}^{\mathrm{Arimoto}}(X \mid Y) = \gamma} H_{\alpha}^{\mathrm{Arimoto}}(X \mid Y) ,
\label{eq:AR_max}
\end{align}
respectively.
As $H_{\beta}^{\mathrm{Arimoto}}(X \mid Y)$ is a strictly monotone function of the minimum average probability of error if $\beta = \infty$, both functions $f_{\min}(\alpha, \infty, \gamma)$ and $f_{\max}(\alpha, \infty, \gamma)$ can be thought of as reverse and forward Fano inequalities on $H_{\alpha}^{\mathrm{Arimoto}}(X \mid Y)$, respectively (cf.\ Section~V in the arXiv paper \cite{sakai_iwata_isit2017}).
Sason--Verd\'{u}~\cite{sason_verdu_2017} also gave generalizations of the forward and reverse Fano's inequalities on $H_{\alpha}^{\mathrm{Arimoto}}(X \mid Y)$.
Moreover, in the forward Fano inequality pertaining to $H_{\alpha}^{\mathrm{Arimoto}}(X \mid Y)$, they generalized in Theorem~8 of \cite{sason_verdu_2017} the decoding rules from unique decoding to list decoding as follows:
\begin{align}
\max_{(X, Y) : P_{\mathrm{e}}^{(L)}(X \mid Y) \le \varepsilon} H_{\alpha}^{\mathrm{Arimoto}}(X \mid Y)
=
\frac{ 1 }{ 1 - \alpha } \log \left( L^{1-\alpha} (1 - \varepsilon) + (M - L)^{1-\alpha} \varepsilon^{\alpha} \right)
\label{eq:Sason-Verdu}
\end{align}
for every $0 \le \varepsilon \le 1 - L/M$ and $\alpha \in (0, 1) \cup (1, \infty)$, where the maximization is taken as with \eqref{eq:Fano_list}.
Similar to \eqref{eq:Fano_list}, the right-hand side of \eqref{eq:Sason-Verdu} coincides with the R\'{e}nyi entropy \cite{renyi_1961} of the extremal distribution of type-$0$.
Note that the reverse Fano inequality proven in \cite{sakai_iwata_isit2017, sason_verdu_2017} does not require that $\mathcal{X}$ is finite.
On the other hand, the forward Fano inequality proven in \cite{sakai_iwata_isit2017, sason_verdu_2017} is applicable \emph{only} when $\mathcal{X}$ is finite.

\subsubsection{Lower Bounds on Mutual Information}

Han--Verd\'{u} \cite{han_verdu_1994} generalized Fano's inequality on a countably infinite alphabet $\mathcal{X}$ by investigating lower bounds on the mutual information, i.e.,
\begin{align}
I(X \wedge Y)
\ge
\mathbb{P}\{ X \neq Y \} \log \frac{ \mathbb{P}\{ X \neq Y \} }{ \mathbb{P}\{ \bar{X} \neq \bar{Y} \} } + \mathbb{P}\{ X = Y \} \log \frac{ \mathbb{P}\{ X = Y \} }{ \mathbb{P}\{ \bar{X} = \bar{Y} \} } ,
\label{eq:Han-Verdu-Fano}
\end{align}
via the data processing lemma without additional constraints on the r.v.'s $X$ and $Y$, where $\bar{X}$ and $\bar{Y}$ are independent r.v.'s having marginals as $X$ and $Y$ respectively.
Polyanskiy--Verd\'{u} \cite{polyanskiy_verdu_allerton2010} showed a lower bound on Sibson's $\alpha$-mutual information by using the data processing lemma for the R\'{e}nyi divergence.
Recently, Sason \cite{sason_2019} generalized Fano's inequality with list decoding via the \emph{strong} data processing lemma for the $f$-divergences.

Liu--Verd\'{u} \cite{liu_verdu_isit2017} showed that
\begin{align}
I(X^{n} \wedge Y^{n})
\ge
\log M_{n} + \mathrm{O}( \sqrt{ n } )
\end{align}
as $n \to \infty$, provided that the \emph{geometric} average probability of error, which is a weaker and a stronger criteria than the maximum and the average error criteria, respectively, satisfies
\begin{align}
\left( \prod_{m = 1}^{M_{n}} \mathbb{P}\{ Y^{n} \in \mathcal{D}_{m, n} \mid X^{n} = \bvec{c}_{m, n} \} \right)^{1/M_{n}}
\ge
1 - \varepsilon
\end{align}
for sufficiently large $n$, where
$X^{n}$ is a r.v.\ uniformly distributed on the codeword set $\{ \bvec{c}_{m, n} \}_{m = 1}^{M_{n}}$,
$Y^{n}$ is a r.v.\ induced by the $n$-fold product of a discrete memoryless channel with the input $X^{n}$,
$M_{n}$ is a positive integer denoting the message size,
$\{ D_{m, n} \}_{m = 1}^{M_{n}}$ is a collection of disjoint subsets playing the role of decoding regions, and
$0 < \varepsilon < 1$ is a tolerated probability of error.
This is a second-order asymptotic estimate on the mutual information, and is derived by using the Donsker--Varadhan lemma (cf.\ Equation~(3.4.67) of \cite{raginsky_sason_2014}) and the so-called pumping-up argument.

\subsection{Paper Organization}

The rest of this paper is organized as follows.
\sectref{sect:Fano-type} introduces basic notations and definitions to understand our generalized conditional information measure $\mathsf{H}_{\phi}(X \mid Y)$ and the principal maximization problem $\mathbb{H}_{\phi}(Q, L, \varepsilon, \mathcal{Y})$.
\sectref{sect:Fano_h} presents the main results: our Fano-type inequalities.
\sectref{sect:Fano_Renyi} specializes our Fano-type inequalities to Shannon's and R\'{e}nyi's information measures, and discusses generalizations of Erokhin's function from the ordinary mutual information to Sibson's and Arimoto's $\alpha$-mutual information.
\sectref{sect:vanishing} investigates several conditions on a general source in which the vanishing error probabilities implies the vanishing equivocations; a novel characterization of the AEP via Fano's inequality is also presented.
\sectref{sect:proof_Fano} proves our Fano-type inequalities stated in \sectref{sect:Fano_h}, and contains most technical contributions in this study.
\sectref{sect:proof_equivocation} proves the asymptotic behaviors stated in \sectref{sect:vanishing}.
Finally, \sectref{sect:conclusion} concludes this study with some remarks.

\section{Preliminaries}
\label{sect:Fano-type}

\subsection{A General Class of Conditional Information Measures}
\label{sect:majorization}

This subsection introduces some notions in majorization theory \cite{marshall_olkin_arnold_majorization} and a rigorous definition of generalized conditional information measure $\mathsf{H}_{\phi}(X \mid Y)$ defined in \eqref{def:A}.
Let $\mathcal{X} = \{ 1, 2, \dots \}$ be a countably infinite alphabet.
A \emph{discrete probability distribution} $P$ on $\mathcal{X}$ is a map $P : \mathcal{X} \to [0, 1]$ satisfying $\sum_{x \in \mathcal{X}} P( x ) = 1$. 
In this paper, motivated to consider the joint probability distributions on $\mathcal{X} \times \mathcal{Y}$, it is called an \emph{$\mathcal{X}$-marginal.}
Given an $\mathcal{X}$-marginal $P$, a \emph{decreasing rearrangement} of $P$ is denoted by $P^{\downarrow}$, i.e., it fulfills
\begin{align}
P^{\downarrow}(1) \ge P^{\downarrow}(2) \ge P^{\downarrow}(3) \ge P^{\downarrow}(4) \ge P^{\downarrow}(5) \ge \cdots .
\label{def:decreasing_rearrangement}
\end{align}
The following definition gives us the notion of majorization for $\mathcal{X}$-marginals.

\begin{definition}[{Majorization \cite{marshall_olkin_arnold_majorization}}]
\label{def:majorization}
An $\mathcal{X}$-marginal $P$ is said to \emph{majorize} another $\mathcal{X}$-marginal $Q$ if
\begin{align}
\sum_{i = 1}^{k} P^{\downarrow}( i )
\ge
\sum_{i = 1}^{k} Q^{\downarrow}( i )
\end{align}
for every $k \ge 1$.
This relation is denoted by $P \succ Q$ or $Q \prec P$.
\end{definition}

Let $\mathcal{P}(\mathcal{X})$ be the set of $\mathcal{X}$-marginals.
The following definitions are important postulates on a function $\phi : \mathcal{P}( \mathcal{X} ) \to [0, \infty]$ playing the role of an information measure of an $\mathcal{X}$-marginal.

\begin{definition}
\label{def:symmetry}
A function $\phi : \mathcal{P}( \mathcal{X} ) \to [0, \infty]$ is said to be \emph{symmetric} if it is invariant for any permutation of probability masses, i.e., $\phi( P ) = \phi( P^{\downarrow} )$ for every $P \in \mathcal{P}( \mathcal{X} )$.
\end{definition}

\begin{definition}
\label{def:semicontinuous}
A function $\phi : \mathcal{P}( \mathcal{X} ) \to [0, \infty]$ is said to be \emph{lower semicontinuous} if for any $P \in \mathcal{P}(\mathcal{X})$, it holds that $\liminf_{n} \phi( P_{n} ) \ge \phi( P )$ for every pointwise convergent sequence $P_{n} \to P$, where the pointwise convergence $P_{n} \to P$ means that $P_{n}( x ) \to P( x )$ as $n \to \infty$ for every $x \in \mathcal{X}$.
\end{definition}

\begin{definition}
\label{def:convex}
A function $\phi : \mathcal{P}( \mathcal{X} ) \to [0, \infty]$ is said to be \emph{convex} if $\phi( R ) \le \lambda \phi(P) + (1 - \lambda) \phi(Q)$ with $R = \lambda P + (1-\lambda) Q$ for every $P, Q \in \mathcal{P}( \mathcal{X} )$ and $0 \le \lambda \le 1$.
\end{definition}

\begin{definition}
\label{def:quasiconvex}
A function $\phi : \mathcal{P}( \mathcal{X} ) \to [0, \infty]$ is said to be \emph{quasiconvex} if the sublevel set $\{ P \in \mathcal{P}( \mathcal{X} ) \mid \phi( P ) \le c \}$ is convex for every $P \in \mathcal{P}( \mathcal{X} )$ and $c \in [0, \infty)$.
\end{definition}

\begin{definition}
\label{def:Schur}
A function $\phi : \mathcal{P}( \mathcal{X} ) \to [0, \infty]$ is said to be \emph{Schur-convex} if $P \prec Q$ implies that $\phi(P) \le \phi(Q)$.
\end{definition}

In Definitions~\ref{def:convex}--\ref{def:Schur}, each term or its suffix \emph{convex} is replaced by \emph{concave} if $-\phi$ fulfills the condition.
In \defref{def:semicontinuous}, note that the pointwise convergence of $\mathcal{X}$-marginals is equivalent to the convergence in the variational distance topology (see, e.g., Lemma~3.1 of \cite{topsoe_2001} or Section~III-D of \cite{erven_harremoes_2014}).

Let $X$ be an $\mathcal{X}$-valued r.v.\ and $Y$ a $\mathcal{Y}$-valued r.v., where $\mathcal{Y}$ is an abstract alphabet.
Unless stated otherwise, assume that the measurable space of $\mathcal{Y}$ with a certain $\sigma$-algebra is standard Borel, where a measurable space is said to be \emph{standard Borel} if its $\sigma$-algebra is the Borel $\sigma$-algebra generated by a Polish topology on the space.
Assuming that $\phi : \mathcal{P}(\mathcal{X}) \to [0, \infty]$ is a symmetric, concave, and lower semicontinuous function, the generalized conditional information measure $\mathsf{H}_{\phi}(X \mid Y)$ is defined by \eqref{def:A}.
The postulates on $\phi$ we have imposed here are useful for technical reasons to employ majorization theory; see the following lemma.

\begin{proposition}
\label{prop:Schur_convex}
Every symmetric and quasiconvex function $\phi : \mathcal{P}( \mathcal{X} ) \to [0, \infty]$ is Schur-convex.
\end{proposition}

\begin{IEEEproof}[Proof of \propref{prop:Schur_convex}]
In Proposition~3.C.3 of \cite{marshall_olkin_arnold_majorization}, the assertion of \propref{prop:Schur_convex} was proved in the case where the dimension of the domain of $\phi$ is finite.
Employing Theorem~4.2 of \cite{markus_1964} instead of Corollary~2.B.3 of \cite{marshall_olkin_arnold_majorization}, the proof of Proposition~3.C.3 of \cite{marshall_olkin_arnold_majorization} can be directly extended to infinite-dimensional domains.
\end{IEEEproof}

To employ the Schur-concavity property in the sequel, \propref{prop:Schur_convex} suggests assuming that $\phi$ is \emph{symmetric} and \emph{quasiconcave.}
In addition, to apply Jensen's inequality on the function  $\phi$, it suffices to assume that $\phi$ is \emph{concave} and \emph{lower semicontinuous}, because the domain $\mathcal{P}( \mathcal{X} )$ forms a closed convex bounded set in the variational distance topology (cf.\ Proposition~A-2 of \cite{shirokov_2010}).
Motivated by these properties, we impose the three postulates (corresponding to Definitions~\ref{def:symmetry}--\ref{def:convex}) on $\phi$ in this study.

\subsection{Minimum Average Probability of List Decoding Error}
\label{sect:Pe}

Consider a certain communication model for which a $\mathcal{Y}$-valued r.v.\ $Y$ plays the role of the side-information of an $\mathcal{X}$-valued r.v.\ $X$.
A \emph{list decoding scheme} with a list size $1 \le L < \infty$ is a decoding scheme producing $L$ candidates for realizations of $X$ when we observe a realization of $Y$.
The \emph{minimum average error probability under list decoding} is defined by
\begin{align}
P_{\mathrm{e}}^{(L)}(X \mid Y)
\coloneqq
\min_{f : \mathcal{Y} \to \binom{\mathcal{X}}{L}} \mathbb{P}\{ X \notin f(Y) \} ,
\label{def:Pe}
\end{align}
where the minimization is taken over all set-valued functions $f : \mathcal{Y} \to \binom{\mathcal{X}}{L}$ with the decoding range
\begin{align}
\binom{\mathcal{X}}{L}
& \coloneqq
\{ \mathcal{D} \subset \mathcal{X} \mid |\mathcal{D}| = L \} ,
\end{align}
and $| \cdot |$ stands for the cardinality of a set.
If $\mathcal{S}$ is an infinite set, then we assume that $|\mathcal{S}| = \infty$ as usual.
If $L = 1$, then \eqref{def:Pe} coincides with \emph{the average error probability of the maximum a posteriori (MAP) decoding scheme.}
For the sake of brevity, we write
\begin{align}
P_{\mathrm{e}}(X \mid Y)
\coloneqq
P_{\mathrm{e}}^{(1)}(X \mid Y) .
\end{align}
It is clear that
\begin{align}
\mathbb{P} \{ X \notin f(Y) \} \le \varepsilon
\quad \Longrightarrow \quad
P_{\mathrm{e}}^{(L)}(X \mid Y) \le \varepsilon
\end{align}
for \emph{any} list decoder $f : \mathcal{Y} \to \binom{\mathcal{X}}{L}$ and \emph{any} tolerated probability of error $\varepsilon \ge 0$.
Therefore, it suffices to consider the constraint $P_{\mathrm{e}}^{(L)}(X \mid Y) \le \varepsilon$ rather than $\mathbb{P} \{ X \notin f(Y) \} \le \varepsilon$ in our subsequent analyses.

The following proposition is an elementary formula of $P_{\mathrm{e}}^{(L)}(X \mid Y)$ as in the MAP decoding.

\begin{proposition}
\label{prop:listMAP}
It holds that
\begin{align}
P_{\mathrm{e}}^{(L)}(X \mid Y)
=
1 - \mathbb{E} \Bigg[ \sum_{x = 1}^{L} P_{X|Y}^{\downarrow}( x ) \Bigg] .
\label{eq:listMAPformula}
\end{align}
\end{proposition}

\begin{IEEEproof}[Proof of \propref{prop:listMAP}]
See \appref{app:listMAP}.
\end{IEEEproof}

\begin{remark}
It follows from \propref{prop:listMAP} that $\mathsf{H}_{\phi}(X \mid Y)$ defined in \eqref{def:A} can be specialized to $P_{\mathrm{e}}^{(L)}(X \mid Y)$ with
\begin{align}
\phi( P )
=
1 - \sum_{x = 1}^{L} P^{\downarrow}( x ) .
\end{align}
\end{remark}

The following proposition characterizes the feasible region of systems $(Q, L, \varepsilon, \mathcal{Y})$ considered in our principal maximization problem $\mathbb{H}_{\phi}(Q, L, \varepsilon, \mathcal{Y})$ stated in \eqref{def:main_object}.

\begin{proposition}
\label{prop:boundPe}
If $P_{X} = Q$, then
\begin{align}
1 - \sum_{x = 1}^{L \cdot |\mathcal{Y}|} Q^{\downarrow}( x )
\le
P_{\mathrm{e}}^{(L)}(X \mid Y)
\le
1 - \sum_{x = 1}^{L} Q^{\downarrow}( x ) .
\label{ineq:boundPe}
\end{align}
Moreover, both inequalities are sharp in the sense that there exist pairs of r.v.'s $X$ and $Y$ achieving the equalities while respecting the constraint $P_{X} = Q$.
\end{proposition}

\begin{IEEEproof}[Proof of \propref{prop:boundPe}]
See \appref{app:boundPe}.
\end{IEEEproof}

The minimum average error probability for list decoding concerning $X \sim Q$ \emph{without} any side-information is denoted by
\begin{align}
P_{\mathrm{e}}^{(L)}( Q )
\coloneqq
1 - \sum_{x = 1}^{L} Q^{\downarrow}( x ) .
\label{def:Pe_Q}
\end{align}
Then, the second inequality in \eqref{ineq:boundPe} is obvious, and it is similar to the property that \emph{conditioning reduces uncertainty} (cf.\ \cite{cover_thomas_ElementsofInformationTheory}, Theorem~2.8.1).
\propref{prop:boundPe} ensures that when we have to consider the constraints $P_{\mathrm{e}}^{(L)}(X \mid Y) \le \varepsilon$ and $P_{X} = Q$, it suffices to consider a system $(Q, L, \varepsilon, \mathcal{Y})$ satisfying
\begin{align}
1 - \sum_{x = 1}^{L \cdot |\mathcal{Y}|} Q^{\downarrow}( x )
\le
\varepsilon
\le
1 - \sum_{x = 1}^{L} Q^{\downarrow}( x ) .
\label{eq:range_epsilon_list}
\end{align}

\section{Main Results: Fano-Type Inequalities}
\label{sect:Fano_h}

Let $(Q, L, \varepsilon, \mathcal{Y})$ be a system satisfying \eqref{eq:range_epsilon_list}, and $\phi : \mathcal{P}( \mathcal{X} ) \to [0, \infty]$ a symmetric, concave, and lower semicontinuous function.
The main aim of this study is to find an explicit formula or a tight upper bound on $\mathbb{H}_{\phi}(Q, L, \varepsilon, \mathcal{Y})$ defined in \eqref{def:main_object}.
Now, define the \emph{extremal distribution of type-$1$} by the following $\mathcal{X}$-marginal,
\begin{align}
P_{\operatorname{type-1}}( x )
=
P_{\operatorname{type-1}}^{(Q, L, \varepsilon)}( x )
\coloneqq
\begin{dcases}
Q^{\downarrow}( x )
& \mathrm{if} \ 1 \le x < J \ \mathrm{or} \ K_{1} < x < \infty ,
\\
\mathcal{V}( J )
& \mathrm{if} \ J \le x \le L ,
\\
\mathcal{W}( K_{1} )
& \mathrm{if} \ L < x \le K_{1} ,
\end{dcases}
\label{def:type1}
\end{align}
for each $x \in \mathcal{X}$,
the weight $\mathcal{V}( j )$ is defined by
\begin{align}
\mathcal{V}( j )
=
\mathcal{V}^{(Q, L, \varepsilon)}( j )
\coloneqq
\begin{dcases}
\frac{ (1 - \varepsilon) - \sum_{x = 1}^{j - 1} Q^{\downarrow}( x ) }{ L - j + 1 }
& \mathrm{if} \ 1 \le j \le L ,
\\
1
& \mathrm{if} \ j > L
\end{dcases}
\label{def:V}
\end{align}
for each $j \ge 1$,
the weight $\mathcal{W}( k )$ is defined by
\begin{align}
\mathcal{W}( k )
=
\mathcal{W}^{(Q, L, \varepsilon)}( k )
\coloneqq
\begin{dcases}
-1
& \mathrm{if} \ k = L ,
\\
\frac{ \sum_{x = 1}^{k} Q^{\downarrow}( x ) - (1 - \varepsilon) }{ k - L }
& \mathrm{if} \ L <  k < \infty ,
\\
0
& \mathrm{if} \ k = \infty
\end{dcases}
\label{def:W2}
\end{align}
for each $k \ge L$,
the integer $J$ is chosen so that
\begin{align}
J
=
J(Q, L, \varepsilon)
& \coloneqq
\min \{ 1 \le j < \infty \mid Q^{\downarrow}( j ) < \mathcal{V}( j ) \} ,
\label{def:J}
\end{align}
and $K_{1}$ is chosen so that
\begin{align}
K_{1}
=
K_{1}(Q, L, \varepsilon)
& \coloneqq
\sup \{ L \le k < \infty \mid \mathcal{W}( k ) < Q^{\downarrow}( k ) \} .
\label{def:K3}
\end{align}
A graphical representation of $P_{\operatorname{type-1}}$ is shown in \figref{fig:type1_form}.
Under some mild conditions, the following theorem gives an explicit formula of $\mathbb{H}_{\phi}(Q, L, \varepsilon, \mathcal{Y})$.

\begin{figure}[!t]
\centering
\begin{tikzpicture}
\draw [very thick, -latex] (0, -1) -- (0, 6);
\draw [very thick] (-1, 0) -- (12, 0);
\draw [fill, gray!90] (1, 3) rectangle (2, 3.725);
\draw [fill, gray!90] (2, 2) rectangle (3, 3.725);
\filldraw [pattern = north east lines] (3, 0.75) rectangle (4, 1.75);
\filldraw [pattern = north east lines] (4, 0.75) rectangle (5, 1.5);
\filldraw [pattern = north east lines] (5, 0.75) rectangle (6, 1.2);
\filldraw [pattern = north east lines] (6, 0.75) rectangle (7, 1);
\draw [thick] (0, 0) rectangle (1, 5);
\draw [thick] (1, 0) rectangle (2, 3);
\draw [thick] (2, 0) rectangle (3, 2);
\draw [thick] (3, 0) rectangle (4, 1.75);
\draw [thick] (4, 0) rectangle (5, 1.5);
\draw [thick] (5, 0) rectangle (6, 1.2);
\draw [thick] (6, 0) rectangle (7, 1);
\draw [thick] (7, 0) rectangle (8, 0.5);
\draw [thick] (8, 0) rectangle (9, 0.3);
\draw [thick] (3, 0.75) -- (7, 0.75);
\draw [thick] (1, 3.725) -- (3, 3.725);
\draw [thick] (3, 2) -- (3, 3.725);
\foreach \i in {1,...,9}{\draw (\i - 0.5, 0) node [below = 0.75em] {\Large $\i$};}
\draw (0, 5) node [left = 0.75em] {$Q^{\downarrow}( 1 )$};
\draw [thick, dotted] (0, 3.725) node [left = 0.75em] {$\mathcal{V}( J = 2 )$} -- (1, 3.725);
\draw [thick, dotted] (0, 0.75) node [left = 0.75em] {$\mathcal{W}( K_{1} = 7 )$} -- (3, 0.75);
\draw [thick, dotted] (2, 3) -- (2, 3.725);
\draw [very thick, loosely dotted] (9.5, -0.5) -- (11, -0.5);
\draw [very thick, loosely dotted] (9.5, 0.25) -- (11, 0.25);
\draw [very thick] (-0.1, 5) -- (0.1, 5);
\draw [very thick] (-0.1, 3.725) -- (0.1, 3.725);
\draw [very thick] (-0.1, 0.75) -- (0.1, 0.75);
\path [ultra thick, ->] (4.5, 1.2) edge [bend right] (2.5, 3);
\draw (4.5, 5.5) node {$Q^{\downarrow}( 1 ) + \mathcal{V}( J ) + \mathcal{V}( J ) = 1 - \varepsilon$};
\path [->] (0.5, 4.5) edge [bend left] (2.25, 5.5);
\path [->] (1.5, 3.5) edge [bend left = 10] (3.75, 5.25);
\path [->] (2.5, 3.5) edge [bend right] (5, 5.25);
\end{tikzpicture}
\caption{Plot of making the extremal distribution of type-1 defined in \eqref{def:type1} from an $\mathcal{X}$-marginal $Q$, where $L = 3$. Each bar represents a probability mass with decreasing rearrangement $Q^{\downarrow}$.}
\label{fig:type1_form}
\end{figure}

\begin{theorem}
\label{th:main_list}
Suppose that $\varepsilon > 0$ and the cardinality of $\mathcal{Y}$ is at least countably infinite.
Then, it holds that
\begin{align}
\mathbb{H}_{\phi}(Q, L, \varepsilon, \mathcal{Y})
=
\phi( P_{\operatorname{type-1}} ) .
\label{eq:main_list}
\end{align}
\end{theorem}

\begin{IEEEproof}[Proof of \thref{th:main_list}]
See \sectref{sect:proof_main_list}.
\end{IEEEproof}

The Fano-type inequality stated in \eqref{eq:main_list} of \thref{th:main_list} is formulated by the extremal distribution $P_{\operatorname{type-1}}$ defined in \eqref{def:type1}.
The following proposition summarizes basic properties of $P_{\operatorname{type-1}}$.

\begin{proposition}
\label{prop:type1}
The extremal distribution of type-$1$ defined in \eqref{def:type1} satisfies the following,
\begin{itemize}
\item
the probability masses are nonincreasing in $x \in \mathcal{X}$, i.e.,
\begin{align}
P_{\operatorname{type-1}}( 1 )
\ge
P_{\operatorname{type-1}}( 2 )
\ge
P_{\operatorname{type-1}}( 3 )
\ge
P_{\operatorname{type-1}}( 4 )
\ge
P_{\operatorname{type-1}}( 5 )
\ge
\cdots ,
\label{eq:sort_type1}
\end{align}
\item
the sum of first $L$ probability masses of is equal to $1 - \varepsilon$, i.e.,
\begin{align}
\sum_{x = 1}^{L} P_{\operatorname{type-1}}( x )
=
1 - \varepsilon ,
\label{eq:type1_first-L-sum}
\end{align}
consequently, it holds that
\begin{align}
P_{\mathrm{e}}^{(L)}( P_{\operatorname{type-1}} )
=
\varepsilon ,
\label{eq:Pe_type1}
\end{align}
\item
the first $J - 1$ probability masses are equal to that of $Q^{\downarrow}$, i.e.,
\begin{align}
P_{\operatorname{type-1}}( x )
=
Q^{\downarrow}( x )
\qquad (\mathrm{for} \ 1 \le x \le J - 1) ,
\label{eq:type1_below-J}
\end{align}
\item
the probability masses for $J \le x \le L$ are equal to $\mathcal{V}( J )$, i.e.,
\begin{align}
P_{\operatorname{type-1}}( x )
=
\mathcal{V}( J )
\qquad (\mathrm{for} \ J \le x \le L) ,
\label{eq:type1-V}
\end{align}
\item
the probability masses for $L + 1 \le x \le K_{1}$ are equal to $\mathcal{W}( K_{1} )$, i.e.,
\begin{align}
P_{\operatorname{type-1}}( x )
=
\mathcal{W}( K_{1} )
\qquad (\mathrm{for} \ L+1 \le x \le K_{1}) ,
\label{eq:type1-W}
\end{align}
\item
the probability masses for $x \ge K_{1} + 1$ are equal to that of $Q^{\downarrow}$, i.e.,
\begin{align}
P_{\operatorname{type-1}}( x )
=
Q^{\downarrow}( x )
\qquad (\mathrm{for} \ x \ge K_{1} + 1) ,
\label{eq:type1_above-K}
\end{align}
and
\item
it holds that $P_{\operatorname{type-1}}$ majorizes $Q$.
\end{itemize}
\end{proposition}

\begin{IEEEproof}[Proof of \propref{prop:type1}]
See \appref{app:type1}.
\end{IEEEproof}

Although positive tolerated probabilities of error (i.e., $\varepsilon > 0$) are highly interesting in most of the lossless communication systems, the scenario in which the error events with positive probabilities are not allowed (i.e., $\varepsilon = 0$) is also important to deal with the error-free communication systems.
The following theorem is an error-free version of \thref{th:main_list}.

\begin{theorem}
\label{th:main_zero}
Suppose that $\varepsilon = 0$ and $\mathcal{Y}$ is at least countably infinite.
Then, it holds that
\begin{align}
\mathbb{H}_{\phi}(Q, L, 0, \mathcal{Y})
\le
\phi( P_{\operatorname{type-1}} )
\label{eq:main_zero}
\end{align}
with equality if $\supp( Q ) \coloneqq \{ x \in \mathcal{X} \mid Q( x ) > 0 \}$ is finite or $J = L$.
Moreover, if the cardinality of $\mathcal{Y}$ is at least the cardinality of the continuum $\mathbb{R}$, then there exists a $\sigma$-algebra on $\mathcal{Y}$ satisfying \eqref{eq:main_zero} with equality.
\end{theorem}

\begin{IEEEproof}[Proof of \thref{th:main_zero}]
See \sectref{sect:proof_main_zero}.
\end{IEEEproof}

\begin{remark}
Note that $J = L$ holds under the unique decoding rule (i.e., $L = 1$); that is, we see from \thref{th:main_zero} that \eqref{eq:main_zero} holds with equality if $L = 1$.
The inequality $J < L$ occurs only if a non-unique decoding rule (i.e., $L > 1$) is considered.
In \thref{th:main_zero}, the existence of a $\sigma$-algebra on an uncountably infinite alphabet $\mathcal{Y}$ in which \eqref{eq:main_zero} holds with equality is due to R\'{e}v\'{e}sz's generalization of the Birkhoff--von~Neumann decomposition via Kolmogorov's extension theorem; see Sections~\ref{sect:proof_main_list} and~\ref{sect:proof_main_zero} for technical details.
\end{remark}

Consider the case where $\mathcal{Y}$ is a finite alphabet.
Define the \emph{extremal distribution of type-$2$} as the following $\mathcal{X}$-marginal,
\begin{align}
P_{\operatorname{type-2}}( x )
=
P_{\operatorname{type-2}}^{(Q, L, \varepsilon, \mathcal{Y})}( x )
\coloneqq
\begin{dcases}
Q^{\downarrow}( x )
& \mathrm{if} \ 1 \le x < J \ \mathrm{or} \ K_{2} < x < \infty ,
\\
\mathcal{V}( J )
& \mathrm{if} \ J \le x \le L ,
\\
\mathcal{W}( K_{2} )
& \mathrm{if} \ L < x \le K_{2}
\end{dcases}
\label{def:type2}
\end{align}
for each $x \in \mathcal{X}$, where the three quantities $\mathcal{V}( \cdot )$, $\mathcal{W}( \cdot )$, and $J$ are defined in \eqref{def:V}, \eqref{def:W2}, and \eqref{def:J}, respectively, and $K_{2}$ is chosen so that
\begin{align}
K_{2}
=
K_{2}(Q, L, \varepsilon, \mathcal{Y})
& \coloneqq
\max\{ L \le  k \le L \cdot |\mathcal{Y}| \mid \mathcal{W}( k ) < Q^{\downarrow}( k ) \} .
\label{def:K4}
\end{align}
Moreover, define the integer $D$ by
\begin{align}
D
=
D(Q, L, \varepsilon, \mathcal{Y})
\coloneqq
\min \bigg\{ \binom{K_{2} - J + 1}{L - J + 1}, (K_{2} - J)^{2} + 1 \bigg\} ,
\label{def:D2}
\end{align}
where $\binom{a}{b} \coloneqq \frac{ a! }{ b! (a - b)! }$ stands for the binomial coefficient for two integers $0 \le b \le a$.
A graphical representation of $P_{\operatorname{type-2}}$ is illustrated in~\figref{fig:type2_form}.
When $\mathcal{Y}$ is finite, the Fano-type inequality stated in Theorems~\ref{th:main_list} and~\ref{th:main_zero} can be tightened as follows:

\begin{theorem}
\label{th:main_finite}
Suppose that $\mathcal{Y}$ is finite.
Then, it holds that
\begin{align}
\mathbb{H}_{\phi}(Q, L, \varepsilon, \mathcal{Y})
\le
\phi( P_{\operatorname{type-2}} )
\label{eq:main_finite}
\end{align}
with equality if $\varepsilon = P_{\mathrm{e}}^{(L)}( Q )$ or $|\mathcal{Y}| \ge D$.
\end{theorem}

\begin{IEEEproof}[Proof of \thref{th:main_finite}]
See \sectref{sect:proof_main_finite}.
\end{IEEEproof}

\begin{figure}[!t]
\centering
\begin{tikzpicture}
\draw [very thick, -latex] (0, -1) -- (0, 6);
\draw [very thick] (-1, 0) -- (12, 0);
\draw [fill, gray!90] (1, 3) rectangle (2, 3.725);
\draw [fill, gray!90] (2, 2) rectangle (3, 3.725);
\filldraw [pattern = north east lines] (3, 2/3) rectangle (4, 1.75);
\filldraw [pattern = north east lines] (4, 2/3) rectangle (5, 1.5);
\filldraw [pattern = north east lines] (5, 2/3) rectangle (6, 1.2);
\draw [thick] (0, 0) rectangle (1, 5);
\draw [thick] (1, 0) rectangle (2, 3);
\draw [thick] (2, 0) rectangle (3, 2);
\draw [thick] (3, 0) rectangle (4, 1.75);
\draw [thick] (4, 0) rectangle (5, 1.5);
\draw [thick] (5, 0) rectangle (6, 1.2);
\draw [thick] (6, 0) rectangle (7, 1);
\draw [thick] (7, 0) rectangle (8, 0.5);
\draw [thick] (8, 0) rectangle (9, 0.3);
\draw [thick] (3, 2/3) -- (6, 2/3);
\draw [thick] (1, 3.725) -- (3, 3.725);
\draw [thick] (3, 2) -- (3, 3.725);
\foreach \i in {1,...,9}{\draw (\i - 0.5, 0) node [below = 0.75em] {\Large $\i$};}
\draw (0, 5) node [left = 0.75em] {$Q^{\downarrow}( 1 )$};
\draw [thick, dotted] (0, 3.725) node [left = 0.75em] {$\mathcal{V}( J = 2 )$} -- (1, 3.725);
\draw [thick, dotted] (0, 2/3) node [left = 0.75em] {$\mathcal{W}( K_{2} = 6 )$} -- (3, 2/3);
\draw [thick, dotted] (2, 3) -- (2, 3.725);
\draw [very thick, loosely dotted] (9.5, -0.5) -- (11, -0.5);
\draw [very thick, loosely dotted] (9.5, 0.25) -- (11, 0.25);
\draw [very thick] (-0.1, 5) -- (0.1, 5);
\draw [very thick] (-0.1, 3.725) -- (0.1, 3.725);
\draw [very thick] (-0.1, 2/3) -- (0.1, 2/3);
\path [ultra thick, ->] (4.5, 1.2) edge [bend right] (2.5, 3);
\draw (4.5, 5.5) node {$Q^{\downarrow}( 1 ) + \mathcal{V}( J ) + \mathcal{V}( J ) = 1 - \varepsilon$};
\path [->] (0.5, 4.5) edge [bend left] (2.25, 5.5);
\path [->] (1.5, 3.5) edge [bend left = 10] (3.75, 5.25);
\path [->] (2.5, 3.5) edge [bend right] (5, 5.25);
\end{tikzpicture}
\caption{Plot of making the extremal distribution of type-2 defined in \eqref{def:type2} from an $\mathcal{X}$-marginal $Q$, where $L = 3$ and $|\mathcal{Y}| = 2$.
Each bar represents a probability mass of the decreasing rearrangement $Q^{\downarrow}$.}
\label{fig:type2_form}
\end{figure}

Similar to Theorems~\ref{th:main_list} and~\ref{th:main_zero}, the Fano-type inequality stated in \eqref{eq:main_finite} of \thref{th:main_finite} is formulated by the extremal distribution $P_{\operatorname{type-2}}$ defined in \eqref{def:type2}.
The difference between $P_{\operatorname{type-1}}$ and $P_{\operatorname{type-2}}$ is only the difference between $K_{1}$ and $K_{2}$ defined in \eqref{def:K3} and \eqref{def:K4}, respectively.

\begin{remark}
\label{rem:finite}
In contrast to Theorems~\ref{th:main_list} and~\ref{th:main_zero}, \thref{th:main_finite} holds in both cases: $\varepsilon > 0$ and $\varepsilon = 0$.
By \lemref{lem:cond_to_marg} stated in \sectref{sect:proof_main_list}, it can be verified that $P_{\operatorname{type-2}}$ majorizes $P_{\operatorname{type-1}}$,
and it follows from \propref{prop:Schur_convex} that
\begin{align}
\phi( P_{\operatorname{type-2}} )
\le
\phi( P_{\operatorname{type-1}} ) .
\end{align}
Namely, the Fano-type inequalities stated in Theorems~\ref{th:main_list} and~\ref{th:main_zero} also holds for finite $\mathcal{Y}$.
In other words, it holds that
\begin{align}
\mathbb{H}_{\phi}(Q, L, \varepsilon, \mathcal{Y})
\le
\phi( P_{\mathrm{type-1}} )
\end{align}
for every nonempty alphabet $\mathcal{Y}$, provided that $(Q, L, \varepsilon, \mathcal{Y})$ satisfies \eqref{eq:range_epsilon_list}.
As $|\mathcal{Y}| \ge D$ if $L = 1$ (see \eqref{def:D2}), another benefit of \thref{th:main_finite} is that the Fano-type inequality is always sharp under a unique decoding rule (i.e., $L = 1$).
\end{remark}

So far, it is assumed that the probability law $P_{X}$ of the $\mathcal{X}$-valued r.v.\ $X$ is fixed to a given $\mathcal{X}$-marginal $Q$.
When we assume that $X$ is supported on a finite subalphabet of $\mathcal{X}$, we can loosen and simplify our Fano-type inequalities by removing the constraint that $P_{X} = Q$.
Let $L$ and $M$ be two integers satisfying $1 \le L < M$, $\varepsilon$ a real number satisfying $0 \le \varepsilon \le 1 - L/M$, and $\mathcal{Y}$ a nonempty alphabet.
Consider the following maximization,
\begin{align}
\mathbb{H}_{\phi}(M, L, \varepsilon, \mathcal{Y})
\coloneqq
\max_{(X, Y) : P_{\mathrm{e}}^{(L)}(X \mid Y) \le \varepsilon} \mathsf{H}_{\phi}(X \mid Y) ,
\label{def:H_M}
\end{align}
where the maximization is taken over the pairs $(X, Y)$ of r.v.'s satisfying
(i) $X$ is $\{ 1, \dots, M \}$-valued,
(ii) $Y$ is $\mathcal{Y}$-valued, and
(iii) $P_{\mathrm{e}}^{(L)}(X \mid Y) \le \varepsilon$.

\begin{theorem}
\label{th:main_M}
It holds that
\begin{align}
\mathbb{H}_{\phi}(M, L, \varepsilon, \mathcal{Y})
=
\phi( P_{\operatorname{type-0}} ) ,
\end{align}
where $P_{\operatorname{type-0}}$ is defined in \eqref{def:type0}.
\end{theorem}

\begin{IEEEproof}[Proof of \thref{th:main_M}]
See \sectref{sect:main_M}.
\end{IEEEproof}

\begin{remark}
Although Theorems~\ref{th:main_list}--\ref{th:main_finite} depend on the cardinality of $\mathcal{Y}$, the Fano-type inequality stated in \thref{th:main_M} does not depend on it whenever $\mathcal{Y}$ is nonempty.
\end{remark}

\section{Special Cases: Fano-Type Inequalities on Shannon's and R\'{e}nyi's Information Measures}
\label{sect:Fano_Renyi}

In this section, we specialize our Fano-type inequalities stated in Theorems~\ref{th:main_list}--\ref{th:main_M} from general conditional information measures $\mathsf{H}_{\phi}(X \mid Y)$ to Shannon's and R\'{e}nyi's information measures.
We then recover several known results such as those in \cite{fano_1952, erokhin_1958, ho_verdu_2010, sakai_iwata_isit2017, sason_verdu_2017} along the way.

\subsection{On Shannon's Information Measures}
\label{sect:Shannon}

The conditional Shannon entropy \cite{shannon_1948} of an $\mathcal{X}$-valued r.v.\ $X$ given a $\mathcal{Y}$-valued r.v.\ $Y$ is defined by
\begin{align}
H(X \mid Y)
\coloneqq
\mathbb{E}[ H( P_{X|Y} ) ]
=
\mathbb{E}\left[ \sum_{x \in \mathcal{X}} P_{X|Y}( x ) \log \frac{ 1 }{ P_{X|Y}( x ) } \right] ,
\label{def:cond_shannon}
\end{align}
where the (unconditional) Shannon entropy of an $\mathcal{X}$-marginal $P$ is defined by
\begin{align}
H( P )
\coloneqq
\sum_{x \in \mathcal{X}} P( x ) \log \frac{ 1 }{ P( x ) } .
\end{align}

\begin{remark}
It can be verified by the monotone convergence theorem (cf.\ \cite[Theorem~10.1.7]{dudley_2002}) that
\begin{align}
H(X \mid Y)
=
\mathbb{E}\left[ \log \frac{ 1 }{ P_{X|Y}( X ) } \right] ,
\label{eq:MCT}
\end{align}
provided that the right-hand side of \eqref{eq:MCT} is finite.
In some cases, it is convenient to define the conditional Shannon entropy $H(X \mid Y)$ by the right-hand side of \eqref{eq:MCT} (see, e.g., \cite{sakai_tan_2019_source-dispersion}).
\end{remark}

The following proposition is a well-known property of Shannon's information measures.

\begin{proposition}[Tops{\o}e \cite{topsoe_2001}]
\label{prop:shannon}
The Shannon entropy $H( \cdot )$ is symmetric, concave, and lower semicontinuous.
\end{proposition}

Namely, the conditional Shannon entropy $H(X \mid Y)$ is a special case of $\mathsf{H}_{\phi}(X \mid Y)$ with $\phi = H$.
Therefore, defining the quantity
\begin{align}
\mathbb{H}(Q, L, \varepsilon, \mathcal{Y})
\coloneqq
\mathbb{H}_{H}(Q, L, \varepsilon, \mathcal{Y})
=
\sup_{(X, Y) : P_{\mathrm{e}}^{(L)}(X \mid Y) \le \varepsilon, P_{X} = Q} H(X \mid Y) ,
\label{def:Shannon-Fano}
\end{align}
we readily observe the following corollary.

\begin{corollary}
\label{cor:Shannon_list}
Suppose that $\varepsilon > 0$ and the cardinality of $\mathcal{Y}$ is at least countably infinite.
Then, it holds that
\begin{align}
\mathbb{H}(Q, L, \varepsilon, \mathcal{Y})
& =
H( P_{\operatorname{type-1}} )
\notag \\
& =
(J-L+1) \, \mathcal{V}(J) \log \frac{ 1 }{ \mathcal{V}(J) } + (K_{1} - L) \, \mathcal{W}(K_{1}) \log \frac{ 1 }{ \mathcal{W}(K_{1}) } + \sum_{\substack{ x = 1 : \\ x < J \, \mathrm{or} \, x > K_{1} }}^{\infty}  Q^{\downarrow}( x ) \log \frac{ 1 }{ Q^{\downarrow}( x ) } .
\label{eq:Shannon_list}
\end{align}
\end{corollary}

\begin{IEEEproof}[Proof of \corref{cor:Shannon_list}]
\corref{cor:Shannon_list} is a direct consequence of \thref{th:main_list} and \propref{prop:shannon}.
\end{IEEEproof}

\begin{remark}
Applying \thref{th:main_zero} instead of \thref{th:main_list}, an error-free version (i.e., $\varepsilon = 0$) of \corref{cor:Shannon_list} can be considered.
\end{remark}

\begin{remark}
\label{rem:Erokhin}
Note that \corref{cor:Shannon_list} coincides with Theorem~1 of \cite{ho_verdu_2010} if $L = 1$ and $\mathcal{Y} = \mathcal{X}$.
Moreover, we observe from \eqref{eq:Erokhin_type2} and \corref{cor:Shannon_list} that
\begin{align}
\mathbb{I}(Q, \varepsilon)
& =
H( Q ) - \mathbb{H}(Q, 1, \varepsilon, \mathcal{X})
\notag \\
& =
\sum_{x = 1}^{K_{1}} Q^{\downarrow}( x ) \log \frac{ 1 }{ Q^{\downarrow}( x ) } + \mathcal{V}(1) \log \mathcal{V}(1) + (K_{1} - 1) \, \mathcal{W}(K_{1}) \log \mathcal{W}(K_{1})
\end{align}
for every $\mathcal{X}$-marginal $Q$ and every tolerated probability of error $0 \le \varepsilon \le 1 - Q^{\downarrow}( 1 )$, where Erokhin's function $\mathbb{I}(Q, \varepsilon)$ is defined in \eqref{eq:Erokhin}.
See \sectref{sect:Erokhin} for details of generalizing of Erokhin's function.
Kostina--Polyanskiy--Verd\'{u} showed in Theorem~4 and Remark~3 of \cite{kostina_polyanskiy_verdu_2015} that
\begin{align}
\mathbb{I}(Q^{n}, \varepsilon)
=
n \, (1 - \varepsilon) \, H(Q) - \sqrt{ \frac{ n \, V(Q) }{ 2 \pi } } \, \mathrm{e}^{-\Phi^{-1}( \varepsilon )^{2}/2} + \mathrm{O}( \log n )
\qquad (\mathrm{as} \ n \to \infty) ,
\end{align}
where $V(P)$ is defined by
\begin{align}
V(P)
\coloneqq
\sum_{x \in \mathcal{X}} P( x ) \left( \log \frac{ 1 }{ P( x ) } - H( P ) \right)^{2}
\end{align}
and $\Phi^{-1}( \cdot )$ stands for the inverse of the Gaussian cumulative distribution function
\begin{align}
\Phi( u )
\coloneqq
\frac{ 1 }{ \sqrt{ 2 \pi } } \int_{-\infty}^{u} \, \mathrm{e}^{-t^{2}/2} \mathrm{d} t . 
\end{align}
\end{remark}

If $\mathcal{Y}$ is finite, then a tighter version of the Fano-type inequality than \corref{cor:Shannon_list} can be obtained as follows:

\begin{corollary}
\label{cor:Shannon_finite}
Suppose that $\mathcal{Y}$ is finite.
Then, it holds that
\begin{align}
\mathbb{H}(Q, L, \varepsilon, \mathcal{Y})
& \le
H( P_{\operatorname{type-2}} )
\notag \\
& =
(J-L+1) \, \mathcal{V}(J) \log \frac{ 1 }{ \mathcal{V}(J) } + (K_{2} - L) \, \mathcal{W}(K_{2}) \log \frac{ 1 }{ \mathcal{W}(K_{2}) } + \sum_{\substack{ x = 1 : \\ x < J \, \mathrm{or} \, x > K_{2} }}^{\infty}  Q^{\downarrow}( x ) \log \frac{ 1 }{ Q^{\downarrow}( x ) } ,
\label{eq:Shannon_finite}
\end{align}
with equality if $\varepsilon = P_{\mathrm{e}}^{(L)}( Q )$ or $|\mathcal{Y}| \ge D$.
\end{corollary}

\begin{IEEEproof}[Proof of \corref{cor:Shannon_finite}]
\corref{cor:Shannon_finite} is a direct consequence of \thref{th:main_finite} and \propref{prop:shannon}.
\end{IEEEproof}

\begin{remark}
The inequality in \eqref{eq:Shannon_finite} holds with equality if $L = 1$ (cf.\ \remref{rem:finite}).
In fact, when $L = 1$, \corref{cor:Shannon_finite} coincides with Ho--Verd\'{u}'s refinement of Erokhin's function $\mathbb{I}(Q, \varepsilon)$ with finite $\mathcal{Y}$ (see Theorem~4 of \cite{ho_verdu_2010}).
\end{remark}

Similar to \eqref{def:H_M} and \eqref{def:Shannon-Fano}, we can define
\begin{align}
\mathbb{H}(M, L, \varepsilon, \mathcal{Y})
\coloneqq
\mathbb{H}_{H}(M, L, \varepsilon, \mathcal{Y})
=
\max_{(X, Y) : P_{\mathrm{e}}^{(L)}(X \mid Y) \le \varepsilon} H(X \mid Y) ,
\end{align}
and can give an explicit formula of $\mathbb{H}(M, L, \varepsilon, \mathcal{Y})$ as follows.

\begin{corollary}
\label{cor:Shannon_M}
It holds that
\begin{align}
\mathbb{H}(M, L, \varepsilon, \mathcal{Y})
=
H( P_{\operatorname{type-0}} )
=
h_{2}( \varepsilon ) + (1 - \varepsilon) \log L + \varepsilon \log (M - L) .
\end{align}
\end{corollary}

\begin{IEEEproof}[Proof of \corref{cor:Shannon_M}]
\corref{cor:Shannon_M} is a direct consequence of \thref{th:main_M} and \propref{prop:shannon}.
\end{IEEEproof}

\begin{remark}
Indeed, \corref{cor:Shannon_M} states the classical Fano inequality with list decoding; see \eqref{eq:Fano_list}.
\end{remark}

\subsection{On R\'{e}nyi's Information Measures}
\label{sect:Renyi}

Although the choices of Shannon's information measures are unique based on a set of axioms (see, e.g., Theorem~3.6 of \cite{csiszar_korner_2011} and Chapter~3 of \cite{yeung_2008}), there are several different definitions of conditional R\'{e}nyi entropies (cf.\ \cite{fehr_berens_2014, iwamoto_shikata_2014, teixeira_matos_antunes_2012}).
Among them, this study focuses on \emph{Arimoto's} and \emph{Hayashi's} conditional R\'{e}nyi entropies \cite{arimoto_1977, hayashi_2011}.
Arimoto's conditional R\'{e}nyi entropy of $X$ given $Y$ is defined by
\begin{align}
H_{\alpha}^{\mathrm{Arimoto}}(X \mid Y)
\coloneqq
\frac{ \alpha }{ 1 - \alpha } \log \mathbb{E}[ \| P_{X|Y} \|_{\alpha} ]
=
\frac{ \alpha }{ 1 - \alpha } \log \mathbb{E}\left[ \left( \sum_{x \in \mathcal{X}} P_{X|Y}( x )^{\alpha} \right)^{1/\alpha} \right]
\label{def:arimoto}
\end{align}
for each order $\alpha \in (0, 1) \cup (1, \infty)$, where the $\ell_{\alpha}$-norm of an $\mathcal{X}$-marginal $P$ is defined by
\begin{align}
\| P \|_{\alpha}
\coloneqq
\left( \sum_{x \in \mathcal{X}} P( x )^{\alpha} \right)^{1/\alpha} .
\label{def:norm}
\end{align}
Here, note that the (unconditional) R\'{e}nyi entropy \cite{renyi_1961} of an $\mathcal{X}$-marginal $P$ can be defined by
\begin{align}
H_{\alpha}( P )
\coloneqq
\frac{ \alpha }{ 1 - \alpha } \log \| P \|_{\alpha}
=
\frac{ 1 }{ 1 - \alpha } \log \sum_{x \in \mathcal{X}} P( x )^{\alpha} ,
\label{def:renyi}
\end{align}
i.e., it is a monotone function of the $\ell_{\alpha}$-norm.
Basic properties of the $\ell_{\alpha}$-norm can be found in the following proposition.

\begin{proposition}
\label{prop:norm}
The $\ell_{\alpha}$-norm $\| \cdot \|_{\alpha}$ is symmetric and lower semicontinuous.
Moreover, it is concave (resp.\ convex) if $0 < \alpha \le 1$ (resp.\ if $\alpha \ge 1$).
\end{proposition}

\begin{IEEEproof}[Proof of \propref{prop:norm}]
The symmetry is obvious.
The lower semicontinuity was proven by Kova\v{c}evi\'{c}--Stanojevi\'{c}--\v{S}enk in Theorem~5 of \cite{kovacevic_stanojevic_senk_2013}.
The concavity (resp.\ convexity) property can be verified by the reverse (resp.\ forward) Minkowski inequality.
\end{IEEEproof}

\propref{prop:norm} implies that $H_{\alpha}^{\mathrm{Arimoto}}(X \mid Y)$ is a monotone function of $\mathsf{H}_{\phi}(X \mid Y)$ with $\phi = \| \cdot \|_{\alpha}$, i.e.,
\begin{align}
H_{\alpha}^{\mathrm{Arimoto}}(X \mid Y)
=
\frac{ \alpha }{ 1 - \alpha } \log \Big( \mathsf{H}_{\| \cdot \|_{\alpha}}(X \mid Y) \Big) .
\label{eq:arimoto_A}
\end{align}
On the other hand, Hayashi's conditional R\'{e}nyi entropy of $X$ given $Y$ is defined by
\begin{align}
H_{\alpha}^{\mathrm{Hayashi}}(X \mid Y)
\coloneqq
\frac{ 1 }{ 1 - \alpha } \log \mathbb{E}[ \| P_{X|Y} \|_{\alpha}^{\alpha} ]
=
\frac{ 1 }{ 1 - \alpha } \log \mathbb{E}\left[ \sum_{x \in \mathcal{X}} P_{X|Y}( x )^{\alpha} \right]
\end{align}
for each order $\alpha \in (0, 1) \cup (1, \infty)$.
It is easy to see that $\| \cdot \|_{\alpha}^{\alpha} : \mathcal{P}(\mathcal{X}) \to [0, \infty]$ also admits the same properties as those stated in~\propref{prop:norm}.
Therefore, Hayashi's conditional R\'{e}nyi entropy $H_{\alpha}^{\mathrm{Hayashi}}(X \mid Y)$ is also a monotone function of $\mathsf{H}_{\phi}(X \mid Y)$ with $\phi = \| \cdot \|_{\alpha}^{\alpha}$, i.e.,
\begin{align}
H_{\alpha}^{\mathrm{Hayashi}}(X \mid Y)
=
\frac{ 1 }{ 1 - \alpha } \log \Big( \mathsf{H}_{\| \cdot \|_{\alpha}^{\alpha}}(X \mid Y) \Big) .
\label{def:hayashi}
\end{align}
It can be verified by Jensen's inequality (see, e.g., Proposition~1 of \cite{iwamoto_shikata_2014}) that
\begin{align}
H_{\alpha}^{\mathrm{Hayashi}}(X \mid Y)
\le
H_{\alpha}^{\mathrm{Arimoto}}(X \mid Y) .
\label{eq:Jensen_HA}
\end{align}

Similar to \eqref{def:main_object}, we now define
\begin{align}
\mathbb{H}_{\alpha}^{\dagger}(Q, L, \varepsilon, \mathcal{Y})
\coloneqq
\sup_{(X, Y) : P_{\mathrm{e}}^{(L)}(X \mid Y) \le \varepsilon, P_{X} = Q} H_{\alpha}^{\dagger}(X \mid Y)
\label{def:Renyi-Fano}
\end{align}
for each $\dagger \in \{ \mathrm{Arimoto}, \mathrm{Hayashi} \}$ and each $\alpha \in (0, 1) \cup (1, \infty)$.
Then, we can establish the Fano-type inequality on R\'{e}nyi's information measures as follows.

\begin{corollary}
\label{cor:Fano_Renyi_5}
Suppose that $\varepsilon > 0$ and the cardinality of $\mathcal{Y}$ is at least countably infinite.
For every $\dagger \in \{ \mathrm{Arimoto}, \mathrm{Hayashi} \}$ and $\alpha \in (0, 1) \cup (1, \infty)$, it holds that
\begin{align}
\mathbb{H}_{\alpha}^{\dagger}(Q, L, \varepsilon, \mathcal{Y})
& =
H_{\alpha}( P_{\operatorname{type-1}} )
\notag \\
& =
\frac{ 1 }{ 1 - \alpha } \log \left( (J-L+1) \, \mathcal{V}(J)^{\alpha} + (K_{1} - L) \, \mathcal{W}(K_{1})^{\alpha} + \sum_{\substack{ x = 1 : \\ x < J \, \mathrm{or} \, x > K_{1} }}^{\infty} Q^{\downarrow}( x )^{\alpha} \right) .
\label{eq:Fano-type_arimoto_type5}
\end{align}
\end{corollary}

\begin{IEEEproof}[Proof of \corref{cor:Fano_Renyi_5}]
Let $\dagger = \mathrm{Arimoto}$.
It follows from \thref{th:main_list} and \propref{prop:norm} that
\begin{align}
0 < \alpha \le 1
\quad & \Longrightarrow \quad
\sup_{(X, Y) : P_{\mathrm{e}}^{(L)}(X \mid Y) \le \varepsilon, P_{X} = Q} \mathbb{E}\big[ \| P_{X|Y} \|_{\alpha} \big]
=
\| P_{\operatorname{type-1}} \|_{\alpha} ,
\label{eq:Fano-norm1} \\
\alpha \ge 1
\quad & \Longrightarrow \quad
\inf_{(X, Y) : P_{\mathrm{e}}^{(L)}(X \mid Y) \le \varepsilon, P_{X} = Q} \mathbb{E}\big[ \| P_{X|Y} \|_{\alpha} \big]
=
\| P_{\operatorname{type-1}} \|_{\alpha} .
\label{eq:Fano-norm2}
\end{align}
As the mapping $u \mapsto (\alpha / (1 - \alpha)) \log u$ is strictly increasing (resp.\ strictly decreasing) if $0 < \alpha < 1$ (resp.\ if $\alpha > 1$), it follows from \eqref{def:renyi}, \eqref{eq:arimoto_A}, \eqref{eq:Fano-norm1}, and \eqref{eq:Fano-norm2} that
\begin{align}
\sup_{(X, Y) : P_{\mathrm{e}}^{(L)}(X \mid Y) \le \varepsilon, P_{X} = Q} H_{\alpha}^{\mathrm{Arimoto}}(X \mid Y)
=
H_{\alpha}( P_{\operatorname{type-1}} ) .
\label{eq:Fano_Renyi_type5}
\end{align}
The proof for the case when $\dagger = \mathrm{Hayashi}$ is the same as above, proving \corref{cor:Fano_Renyi_5}.
\end{IEEEproof}

\begin{remark}
Applying \thref{th:main_zero} instead of \thref{th:main_list}, an error-free version (i.e., $\varepsilon = 0$) of \corref{cor:Fano_Renyi_5} can be considered.
\end{remark}

\begin{remark}
Although Hayashi's conditional R\'{e}nyi entropy is smaller than Arimoto's one in general (see \eqref{eq:Jensen_HA}), \corref{cor:Fano_Renyi_5} implies that the maximization problem $\mathbb{H}_{\alpha}^{\dagger}(Q, L, \varepsilon, \mathcal{Y})$ results in the same R\'{e}nyi entropy $H_{\alpha}( P_{\operatorname{type-1}} )$ for each $\dagger \in \{ \mathrm{Arimoto}, \mathrm{Hayashi} \}$.
\end{remark}

When $\mathcal{Y}$ is finite, a tighter Fano-type inequality than \corref{cor:Fano_Renyi_5} can be obtained as follows.

\begin{corollary}
\label{cor:Fano_Renyi_6}
Suppose that $\mathcal{Y}$ is finite.
For any $\dagger \in \{ \mathrm{Arimoto}, \mathrm{Hayashi} \}$ and $\alpha \in (0, 1) \cup (1, \infty)$, it holds that
\begin{align}
\mathbb{H}_{\alpha}^{\dagger}(Q, L, \varepsilon, \mathcal{Y})
& \le
H_{\alpha}( P_{\operatorname{type-2}} )
\notag \\
& =
\frac{ 1 }{ 1 - \alpha } \log \left( (J-L+1) \, \mathcal{V}(J)^{\alpha} + (K_{2} - L) \, \mathcal{W}(K_{2})^{\alpha} + \sum_{\substack{ x = 1 : \\ x < J \, \mathrm{or} \, x > K_{2} }}^{\infty} Q^{\downarrow}( x )^{\alpha} \right) ,
\label{eq:Fano-type_arimoto_type6}
\end{align}
with equality if $\varepsilon = P_{\mathrm{e}}^{(L)}( Q )$ or $|\mathcal{Y}| \ge D$.
\end{corollary}

\begin{IEEEproof}[Proof of \corref{cor:Fano_Renyi_6}]
The proof is the same as the proof of \corref{cor:Fano_Renyi_5} by replacing \thref{th:main_list} by \thref{th:main_finite}.
\end{IEEEproof}

Similar to \eqref{def:H_M} and \eqref{def:Renyi-Fano}, define
\begin{align}
\mathbb{H}_{\alpha}^{\dagger}(M, L, \varepsilon, \mathcal{Y})
\coloneqq
\max_{(X, Y) : P_{\mathrm{e}}^{(L)}(X \mid Y) \le \varepsilon} H_{\alpha}^{\dagger}(X \mid Y)
\end{align}
for each $\dagger \in \{ \mathrm{Arimoto}, \mathrm{Hayashi} \}$ and each $\alpha \in (0, 1) \cup (1, \infty)$.

\begin{corollary}
\label{cor:Renyi_M}
For every $\dagger \in \{ \mathrm{Arimoto}, \mathrm{Hayashi} \}$ and $\alpha \in (0, 1) \cup (1, \infty)$, it holds that
\begin{align}
\mathbb{H}_{\alpha}^{\dagger}(M, L, \varepsilon, \mathcal{Y})
=
H_{\alpha}( P_{\operatorname{type-0}} )
=
\frac{ 1 }{ 1 - \alpha } \log \Big( L^{1-\alpha} (1 - \varepsilon) + (M - L)^{1-\alpha} \varepsilon^{\alpha} \Big) .
\end{align}
\end{corollary}

\begin{IEEEproof}[Proof of \corref{cor:Renyi_M}]
The proof is the same as the proof of \corref{cor:Fano_Renyi_5} by replacing \thref{th:main_list} by \thref{th:main_M}.
\end{IEEEproof}

\begin{remark}
\label{rem:reduction2}
When $\dagger = \mathrm{Arimoto}$, \corref{cor:Renyi_M} coincides with Sason--Verd\'{u}'s generalization (cf.\ Theorem~8 of \cite{sason_verdu_2017}) of Fano's inequality for R\'{e}nyi's information measures with list decoding (see \eqref{eq:Sason-Verdu}).
\end{remark}

\begin{remark}
It follows by l'H\^{o}pital's rule that
\begin{align}
\lim_{\alpha \to 1} H_{\alpha}( P_{\operatorname{type-0}} )
& =
H( P_{\operatorname{type-0}} ) ,
\\
\lim_{\alpha \to 1} H_{\alpha}( P_{\operatorname{type-1}} )
& =
H( P_{\operatorname{type-1}} ) ,
\\
\lim_{\alpha \to 1} H_{\alpha}( P_{\operatorname{type-2}} )
& =
H( P_{\operatorname{type-2}} ) .
\end{align}
Therefore, our Fano-type inequalities stated in Corollaries~\ref{cor:Shannon_list}--\ref{cor:Renyi_M} satisfy the continuity of Shannon's and R\'{e}nyi's information measures with respect to the order $0 < \alpha < \infty$.
\end{remark}

\subsection{Generalization of Erokhin's Function to $\alpha$-Mutual Information}
\label{sect:Erokhin}

Erokhin's function $\mathbb{I}(Q, \varepsilon)$ defined in \eqref{eq:Erokhin} can be generalized to the $\alpha$-mutual information (cf.\ \cite{verdu_ita2015}) as follows:
Let $X$ be an $\mathcal{X}$-valued r.v.\ and $Y$ a $\mathcal{Y}$-valued r.v.
Sibson's $\alpha$-mutual information \cite{sibson_1969} (see also Equation~(32) of \cite{verdu_ita2015}, Equation~(13) of \cite{csiszar_1995}, and Definition~7 of \cite{ho_verdu_isit2015}) is defined by
\begin{align}
I_{\alpha}^{\mathrm{Sibson}}(X \wedge Y)
\coloneqq
\inf_{Q_{Y}} D_{\alpha}(P_{X, Y} \, \| \, P_{X} \times Q_{Y})
\label{def:alpha-mutual}
\end{align}
for each $0 < \alpha < \infty$, where $P_{X, Y}$ (resp.\ $P_{X}$) denotes the probability measure on $\mathcal{X} \times \mathcal{Y}$ (resp.\ $\mathcal{X}$) induced by the pair $(X, Y)$ of r.v.'s (resp.\ the r.v.\ $X$), the infimum is taken over the probability measures $Q_{Y}$ on $\mathcal{Y}$, and the R\'{e}nyi divergence \cite{renyi_1961} between two probability measures $\mu$ and $\nu$ on $\mathcal{A}$ is defined by
\begin{align}
D_{\alpha}(\mu \, \| \, \nu)
\coloneqq
\begin{dcases}
\frac{ 1 }{ \alpha - 1 } \log \left( \int_{\mathcal{A}} \left( \frac{ \mathrm{d} \mu }{ \mathrm{d} \nu } \right)^{\alpha} \, \mathrm{d} \nu \right)
& \mathrm{if} \ \mu \ll \nu \ \mathrm{and} \ \alpha \neq 1 ,
\\
\int_{\mathcal{A}} \left( \log \frac{ \mathrm{d} \mu }{ \mathrm{d} \nu } \right) \mathrm{d} \mu
& \mathrm{if} \ \mu \ll \nu \ \mathrm{and} \ \alpha = 1 ,
\\
\infty
& \mathrm{otherwise}
\end{dcases}
\end{align}
for each $0 < \alpha < \infty$.
Note that Sibson's $\alpha$-mutual information coincides with the ordinary mutual information when $\alpha = 1$, i.e., it holds that $I(X \wedge Y) = I_{1}(X \wedge Y)$.
Similar to \eqref{def:main_object} and \eqref{eq:Erokhin}, given a system $(Q, L, \varepsilon, \mathcal{Y})$ satisfying \eqref{eq:range_epsilon_list}, define
\begin{align}
\mathbb{I}_{\alpha}^{\mathrm{Sibson}}(Q, L, \varepsilon, \mathcal{Y})
\coloneqq
\inf_{(X, Y) : P_{\mathrm{e}}^{(L)}(X \mid Y) \le \varepsilon, P_{X} = Q} I_{\alpha}^{\mathrm{Sibson}}(X \wedge Y) ,
\label{def:Sibson-Erokhin}
\end{align}
where the infimum is taken over the pairs of r.v.'s $X$ and $Y$ in which (i) $X$ is $\mathcal{X}$-valued, (ii) $Y$ is $\mathcal{Y}$-valued, (iii) $P_{\mathrm{e}}^{(L)}(X \mid Y) \le \varepsilon$, and (iv) $P_{X} = Q$.
By convention, we denote by
\begin{align}
\mathbb{I}(Q, L, \varepsilon, \mathcal{Y})
\coloneqq
\mathbb{I}_{1}^{\mathrm{Sibson}}(Q, L, \varepsilon, \mathcal{Y}) .
\end{align}
It is clear that this definition can be specialized to Erokhin's function $\mathbb{I}(Q, \varepsilon)$ defined in \eqref{eq:Erokhin};
in other words, it holds that
\begin{align}
\mathbb{I}(Q, 1, \varepsilon, \mathcal{X})
=
\mathbb{I}(Q, \varepsilon) ;
\end{align}
see \remref{rem:Erokhin}.

\begin{corollary}[When $\alpha = 1$]
\label{cor:Shannon-Erokhin}
Suppose that $\varepsilon > 0$ and the cardinality of $\mathcal{Y}$ is at least countably infinite. Then, it holds that
\begin{align}
\mathbb{I}(Q, L, \varepsilon, \mathcal{Y})
& =
H( Q ) - \mathbb{H}(Q, L, \varepsilon, \mathcal{Y})
\notag \\
& =
\sum_{x = J}^{K_{1}} Q^{\downarrow}( x ) \log \frac{ 1 }{ Q^{\downarrow}( x ) } + (J-L+1) \, \mathcal{V}(J) \log \mathcal{V}(J) + (K_{1} - L) \, \mathcal{W}(K_{1}) \log \mathcal{W}(K_{1}) .
\label{eq:Shannon-Erokhin}
\end{align}
\end{corollary}

\begin{IEEEproof}[Proof of \corref{cor:Shannon-Erokhin}]
The equality in \eqref{eq:Shannon-Erokhin} is trivial from the well-known identity $I(X \wedge Y) = H(X) - H(X \mid Y)$.
The inequality in \eqref{eq:Shannon-Erokhin} follows from \corref{cor:Shannon_list}, completing the proof.
\end{IEEEproof}

\begin{corollary}[Sibson, when $\alpha \neq 1$]
\label{cor:Sibson-Erokhin}
Suppose that $\varepsilon > 0$ and $\mathcal{Y}$ is countably infinite.
For every $\alpha \in (0, 1) \cup (1, \infty)$, it holds that
\begin{align}
\mathbb{I}_{\alpha}^{\mathrm{Sibson}}(Q, L, \varepsilon, \mathcal{Y})
& =
H_{\alpha}( Q^{(1/\alpha)} ) - \mathbb{H}_{\alpha}^{\mathrm{Arimoto}}(Q^{(1/\alpha)}, L, \varepsilon, \mathcal{Y})
\notag \\
& =
\frac{ 1 }{ \alpha - 1 } \log \left( 1 - \sum_{x = J^{(1/\alpha)}}^{K_{1}^{(1/\alpha)}} Q^{\downarrow}( x ) \right.
\notag \\
& \quad\left. \vphantom{\sum_{\substack{ x = 1 : \\ x < J \, \mathrm{or} \, x > K_{1} }}^{\infty}}
{} + \left( (J^{(1/\alpha)}-L+1) \, \mathcal{V}^{(1/\alpha)}(J^{(1/\alpha)})^{\alpha} + (K_{1}^{(1/\alpha)} - L) \, \mathcal{W}^{(1/\alpha)}(K_{1}^{(1/\alpha)})^{\alpha} \right) \left( \sum_{x \in \mathcal{X}} Q( x )^{1/\alpha} \right)^{\alpha} \right) ,
\label{eq:Sibson-Erokhin}
\end{align}
where $Q^{(s)}$ stands for the $s$-tilted distribution of $Q$ with real parameter $0 < s < \infty$, i.e.,
\begin{align}
Q^{(s)}( x )
\coloneqq
\frac{ Q( x )^{s} }{ \sum_{x^{\prime} \in \mathcal{X}} Q( x^{\prime} )^{s} }
\end{align}
for each $x \in \mathcal{X}$, and $\mathcal{V}^{(s)}( \cdot )$, $W^{(s)}( \cdot )$, $J^{(s)}$, and $K_{1}^{(s)}$ are defined as in \eqref{def:V}, \eqref{def:W2}, \eqref{def:J}, and \eqref{def:K3}, respectively, by replacing the $\mathcal{X}$-marginal $Q$ by the $s$-tilted distribution $Q^{(s)}$.
\end{corollary}

\begin{IEEEproof}[Proof of \corref{cor:Sibson-Erokhin}]
As Sibson's identity \cite{sibson_1969} (see also \cite{csiszar_1995}, Equation~(12)) states that
\begin{align}
D_{\alpha}( P_{X, Y} \, \| \, P_{X} \times Q_{Y} )
=
D_{\alpha}( P_{X, Y} \, \| \, P_{X} \times Q_{\alpha} ) + D_{\alpha}( Q_{\alpha} \, \| \, Q_{Y} ) ,
\end{align}
where $Q_{\alpha}$ stands for the probability distribution on $\mathcal{Y}$ given as
\begin{align}
Q_{\alpha}( y )
=
\left( \sum_{x \in \mathcal{X}} P_{X, Y}(x, y)^{\alpha} \, P_{X}( x )^{1-\alpha} \right)^{1/\alpha} \left( \sum_{y^{\prime} \in \mathcal{Y}} \left( \sum_{x^{\prime} \in \mathcal{X}} P_{X, Y}(x^{\prime}, y^{\prime})^{\alpha} \, P_{X}( x^{\prime} )^{1-\alpha} \right)^{1/\alpha} \right)^{-1}
\end{align}
for each $y \in \mathcal{Y}$, we observe that
\begin{align}
I_{\alpha}^{\mathrm{Sibson}}(X \wedge Y)
=
\frac{ \alpha }{ \alpha - 1 } \log \sum_{y \in \mathcal{Y}} \left( \sum_{x \in \mathcal{X}} P_{X, Y}(x, y)^{\alpha} \, P_{X}( x )^{1-\alpha} \right)^{1/\alpha}
\label{eq:alpha-mutual_formula1}
\end{align}
for every $\alpha \in (0, 1) \cup (1, \infty)$, provided that $\mathcal{Y}$ is countable.
On the other hand, it follows from \cite[Equation~(13)]{arimoto_1977} that
\begin{align}
I_{\alpha}^{\mathrm{Arimoto}}(X \wedge Y)
=
\frac{ \alpha }{ \alpha - 1 } \log \sum_{y \in \mathcal{Y}} \left( \sum_{x \in \mathcal{X}} \frac{ P_{X, Y}(x, y)^{\alpha} }{ \sum_{x^{\prime} \in \mathcal{X}} P_{X}( x^{\prime} )^{\alpha} } \right)^{1/\alpha}
\label{eq:alpha-mutual_formula2}
\end{align}
for every $\alpha \in (0, 1) \cup (1, \infty)$, provided that $\mathcal{Y}$ is countable.
Combining \eqref{eq:alpha-mutual_formula1} and \eqref{eq:alpha-mutual_formula2}, we have the first equality in \eqref{eq:Sibson-Erokhin}.
Finally, the second equality in \eqref{eq:Sibson-Erokhin} follows from \corref{cor:Fano_Renyi_5} after some algebra.
This completes the proof of \corref{cor:Sibson-Erokhin}.
\end{IEEEproof}

In contrast to \eqref{def:alpha-mutual}, Arimoto defined the $\alpha$-mutual information (\cite{arimoto_1977}, Equation~(15)) by
\begin{align}
I_{\alpha}^{\mathrm{Arimoto}}(X \wedge Y)
\coloneqq
H_{\alpha}( X ) - H_{\alpha}^{\mathrm{Arimoto}}(X \mid Y)
\end{align}
for every $\alpha \in (0, 1) \cup (1, \infty)$.
Similar to \eqref{def:Sibson-Erokhin}, one can define
\begin{align}
\mathbb{I}_{\alpha}^{\mathrm{Arimoto}}(Q, L, \varepsilon, \mathcal{Y})
& \coloneqq
\inf_{(X, Y) : P_{\mathrm{e}}^{(L)}(X \mid Y) \le \varepsilon, P_{X} = Q} I_{\alpha}^{\mathrm{Arimoto}}(X \wedge Y) ,
\end{align}
and a counterpart of \corref{cor:Sibson-Erokhin} can be stated as follows.

\begin{corollary}[Arimoto, when $\alpha \neq 1$]
\label{cor:Arimoto-Erokhin}
Suppose that $\varepsilon > 0$ and the cardinality of $\mathcal{Y}$ is at least countably infinite.
For every $\alpha \in (0, 1) \cup (1, \infty)$, it holds that
\begin{align}
\mathbb{I}_{\alpha}^{\mathrm{Arimoto}}(Q, L, \varepsilon, \mathcal{Y})
& =
H_{\alpha}( Q ) - \mathbb{H}_{\alpha}^{\mathrm{Arimoto}}(Q, L, \varepsilon, \mathcal{Y})
\notag \\
& =
\frac{ 1 }{ \alpha - 1 } \log \left( 1 - \sum_{x = J}^{K_{1}} (Q^{(\alpha)})^{\downarrow}( x )
\right. \notag \\
& \left. \qquad \qquad \qquad
{} + \Big( (J-L+1) \, \mathcal{V}(J)^{\alpha} + (K_{1} - L) \, \mathcal{W}(K_{1})^{\alpha} \Big) \left( \sum_{x \in \mathcal{X}} Q( x )^{\alpha} \right)^{-1} \right) .
\label{eq:Arimoto-Erokhin}
\end{align}
\end{corollary}

\begin{IEEEproof}[Proof of \corref{cor:Arimoto-Erokhin}]
The first equality in \eqref{eq:Arimoto-Erokhin} is obvious from the definition.
The second equality in \eqref{eq:Arimoto-Erokhin} follows from \corref{cor:Fano_Renyi_5} after some algebra, completing the proof.
\end{IEEEproof}

When $\mathcal{Y}$ is finite, then the inequalities stated in Corollaries~\ref{cor:Shannon-Erokhin}--\ref{cor:Arimoto-Erokhin} can be tightened by \thref{th:main_finite} as in Corollaries~\ref{cor:Shannon_finite} and~\ref{cor:Fano_Renyi_6}.
We omit to explicitly state these tightened inequalities in this paper.

\section{Asymptotic Behaviors on Equivocations}
\label{sect:vanishing}

In information theory, the \emph{equivocation} or the \emph{remaining uncertainty} of an r.v.\ $X$ relative to a correlated r.v.\ $Y$ has an important role in establishing  fundamental limits of the optimal transmission ratio and/or rate in several communication models.
Shannon's equivocation $H(X \mid Y)$ is  a well-known measure in the formulation of the notion of  perfect secrecy of symmetric-key encryption in information-theoretic cryptography \cite{shannon_1949}. Iwamoto--Shikata \cite{iwamoto_shikata_2014} considered the extension of such a secrecy criterion by generalizing Shannon's equivocation  to R\'{e}nyi's equivocation by showing various desired properties of the latter.
Recently, Hayashi--Tan \cite{hayashi_tan_2017} and Tan--Hayashi \cite{tan_hayashi_2018} studied the asymptotics of Shannon's and R\'{e}nyi's equivocations when the side-information about the source is given via a various class of random hash functions with a fixed rate.

In this section, we assume that certain error probabilities vanish and we then establish asymptotic behaviors on Shannon's, or sometimes on R\'{e}nyi's, equivocations via the Fano-type inequalities stated in \sectref{sect:Fano_Renyi}.

\subsection{Fano's Inequality Meets  the AEP}
\label{sect:normalized}

We consider a  general form of the asymptotic equipartition property (AEP) as follows.

\begin{definition}[{\cite{verdu_han_1997}}]
\label{def:AEP}
We say that a sequence of $\mathcal{X}$-valued r.v.'s $\mathbf{X} = \{ X_{n} \}_{n = 1}^{\infty}$ satisfies the \emph{AEP} if
\begin{align}
\lim_{n \to \infty} \mathbb{P} \bigg\{ \log \frac{ 1 }{ P_{X_{n}}( X_{n} ) } \le (1 - \delta) \, H( X_{n} ) \bigg\}
=
0
\label{eq:AEP}
\end{align}
for every fixed $\delta > 0$.
\end{definition}

In the literature, the r.v.\ $X_{n}$ is  commonly represented as a random vector $X_{n} = (Z_{1}^{(n)}, \dots, Z_{n}^{(n)})$. The formulation without reference to random vectors means that $\mathbf{X} = \{ X_{n} \}_{n=1}^{\infty}$ is a \emph{general source} in the sense of Page~100 of \cite{han_InformationSpectrum}.

Let $\{ L_{n} \}_{n = 1}^{\infty}$ be a sequence of positive integers, $\{ \mathcal{Y}_{n} \}_{n = 1}^{\infty}$   a sequence of nonempty alphabets, and $\{ (X_{n}, Y_{n}) \}_{n = 1}^{\infty}$ a sequence of pairs of r.v.'s, where $X_{n}$ (resp.\ $Y_{n}$) is $\mathcal{X}$-valued (resp.\ $\mathcal{Y}_{n}$-valued) for each $n \ge 1$.
As
\begin{align}
\lim_{n \to \infty} \mathbb{P}\{ X_{n} \notin f_{n}(Y_{n}) \} = 0
\quad \Longrightarrow \quad
\lim_{n \to \infty} P_{\mathrm{e}}^{(L_{n})}(X_{n} \mid Y_{n}) = 0
\end{align}
for \emph{any} sequence of list decoders $\{ f_{n} : \mathcal{Y} \to \binom{\mathcal{X}}{L_{n}} \}_{n = 1}^{\infty}$, it suffices to assume that $P_{\mathrm{e}}^{(L_{n})}(X_{n} \mid Y_{n}) = \mathrm{o}( 1 )$ as $n \to \infty$ in our analysis.
The following theorem is a novel characterization of the AEP via Fano's inequality.

\begin{theorem}
\label{th:Fano_meets_AEP}
Suppose that a general source $\mathbf{X} = \{ X_{n} \}_{n = 1}^{\infty}$ satisfies the AEP, and $H( X_{n} ) = \Omega( 1 )$ as $n \to \infty$.
Then, it holds that
\begin{align}
\lim_{n \to \infty} P_{\mathrm{e}}^{(L_{n})}(X_{n} \mid Y_{n}) = 0
\quad \Longrightarrow \quad
\big| H(X_{n} \mid Y_{n}) - \log L_{n} \big|^{+} = \mathrm{o}\big( H(X_{n}) \big) ,
\label{eq:vanishing_normalizedRenyi_list1}
\end{align}
where $| u |^{+} \coloneqq \max\{ 0, u \}$ for $u \in \mathbb{R}$.
Consequently, it holds that
\begin{align}
\lim_{n \to \infty} P_{\mathrm{e}}^{(L_{n})}(X_{n} \mid Y_{n}) = \lim_{n \to \infty} \frac{ \log L_{n} }{ H( X_{n} ) } = 0
\quad \Longrightarrow \quad
\lim_{n \to \infty} \frac{ H(X_{n} \mid Y_{n}) }{ H( X_{n} ) } = 0 .
\label{eq:vanishing_normalizedRenyi_list2}
\end{align}
\end{theorem}

\begin{IEEEproof}[Proof of \thref{th:Fano_meets_AEP}]
See \appref{sect:Fano_meets_AEP}.
\end{IEEEproof}

The following three examples are particularizations of \thref{th:Fano_meets_AEP}.

\begin{example}
\label{ex:iid_case}
Let $\{ Z_{n} \}_{n = 1}^{\infty}$ be an i.i.d.\ source on a countably infinite alphabet $\mathcal{X}$ with finite Shannon entropy $H( Z_{1} ) < \infty$.
Suppose that $X_{n} = (Z_{1}, \dots, Z_{n})$ and $\mathcal{Y}_{n} = \mathcal{X}^{n}$ for each $n \ge 1$.
Then, \thref{th:Fano_meets_AEP} states that
\begin{align}
\lim_{n \to \infty} \mathbb{P}\{ X_{n} \neq Y_{n} \} = 0
\quad \Longrightarrow \quad
\lim_{n \to \infty} \frac{ 1 }{ n } H(X_{n} \mid Y_{n}) = 0 .
\end{align}
This result is commonly referred to as the  weak converse property of  the source $\{ Z_{n} \}_{n = 1}^{\infty}$ in the unique decoding setting.
\end{example}

\begin{example}
\label{ex:list}
Let $\mathbf{X} = \{ X_{n} \}_{n = 1}^{\infty}$ be a source as described in \exref{ex:iid_case}.
Even if the list decoding setting, \thref{th:Fano_meets_AEP} states that
\begin{align}
\lim_{n \to \infty} P_{\mathrm{e}}^{(L_{n})}(X_{n} \mid Y_{n}) = \lim_{n \to \infty} \frac{ 1 }{ n } \log L_{n} = 0
\quad \Longrightarrow \quad
\lim_{n \to \infty} \frac{ 1 }{ n } H(X_{n} \mid Y_{n}) = 0 ,
\end{align}
similarly to~\exref{ex:iid_case}.
This is a key observation in Ahlswede--G\'{a}cs--K\"{o}rner's proof of the strong converse property of degraded broadcast channels; see Chapter~5 of \cite{ahlswede_gacs_korner_1976} (see also Section~3.6.2 of \cite{raginsky_sason_2014} and Lemma~1 of \cite{kim_sutivong_cover_2008}).
\end{example}

\begin{example}
\label{ex:poisson}
Consider the Poisson source $\mathbf{X} = \{ X_{n} \}_{n = 1}^{\infty}$ with growing mean $\lambda_{n} = \omega( 1 )$ as $n \to \infty$, i.e.,
\begin{align}
P_{X_{n}}( k )
=
\frac{ \lambda_{n}^{k-1} \, \mathrm{e}^{- \lambda_{n}} }{ (k - 1)! }
\qquad \mathrm{for} \ k \in \mathcal{X} = \{ 1, 2, \dots \} .
\end{align}
It is known that
\begin{align}
\lim_{n \to \infty} \frac{ H( X_{n} ) }{ (1/2) \log \lambda_{n} }
=
1 ,
\end{align}and the Poisson source $\mathbf{X}$ satisfies the AEP (see \cite{verdu_han_1997}).
Therefore, it follows from \thref{th:Fano_meets_AEP} that
\begin{align}
\lim_{n \to \infty} P_{\mathrm{e}}^{(L_{n})}(X_{n} \mid Y_{n}) = 0
\quad \Longrightarrow \quad
|H(X_{n} \mid Y_{n}) - \log L_{n}|^{+} = \mathrm{o}( \log \lambda_{n} ) .
\end{align}
\end{example}

The following example shows a general source that satisfies neither the AEP nor \eqref{eq:vanishing_normalizedRenyi_list1}.

\begin{example}
\label{ex:counter_AEP}
Let $L \ge 1$ be an integer, $\gamma > 0$ a  positive real, and $\{ \delta_{n} \}_{n = 1}^{\infty}$ a sequence of reals satisfying $\delta_{n} = \mathrm{o}( 1 )$ and $0 < \delta_{n} < 1$ for each $n \ge 1$.
As $p \mapsto h_{2}( p ) / p$ is continuous on $(0, 1]$ and $h_{2}( p ) / p \to \infty$ as $p \to 0^{+}$, one can find a sequence of reals $\{ p_{n} \}_{n = 1}^{\infty}$ satisfying $0 < p_{n} \le \min\{ 1, (1 - \delta_{n})/(\delta_{n} \, L) \}$ for each $n \ge 1$ and
\begin{align}
\frac{ \delta_{n} \, h_{2}( p_{n} ) }{ p_{n} }
=
\gamma
\quad \mathrm{for} \ \mathrm{sufficiently} \ \mathrm{large} \ n .
\end{align}
Consider a general source $\mathbf{X} = \{ X_{n} \}_{n = 1}^{\infty}$ whose component distributions are given  by
\begin{align}
P_{X_{n}}( x )
=
\begin{dcases}
\frac{ 1 - \delta_{n} }{ L }
& \mathrm{if} \ 1 \le x \le L ,
\\
\delta_{n} \, p_{n} (1 - p_{n})^{x - (L + 1)}
& \mathrm{if} \ x \ge L + 1
\end{dcases}
\end{align}
for each $n \ge 1$.
Suppose that $X_{n} \Perp Y_{n}$ for each $n \ge 1$.
After some algebra, we have
\begin{align}
P_{\mathrm{e}}^{(L)}(X_{n} \mid Y_{n})
& =
P_{\mathrm{e}}^{(L)}( X_{n} )
=
\delta_{n} ,
\\
H( X_{n} \mid Y_{n} )
& =
H( X_{n} )
=
h_{2}( \delta_{n} ) + (1 - \delta_{n}) \log L + \frac{ \delta_{n} \, h_{2}( p_{n} ) }{ p_{n} }
\end{align}
for each $n \ge 1$.
Therefore, we observe that
\begin{align}
\lim_{n \to \infty} P_{\mathrm{e}}^{(L)}(X_{n} \mid Y_{n})
=
0
\end{align}
holds, but
\begin{align}
\lim_{n \to \infty} \frac{ | H(X_{n} \mid Y_{n}) - \log L |^{+} }{ H( X_{n} ) } = 0
\end{align}
does not hold.
In fact, it holds that $H(X_{n}) \to \gamma + \log L$ as $n \to \infty$ and
\begin{align}
\lim_{n \to \infty} P_{X_{n}}( x )
=
\begin{dcases}
\frac{ 1 }{ L }
& \mathrm{if} \ 1 \le x \le L ,
\\
0
& \mathrm{if} \ x \ge L .
\end{dcases}
\end{align}
Consequently, we also see that $\mathbf{X} = \{ X_{n} \}_{n = 1}^{\infty}$ does not satisfy the AEP.
\end{example}

\exref{ex:counter_AEP} implies that the AEP has an important role in \thref{th:Fano_meets_AEP}.

\subsection{Vanishing Unnormalized R\'{e}nyi's Equivocations}
\label{sect:unnormalized}

Let $X$ be an $\mathcal{X}$-valued r.v.\ satisfying $H( X ) < \infty$, $\{ L_{n} \}_{n = 1}^{\infty}$  a sequence of positive integers, $\{ \mathcal{Y}_{n} \}_{n = 1}^{\infty}$ a sequence of nonempty alphabets, and $\{ (X_{n}, Y_{n}) \}_{n = 1}^{\infty}$ a sequence of $\mathcal{X} \times \mathcal{Y}_{n}$-valued r.v.'s.
The following theorem provides  four conditions on a general source $\mathbf{X} = \{ X_{n} \}_{n = 1}^{\infty}$ such that vanishing error probabilities implies vanishing \emph{unnormalized} Shannon's and R\'{e}nyi's equivocations.

\begin{theorem}
\label{th:vanishing_conditionalRenyi_list}
Let $\alpha \ge 1$ be an order.
Suppose that any one of the following four conditions hold,
\begin{itemize}
\item[(a)]
the order $\alpha$ is strictly larger than $1$, i.e., $\alpha > 1$,
\item[(b)]
the sequence $\{ X_{n} \}_{n = 1}^{\infty}$ satisfies the AEP and $H( X_{n} ) = \mathrm{O}( 1 )$ as $n \to \infty$,
\item[(c)]
there exists an $n_{0} \ge 1$ such that $P_{X_{n}}$ majorizes $P_{X}$ for every $n \ge n_{0}$,
\item[(d)]
the sequence $\{ X_{n} \}_{n = 1}^{\infty}$ converges in distribution to $X$ and 
$H( X_{n} ) \to H( X )$ as $n \to \infty$.
\end{itemize}
Then, it holds that for each $\dagger \in \{ \mathrm{Arimoto}, \mathrm{Hayashi} \}$,
\begin{align}
\lim_{n \to \infty} P_{\mathrm{e}}^{(L_{n})}(X_{n} \mid Y_{n}) = 0
\quad \Longrightarrow \quad
\lim_{n \to \infty} \big| H_{\alpha}^{\dagger}(X_{n} \mid Y_{n}) - \log L_{n} \big|^{+}
=
0 .
\label{eq:vanishing_conditionalRenyi_list1}
\end{align}
\end{theorem}

\begin{IEEEproof}[Proof of \thref{th:vanishing_conditionalRenyi_list}]
See \appref{sect:vanishing_conditionalRenyi_list}.
\end{IEEEproof}

In contrast to Condition (b) of \thref{th:vanishing_conditionalRenyi_list}, Conditions (a), (c), and (d) of \thref{th:vanishing_conditionalRenyi_list} do  not require the AEP to hold.
Interestingly, Condition (a) of \thref{th:vanishing_conditionalRenyi_list} states that \eqref{eq:vanishing_conditionalRenyi_list1} holds for every $\alpha > 1$ and $\dagger \in \{ \mathrm{Arimoto}, \mathrm{Hayashi} \}$ without any other conditions on the general source $\mathbf{X} = \{ X_{n} \}_{n = 1}^{\infty}$.

\begin{remark}
If $L_{n} = 1$ for each $n \ge 1$, then Conditions (c) and (d) of \thref{th:vanishing_conditionalRenyi_list} coincide with Ho--Verd\'{u}'s result stated in Theorem~18 of \cite{ho_verdu_2010}.
Moreover, if $L_{n} = 1$ for each $n \ge 1$, and if $X_{n}$ is $\{ 1, \dots, M^{n} \}$-valued for each $n \ge 1$, then Condition (a) of \thref{th:vanishing_conditionalRenyi_list} coincides with Sason--Verd\'{u}'s result stated in Assertion (a) of Theorem~4 of \cite{sason_verdu_2017}.
\end{remark}

\subsection{Under the Symbol-Wise Error Criterion}
\label{sect:sym}

Let $\mathbf{L} = \{ L_{n} \}_{n = 1}^{\infty}$ be a sequence of positive integers, $\{ \mathcal{Y}_{n} \}_{n = 1}^{\infty}$ a sequence of nonempty alphabets, and $\{ (X_{n}, Y_{n}) \}_{n = 1}^{\infty}$ a sequence of $\mathcal{X} \times \mathcal{Y}_{n}$-valued r.v.'s satisfying $H( X_{n} ) < \infty$ for every $n \ge 1$.
In this subsection, we focus on the \emph{minimum arithmetic-mean probability of symbol-wise list decoding error} defined~as
\begin{align}
P_{\mathrm{e, sym.}}^{(\mathbf{L})}(X^{n} \mid Y^{n})
\coloneqq
\frac{ 1 }{ n } \sum_{i = 1}^{n} P_{\mathrm{e}}^{(L_{i})}(X_{i} \mid Y_{i}) ,
\label{def:sym}
\end{align}
where $X^{n} = (X_{1}, X_{2}, \dots, X_{n})$ and $Y^{n} = (Y_{1}, Y_{2}, \dots, Y_{n})$.
Now, let $X$ be an $\mathcal{X}$-valued r.v.\ satisfying $H(X) < \infty$.
Under this symbol-wise error criterion, the following theorem holds.

\begin{theorem}
\label{th:VanishingAMsymbolerror}
Suppose that $P_{X_{n}}$ majorizes $P_{X}$ for sufficiently large $n$.
Then, it holds that
\begin{align}
\lim_{n \to \infty} P_{\mathrm{e, sym.}}^{(\mathbf{L})}(X^{n} \mid Y^{n}) = 0
\quad \Longrightarrow \quad
\limsup_{n \to \infty} \frac{ 1 }{ n } H(X^{n} \mid Y^{n}) \le \limsup_{n \to \infty} \log L_{n} .
\label{eq:VanishingAMsymbolerror}
\end{align}
\end{theorem}

\begin{IEEEproof}[Proof of \thref{th:VanishingAMsymbolerror}]
See \appref{sect:VanishingAMsymbolerror}.
\end{IEEEproof}

It is known that the classical Fano inequality stated in \eqref{eq:Fano} can be extended from the average error criterion $\mathbb{P}\{ X^{n} \neq Y^{n} \}$ to the symbol-wise error criterion $(1/n) \mathbb{E}[ d_{\mathrm{H}}(X^{n}, Y^{n}) ]$ (see Corollary~3.8 of \cite{csiszar_korner_2011}), where
\begin{align}
d_{\mathrm{H}}( x^{n}, y^{n} )
\coloneqq
|\{ 1 \le i \le n \mid x_{i} \neq y_{i} \}|
\end{align}
stands for the Hamming distance between two strings $x^{n} = (x_{1}, \dots, n_{n})$ and $y^{n} = (y_{1}, \dots, y_{n})$.
In fact, \thref{th:VanishingAMsymbolerror} states that
\begin{align}
\lim_{n \to \infty} \frac{ 1 }{ n } \mathbb{E}[ d_{\mathrm{H}}(X^{n}, Y^{n}) ] = 0
\quad \Longrightarrow \quad
\lim_{n \to \infty} \frac{ 1 }{ n } H(X^{n} \mid Y^{n}) = 0 ,
\end{align}
provided that $P_{X_{n}}$ majorizes $P_{X}$ for sufficiently large $n$.

However, in the list decoding setting, we observe that $P_{\mathrm{e, sym.}}^{(\mathbf{L})}(X^{n} \mid Y^{n}) = \mathrm{o}( 1 )$ does not imply $H(X^{n} \mid Y^{n}) = \mathrm{o}( n )$ in general.
A counterexample can be readily constructed.

\begin{example}
\label{ex:non_vanishing_AM}
Let $\{ X_{n} \}_{n = 1}^{\infty}$ be uniformly distributed Bernoulli r.v.'s, and $\{ Y_{n} \}_{n = 1}^{\infty}$ arbitrary r.v.'s.
Suppose that $(X_{n}, Y_{n}) \Perp (X_{m}, Y_{m})$ if $n \neq m$, $X_{n} \Perp Y_{n}$ for each $n \ge 1$, and $L_{n} = 2$ for each $n \ge 1$.
Then, we observe that
\begin{align}
P_{\mathrm{e, sym.}}^{(\mathbf{L})}(X^{n} \mid Y^{n})
=
0
\end{align}
for every $n \ge 1$, but
\begin{align}
H(X^{n} \mid Y^{n}) = n \log 2
\end{align}
for every $n \ge 1$.
\end{example}

\section{Proofs of Fano-Type Inequalities}
\label{sect:proof_Fano}

In this section, we prove Theorems~\ref{th:main_list}--\ref{th:main_M} via majorization theory \cite{marshall_olkin_arnold_majorization}.

\subsection{Proof of \thref{th:main_list}}
\label{sect:proof_main_list}

We shall relax the feasible regions of the supremum in \eqref{def:main_object} via some lemmas, i.e., our preliminary results.
Define a notion of symmetry for the conditional distribution $P_{X|Y}$ as follows.

\begin{definition}
\label{def:UD}
A jointly distributed pair $(X, Y)$ is said to be \emph{connected uniform-dispersively} if $P_{X|Y}^{\downarrow}$ is almost surely constant.
\end{definition}

\begin{remark}
\label{rem:uniformly-dispersive}
The term introduced in \defref{def:UD} is inspired by \emph{uniformly dispersive channels} named by Massey (see Page~77 of \cite{massey_1996}).
In fact, if $\mathcal{Y}$ is countable and $X$ (resp.\ $Y$) denotes the output (resp.\ input) of a channel $P_{X|Y}$, then the channel $P_{X|Y}$ can be thought of as a uniformly dispersive channel, provided that $(X, Y)$ is connected uniform-dispersively.
Initially, Fano said such channels to be \emph{uniform from the input;} see Page~127 of \cite{fano_1961}.
Refer to Section~II-A of \cite{sakai_iwata_2018} for several symmetry notions of channels.
\end{remark}

Although an almost surely constant $P_{X|Y}$ implies the independence $X \Perp Y$, note also that an almost surely constant $P_{X|Y}^{\downarrow}$ does not imply the independence.
We now give the following lemma.

\begin{lemma}
\label{lem:convex_combination_majorization}
If a jointly distributed pair $(X, Y)$ is connected uniform-dispersively, then $P_{X|Y}$ majorizes $P_{X}$ a.s.
\end{lemma}

\begin{IEEEproof}[Proof of \lemref{lem:convex_combination_majorization}]
Let $k$ be a positive integer.
Choose a collection $\{ x_{i} \}_{i = 1}^{k}$ of $k$ distinct elements in $\mathcal{X}$ so that
\begin{align}
P_{X}( x_{i} )
=
P_{X}^{\downarrow}( i )
\end{align}
for every $1 \le i \le k$.
As
\begin{align}
\sum_{i = 1}^{k} P_{X|Y}( x_{i} )
\le
\sum_{x = 1}^{k} P_{X|Y}^{\downarrow}( x )
\qquad (\mathrm{a.s.})
\end{align}
and
\begin{align}
P_{X}( x )
=
\mathbb{E}[ P_{X|Y}( x ) ]
\end{align}
for each $x \in \mathcal{X}$, we observe that
\begin{align}
\sum_{x = 1}^{k} P_{X}^{\downarrow}( x )
=
\mathbb{E} \Bigg[ \sum_{i = 1}^{k} P_{X|Y}( x_{i} ) \Bigg]
\le
\mathbb{E} \Bigg[ \sum_{x = 1}^{k} P_{X|Y}^{\downarrow}( x ) \Bigg] .
\label{ineq:majorization_PXdr_PXYdr}
\end{align}
If $(X, Y)$ is connected uniform-dispersively (see \defref{def:UD}), then \eqref{ineq:majorization_PXdr_PXYdr} implies that
\begin{align}
\sum_{x = 1}^{k} P_{X}^{\downarrow}( x )
\le
\sum_{x = 1}^{k} P_{X|Y}^{\downarrow}( x )
\qquad \mathrm{(a.s.)} ,
\end{align}
which is indeed the majorization relation stated \defref{def:majorization}, completing the proof of \lemref{lem:convex_combination_majorization}.
\end{IEEEproof}

\begin{remark}
\lemref{lem:convex_combination_majorization} is can be thought of as a novel characterization of uniformly dispersive channels via the majorization relation; see \remref{rem:uniformly-dispersive}.
More precisely, given an input distribution $P$ on $\mathcal{X}$ and a uniformly dispersive channel $W : \mathcal{X} \to \mathcal{Y}$ with countable output alphabet $\mathcal{Y}$, it holds that $W(\cdot \mid x)$ majorizes the output distribution $PW$ for every $x \in \mathcal{X}$, where $PW$ is given by
\begin{align}
PW( y )
\coloneqq
\sum_{x \in \mathcal{X}} P( x ) \, W(y \mid x)
\end{align}
for each $y \in \mathcal{Y}$.
\end{remark}

\begin{definition}
Let $\mathcal{A}$ be a collection of jointly distributed pairs of an $\mathcal{X}$-valued r.v.\ and a $\mathcal{Y}$-valued r.v.
We say that $\mathcal{A}$ has \emph{balanced conditional distributions} if $(X ,Y) \in \mathcal{A}$ implies that there exists $(U, V) \in \mathcal{A}$ satisfying
\begin{align}
P_{U|V}^{\downarrow}( x )
=
\mathbb{E} \big[ P_{X|Y}^{\downarrow}( x ) \big]
\qquad \mathrm{(a.s.)}
\label{eq:balanced_X|Y}
\end{align}
for every $x \in \mathcal{X}$.
\end{definition}

For such a collection $\mathcal{A}$, the following lemma holds.

\begin{lemma}
\label{lem:max_necessity}
Suppose that $\mathcal{A}$ has balanced conditional distributions.
For any $(X, Y) \in \mathcal{A}$, there exists a pair $(U, V) \in \mathcal{A}$ connected uniform-dispersively such that
\begin{align}
\mathsf{H}_{\phi}(U \mid V)
\ge
\mathsf{H}_{\phi}(X \mid Y) .
\label{eq:maximum_A}
\end{align}
\end{lemma}

\begin{IEEEproof}[Proof of \lemref{lem:max_necessity}]
For any $(X, Y) \in \mathcal{A}$, it holds that
\begin{align}
\mathsf{H}_{\phi}(X \mid Y)
& \overset{\mathclap{\text{(a)}}}{=}
\mathbb{E} \big[ \phi\big( P_{X|Y}^{\downarrow} \big) \big]
\notag \\
& \overset{\mathclap{\text{(b)}}}{\le}
\phi\big( \mathbb{E} \big[ P_{X|Y}^{\downarrow} \big] \big)
\notag \\
& \overset{\mathclap{\text{(c)}}}{=}
\phi \big( P_{U|V}^{\downarrow} \big)
\qquad \mathrm{(a.s.)}
\notag \\
& \overset{\mathclap{\text{(d)}}}{=}
\mathbb{E} \big[ \phi( P_{U|V} ) \big]
\qquad \mathrm{(a.s.)}
\notag \\
& =
\mathsf{H}_{\phi}(U \mid V) ,
\label{ineq:Af_S_XY}
\end{align}
where
\begin{itemize}
\item
(a) follows by the symmetry of $\phi$,
\item
(b) follows by Jensen's inequality (see \cite{shirokov_2010}, Proposition~A-2),
\item
(c) follows by the existence of a pair $(U, V) \in \mathcal{A}$ connected uniform-dispersively (see \eqref{eq:balanced_X|Y}), and
\item
(d) follows by the symmetry of $\phi$ again.
\end{itemize}
This completes the proof of \lemref{lem:max_necessity}.
\end{IEEEproof}

For a system $(Q, L, \varepsilon, \mathcal{Y})$ satisfying \eqref{eq:range_epsilon_list}, we now define a collection of pairs of r.v.'s as follows,
\begin{align}
\mathcal{R}(Q, L, \varepsilon, \mathcal{Y})
\coloneqq
\left\{ (X, Y) \ \middle|
\begin{array}{l}
\text{$X$ is $\mathcal{X}$-valued} , \\
\text{$Y$ is $\mathcal{Y}$-valued} , \\
P_{\mathrm{e}}^{(L)}(X \mid Y) \le \varepsilon , \\
P_{X} = Q
\end{array}
\right\} .
\label{def:region_R}
\end{align}
Note that this is the feasible region of the supremum in \eqref{def:main_object}.
The main idea of proving \thref{th:main_list} is to apply \lemref{lem:max_necessity} for this collection.
The collection $\mathcal{R}(Q, L, \varepsilon, \mathcal{Y})$ does not, however, have balanced conditional distributions in general.
More specifically, there exists a measurable space $\mathcal{Y}$ such that $\mathcal{R}(Q, L, \varepsilon, \mathcal{Y})$ does not have balanced conditional distributions even if $\mathcal{Y}$ is standard Borel.
Fortunately, the following lemma can avoid this issue by blowing-up the collection $\mathcal{R}(Q, L, \varepsilon, \mathcal{Y})$ via the infinite-dimensional version of Birkhoff's theorem \cite{revesz_1962}.

\begin{lemma}
\label{lem:balanced_Pe}
If the cardinality of $\mathcal{Y}$ is at least the cardinality of the continuum $\mathbb{R}$, then there exists a $\sigma$-algebra on $\mathcal{Y}$ such that the collection $\mathcal{R}(Q, L, \varepsilon, \mathcal{Y})$ has balanced conditional distributions.
\end{lemma}

\begin{IEEEproof}[Proof of \lemref{lem:balanced_Pe}]
First, we shall choose an appropriate alphabet $\mathcal{Y}$ so that its cardinality is the cardinality of the continuum.
Denote by $\Psi$ the set of $\infty \times \infty$ permutation matrices, where an $\infty \times \infty$ permutation matrix is a real matrix $\Pi = \{ \pi_{i, j} \}_{i, j = 1}^{\infty}$ satisfying either $\pi_{i, j} = 0$ or $\pi_{i, j} = 1$ for each $1 \le i, j < \infty$, and
\begin{align}
\sum_{j = 1}^{\infty} \pi_{i, j}
& =
1
\qquad \mathrm{for} \ \mathrm{each} \ 1 \le i < \infty ,
\\
\sum_{i = 1}^{\infty} \pi_{i, j}
& =
1
\qquad \mathrm{for} \ \mathrm{each} \ 1 \le j < \infty .
\end{align}
For an $\infty \times \infty$ permutation matrix $\Pi = \{ \pi_{i, j} \}_{i, j} \in \Psi$, define the permutation $\psi_{\Pi}$ on $\mathcal{X} = \{ 1, 2, \dots \}$ by
\begin{align}
\psi_{\Pi}( i )
\coloneqq
\sum_{j = 1}^{\infty} \pi_{i, j} \, j .
\label{def:psi}
\end{align}
It is known that there is a one-to-one correspondence between the permutation matrices $\Pi$ and the bijections $\psi_{\Pi}$; and thus, the cardinality of $\Psi$ is the cardinality of the continuum.
Therefore, in this proof, we may assume without loss of generality that $\mathcal{Y} = \Psi$.

Second, we shall construct an appropriate $\sigma$-algebra on $\mathcal{Y}$ via the infinite-dimensional version of Birkhoff's theorem (cf.\ Theorem~2 of \cite{revesz_1962}) for $\infty \times \infty$ doubly stochastic matrices, where an $\infty \times \infty$ doubly stochastic matrix is a real matrix $\mathbf{M} = \{ m_{i, j} \}_{i, j = 1}^{\infty}$ satisfying $0 \le m_{i, j} \le 1$ for each $1 \le i, j < \infty$, and
\begin{align}
\sum_{j = 1}^{\infty} m_{i, j}
& =
1
\qquad \mathrm{for} \ \mathrm{each} \ 1 \le i < \infty ,
\\
\sum_{i = 1}^{\infty} m_{i, j}
& =
1
\qquad \mathrm{for} \ \mathrm{each} \ 1 \le j < \infty .
\end{align}
Similar to $\Psi$, denote by $\Psi_{i, j}$ the set of $\infty \times \infty$ permutation matrices in which the entry in the $i$th row and the $j$th column is $1$, where note that $\Psi_{i, j} \subset \mathcal{Y}$.
Then, the following lemma holds.

\begin{lemma}[{infinite-dimensional version of Birkhoff's theorem; cf.\ Theorem~2 of \cite{revesz_1962}}]
\label{lem:infinite-Birkhoff}
There exists a $\sigma$-algebra $\Gamma$ on $\mathcal{Y}$ such that (i) $\Psi_{i, j} \in \Gamma$ for every $1 \le i, j < \infty$ and (ii) for any $\infty \times \infty$ doubly stochastic matrix $\mathbf{M} = \{ m_{i, j} \}_{i, j = 1}^{\infty}$, there exists a probability measure $\mu$ on $(\mathcal{Y}, \Gamma)$ such that $\mu( \Psi_{i, j} ) = m_{i, j}$ for every $1 \le i, j < \infty$.
\end{lemma}

\begin{remark}
In the original statement of Theorem~2 of \cite{revesz_1962}, it is written that a probability space $(\mathcal{Y}, \Gamma, \mu)$ exists for a given $\infty \times \infty$ doubly stochastic matrix $\mathbf{M}$, namely, the $\sigma$-algebra $\Gamma$ may depend on $\mathbf{M}$.
However, the construction of $\Gamma$ is independent of $\mathbf{M}$ (see Page~196 of \cite{revesz_1962}); and we can restate Theorem~2 of \cite{revesz_1962} as \lemref{lem:infinite-Birkhoff}.
\end{remark}

This is a probabilistic description of an $\infty \times \infty$ doubly stochastic matrix via a probability measure on the $\infty \times \infty$ permutation matrices.
The existence of the probability measure $\mu$ is due to Kolmogorov's extension theorem.
We employ this $\sigma$-algebra $\Gamma$ on $\mathcal{Y}$ in the proof.

Thirdly, we shall show that under this measurable space $(\mathcal{Y}, \Gamma)$, the collection $\mathcal{R}(Q, L, \varepsilon, \mathcal{Y})$ has balanced conditional distributions defined in \eqref{eq:balanced_X|Y}.
In other words, for a given pair $(X, Y) \in \mathcal{R}(Q, L, \varepsilon, \mathcal{Y})$, it suffices to construct another pair $(U, V)$ of r.v.'s satisfying \eqref{eq:balanced_X|Y} and $(U, V) \in \mathcal{R}(Q, L, \varepsilon, \mathcal{Y})$.
At first, construct its conditional distribution $P_{U|V}$ by
\begin{align}
P_{U|V}( x )
=
\mathbb{E} \big[ P_{X|Y}^{\downarrow}( \psi_{V}( x ) ) \ \big| \ V \big]
\qquad \mathrm{(a.s.)}
\label{eq:cond_Q}
\end{align}
for each $x \in \mathcal{X}$, where $\mathbb{E}[Z \mid W]$ stands for the conditional expectation of a real-valued r.v.\ $Z$ given the sub-$\sigma$-algebra $\sigma(W)$ generated by a r.v.\ $W$, and $\phi_{V}$ is given as in \eqref{def:psi}.
As $\psi_{V}( x )$ is $\sigma(V)$-measurable for each $x \in \mathcal{X}$, it is clear that
\begin{align}
P_{U|V}^{\downarrow}( x )
=
\mathbb{E} \big[ P_{X|Y}^{\downarrow}( x ) \ \big| \ V \big]
=
\mathbb{E} \big[ P_{X|Y}^{\downarrow}( \psi_{V}( \psi_{V}^{-1}( x ) ) ) \ \big| \ V \big]
=
P_{U|V}( \psi_{V}^{-1}( x ) )
\qquad (\mathrm{a.s.})
\label{eq:V-measurable}
\end{align}
for every $x \in \mathcal{X}$.
Thus, we readily see that \eqref{eq:balanced_X|Y} holds, and $(U, V)$ is connected uniform-dispersively.
Thus, by \eqref{ineq:majorization_PXdr_PXYdr} and the hypothesis that $P_{X} = Q$, we see that $P_{U|V}$ majorizes $Q$ a.s.
Therefore, it follows from the well-known characterization of the majorization relation via $\infty \times \infty$ doubly stochastic matrices (see Lemma~3.1 of \cite{markus_1964} or Page~25 of \cite{marshall_olkin_arnold_majorization}) that one can find an $\infty \times \infty$ doubly stochastic matrix $\mathbf{M} = \{ m_{i, j} \}_{i, j = 1}^{\infty}$ satisfying
\begin{align}
Q( i )
=
\sum_{j = 1}^{\infty} m_{i, j} \, P_{U|V}^{\downarrow}( j )
\qquad \mathrm{(a.s.)}
\label{eq:HLP}
\end{align}
for every $i \ge 1$.
By \lemref{lem:infinite-Birkhoff}, we can construct an induced probability measure $P_{V}$ so that $P_{V}( \Psi_{i, j} ) = m_{i, j}$ for each $1 \le i, j < \infty$.
Now, the pair of $P_{U|V}$ and $P_{V}$ can define the probability law of $(U, V)$.
To ensure that $(U, V)$ belongs to $\mathcal{R}(Q, L, \varepsilon, \mathcal{Y})$, it remains to verity that $P_{\mathrm{e}}^{(L)}(U \mid V) \le \varepsilon$ and $P_{U} = Q$.

As $\psi_{\Pi}$ is a permutation defined in \eqref{def:psi}, we have
\begin{align}
P_{\mathrm{e}}^{(L)}(X \mid Y) \,
& \overset{\mathclap{\text{(a)}}}{=}
1 - \mathbb{E}\bigg[ \sum_{x = 1}^{L} P_{X|Y}^{\downarrow}( x ) \bigg]
\notag \\
& =
1 - \mathbb{E}\bigg[ \sum_{x = 1}^{L} \mathbb{E}[ P_{X|Y}^{\downarrow}( x ) \mid V ] \bigg]
\notag \\
& \overset{\mathclap{\text{(b)}}}{=}
1 - \mathbb{E}\bigg[ \sum_{x = 1}^{L} P_{U|V}^{\downarrow}( x ) \bigg]
\notag \\
& \overset{\mathclap{\text{(c)}}}{=} \,
P_{\mathrm{e}}^{(L)}(U \mid V) ,
\label{eq:same_Pe_P-Q}
\end{align}
where
\begin{itemize}
\item
(a) and (c) follow from \propref{prop:listMAP}, and
\item
and (b) follows from \eqref{eq:V-measurable}.
\end{itemize}
Therefore, we see that $P_{\mathrm{e}}^{(L)}(X \mid Y) \le \varepsilon$ is equivalent to $P_{\mathrm{e}}^{(L)}(U \mid V) \le \varepsilon$.
Furthermore, we observe that
\begin{align}
Q( i ) \,
& \overset{\mathclap{\text{(a)}}}{=} \,
\sum_{j = 1}^{\infty} m_{i, j} \, P_{U|V}^{\downarrow}( j )
\qquad \mathrm{(a.s.)}
\notag \\
& \overset{\mathclap{\text{(b)}}}{=}
\sum_{j = 1}^{\infty} \mathbb{E} \Big[ \boldsymbol{1}_{\{ V \in \Psi_{i, j} \}} \Big] \, P_{U|V}^{\downarrow}( j )
\qquad \mathrm{(a.s.)}
\notag \\
& \overset{\mathclap{\text{(c)}}}{=}
\sum_{j = 1}^{\infty} \mathbb{E} \Big[ \boldsymbol{1}_{\{ V \in \Psi_{i, j} \}} \, P_{U|V}^{\downarrow}( j ) \Big]
\qquad \mathrm{(a.s.)}
\notag \\
& =
\sum_{j = 1}^{\infty} \mathbb{E} \Big[ \mathbb{E} \Big[ \boldsymbol{1}_{\{ V \in \Psi_{i, j} \}} \, P_{U|V}^{\downarrow}( j ) \ \Big| \ V \Big] \Big]
\notag \\
& \overset{\mathclap{\text{(d)}}}{=}
\sum_{j = 1}^{\infty} \mathbb{E} \Big[ \mathbb{E} \Big[ \boldsymbol{1}_{\{ V \in \Psi_{i, j} \}} \, P_{U|V}( \psi_{V}^{-1}( j ) ) \ \Big| \ V \Big] \Big]
\notag \\
& \overset{\mathclap{\text{(e)}}}{=}
\sum_{j = 1}^{\infty} \mathbb{E} \Bigg[ \mathbb{E} \Bigg[ \boldsymbol{1}_{\{ V \in \Psi_{i, j} \}} \, P_{U|V}\Bigg( \sum_{k = 1}^{\infty} \boldsymbol{1}_{\{ V \in \Psi_{k, j} \}} \, k \Bigg) \ \Bigg| \ V \Bigg] \Bigg]
\notag \\
& =
\sum_{j = 1}^{\infty} \mathbb{E} \Bigg[ \boldsymbol{1}_{\{ V \in \Psi_{i, j} \}} \, P_{U|V}\Bigg( \sum_{k = 1}^{\infty} \boldsymbol{1}_{\{ V \in \Psi_{k, j} \}} \, k \Bigg) \Bigg]
\notag \\
& \overset{\mathclap{\text{(f)}}}{=}
\mathbb{E} \Bigg[ \sum_{j = 1}^{\infty} \boldsymbol{1}_{\{ V \in \Psi_{i, j} \}} \, P_{U|V}\Bigg( \sum_{k = 1}^{\infty} \boldsymbol{1}_{\{ V \in \Psi_{k, j} \}} \, k \Bigg) \Bigg]
\notag \\
& \overset{\mathclap{\text{(g)}}}{=}
\mathbb{E} \big[ P_{U|V}( i ) \big]
\notag \\
& =
P_{U}( i )
\label{eq:marginal_PxQx}
\end{align}
for every $i \ge 1$, where
\begin{itemize}
\item
(a) follows from \eqref{eq:HLP},
\item
(b) follows by the identity $m_{i, j} = \mathbb{P}\{ V \in \Psi_{i, j} \}$,
\item
(c) follows from the fact that $(X, Y)$ is connected uniform-dispersively,
\item
(d) follows from \eqref{eq:V-measurable},
\item
(e) follows by the definition of $\Psi_{i, j}$,
\item
(f) follows by the Fubini--Tonelli theorem, and
\item
(f) follows from the fact that the inverse of a permutation matrix is its transpose.
\end{itemize}
Therefore, we have $P_{U} = Q$, and the assertion of \lemref{lem:balanced_Pe} is proved in the case where the cardinality of $\mathcal{Y}$ is the cardinality of the continuum.

Finally, even if the cardinality of $\mathcal{Y}$ is larger than the cardinality of continuum, the assertion of \lemref{lem:balanced_Pe} can be immediately proved by considering the trace of the space $\mathcal{Y}$ on $\Psi$ (cf.\ \cite{chung_2000}, p.~23).
This completes the proof of \lemref{lem:balanced_Pe}.
\end{IEEEproof}

Finally, we show that the Fano-type distribution of type-1 defined in \eqref{def:type1} is the infimum of a certain class of $\mathcal{X}$-marginals with respect to the majorization relation $\prec$.

\begin{lemma}
\label{lem:cond_to_marg}
Suppose that the system $(Q, L, \varepsilon)$ satisfies the right-hand inequality in \eqref{eq:range_epsilon_list}.
For every $\mathcal{X}$-marginal $R$ in which $R$ majorizes $Q$ and $P_{\mathrm{e}}^{(L)}( R ) \le \varepsilon$, it holds that $R$ majorizes $P_{\operatorname{type-1}}$ as well.
\end{lemma}

\begin{IEEEproof}[Proof of \lemref{lem:cond_to_marg}]
We first give an elementary fact of the weak majorization on the finite-dimensional real vectors.

\begin{lemma}
\label{lem:majorizes_uniform}
Let $\bvec{p} = ( p_{i} )_{i = 1}^{n}$ and $\bvec{q} = ( q_{i} )_{i = 1}^{n}$ be $n$-dimensional real vectors satisfying $p_{1} \ge p_{2} \ge \cdots \ge p_{n} \ge 0$ and $q_{1} \ge q_{2} \ge \cdots \ge q_{n} \ge 0$, respectively.
Consider an integer $1 \le k \le n$ satisfying $q_{k} = q_{i}$ for every $i = k, k+1, \dots, n$.
If
\begin{align}
\sum_{i = 1}^{j} p_{i}
& \ge
\sum_{i = 1}^{j} q_{i}
\qquad \mathrm{for} \ j = 1, 2, \dots, k - 1 ,
\label{eq:majorizes_uniform1} \\
\sum_{i = 1}^{n} p_{i}
& \ge
\sum_{i = 1}^{n} q_{i}
\label{eq:majorizes_uniform2}
\end{align}
then it holds that
\begin{align}
\sum_{i = 1}^{j} p_{i}
\ge
\sum_{i = 1}^{j} q_{i}
\qquad \mathrm{for} \ j = 1, 2, \dots, n .
\label{eq:weak_majorization}
\end{align}
\end{lemma}

\begin{IEEEproof}[Proof of \lemref{lem:majorizes_uniform}]
See \appref{app:majorizes_uniform}.
\end{IEEEproof}

Since $P_{\operatorname{type-1}} = P_{\operatorname{type-1}}^{\downarrow}$ (see \propref{prop:type1}), it suffices to prove that
\begin{align}
\sum_{x = 1}^{k} P_{\operatorname{type-1}}( x )
\le
\sum_{x = 1}^{k} R^{\downarrow}( x )
\label{eq:majorization_type5}
\end{align}
for every $k \ge 1$.

As $P_{\operatorname{type-1}}( x ) = Q^{\downarrow}( x )$ for each $1 \le x < J$ (see \propref{prop:type1}), it follows by the majorization relation $Q \prec R$ that \eqref{eq:majorization_type5} holds for each $1 \le k < J$.
Moreover, as $P_{\mathrm{e}}^{(L)}( P_{\operatorname{type-1}} ) = \varepsilon$ (see \propref{prop:type1}), it follows from \eqref{def:Pe_Q} and the hypothesis $P_{\mathrm{e}}^{(L)}( R ) \le \varepsilon$ that
\begin{align}
\sum_{x = J}^{L} P_{\operatorname{type-1}}( x )
\le
\sum_{x = J}^{L} R^{\downarrow}( x ) .
\label{ineq:sum_J_L}
\end{align}
In addition, as \eqref{eq:majorization_type5} holds for each $1 \le k < J$ and $P_{\operatorname{type-1}}( x ) = \mathcal{V}( J )$ for each $J \le x \le L$ (see \propref{prop:type1}), it follows from \lemref{lem:majorizes_uniform} and \eqref{ineq:sum_J_L} that \eqref{eq:majorization_type5} also holds for each $1 \le k \le L$.

Now, suppose that $K_{1} = \infty$.
Then, it follows that
\begin{align}
P_{\operatorname{type-1}}( x ) = \mathcal{W}( \infty ) = 0
\end{align}
for each $x \ge L+1$ (see \propref{prop:type1}).
Thus, Inequality~\eqref{eq:majorization_type5} holds for every $k \ge 1$; therefore, we have that $R$ majorizes $P_{\operatorname{type-1}}$, provided that $K_{1} = \infty$.

Finally, suppose that $K_{1} < \infty$.
Since $P_{\operatorname{type-1}}( x ) = Q^{\downarrow}( x )$ for each $x \ge K_{1} + 1$ (see \propref{prop:type1}), it follows by the majorization relation $Q \prec R$ that \eqref{eq:majorization_type5} holds for every $k \ge K_{1}$.
Moreover, since \eqref{eq:majorization_type5} holds for every $1 \le k \le L$ and every $k \ge K_{1}$, we observe that
\begin{align}
\sum_{x = L+1}^{K_{1}} P_{\operatorname{type-1}}( x )
\le
\sum_{x = L+1}^{K_{1}} R^{\downarrow}( x ) .
\label{ineq:sum_L_K}
\end{align}
Finally, as \eqref{eq:majorization_type5} holds for $1 \le k \le L$ and $P_{\operatorname{type-1}}( x ) = \mathcal{W}( K_{1} )$ for $L < x \le K_{1}$ (see \propref{prop:type1}), it follows by \lemref{lem:majorizes_uniform} and \eqref{ineq:sum_L_K} that \eqref{eq:majorization_type5} holds for every $1 \le k \le K_{1}$.
Therefore, Inequality~\eqref{eq:majorization_type5} holds for every $k \ge 1$, completing the proof of \lemref{lem:cond_to_marg}.
\end{IEEEproof}

Using the above lemmas, we can prove \thref{th:main_list} as follows.

\begin{IEEEproof}[Proof of \thref{th:main_list}]
Let $\varepsilon > 0$.
For the sake of brevity, we write
\begin{align}
\mathcal{R}
& =
\mathcal{R}(Q, L, \varepsilon, \mathcal{Y})
\end{align}
in the proof.
Let $\Upsilon$ be a $\sigma$-algebra on $\mathcal{Y}$, $\Psi$ an alphabet in which its cardinality is the cardinality of the continuum, and $\Gamma$ a $\sigma$-algebra on $\Psi$ so that $\mathcal{R}(Q, L, \varepsilon, \Psi)$ has balanced conditional distributions (see \lemref{lem:balanced_Pe}).
Now, we define the collection
\begin{align}
{\bar{\mathcal{R}}}
\coloneqq
\mathcal{R}(Q, L, \varepsilon, \mathcal{Y} \cup \Psi) ,
\end{align}
where the $\sigma$-algebra on $\mathcal{Y} \cup \Psi$ is given by the smallest $\sigma$-algebra $\Upsilon \vee \Gamma$ containing $\Upsilon$ and $\Gamma$.
It is clear that $\mathcal{R} \subset \bar{\mathcal{R}}$, and $\bar{\mathcal{R}}$ has balanced conditional distributions as well (see the last paragraph in the proof of \lemref{lem:balanced_Pe}).
Then, we have
\begin{align}
\mathbb{H}_{\phi}(Q, L, \varepsilon, \mathcal{Y})
& \overset{\mathclap{\text{(a)}}}{=}
\sup_{(X, Y) \in \mathcal{R}} \mathsf{H}_{\phi}(X \mid Y)
\notag \\
& \overset{\mathclap{\text{(b)}}}{\le}
\sup_{(X, Y) \in \bar{\mathcal{R}}} \mathsf{H}_{\phi}(X \mid Y)
\notag \\
& \overset{\mathclap{\text{(c)}}}{=}
\sup_{\substack{ (X, Y) \in \bar{\mathcal{R}} : \\ \text{$(X, Y)$ is connected uniform-dispersively} }}
\mathsf{H}_{\phi}(X \mid Y)
\notag \\
& \overset{\mathclap{\text{(d)}}}{=}
\sup_{\substack{ (X, Y) \in \bar{\mathcal{R}} : \\ \text{$P_{\mathrm{e}}^{(L)}( P_{X|Y} ) \le \varepsilon$ a.s.} , \\ \text{$(X, Y)$ is connected uniform-dispersively} }}
\phi( P_{X|Y} )
\qquad \mathrm{(a.s.)}
\notag \\
& \overset{\mathclap{\text{(e)}}}{\le}
\sup_{\substack{ R \in \mathcal{P}( \mathcal{X} ) : \\ \text{$Q \prec R$ and $P_{\mathrm{e}}^{(L)}(R) \le \varepsilon$} }} \phi( R )
\qquad \mathrm{(a.s.)}
\notag \\
& \overset{\mathclap{\text{(f)}}}{\le}
\phi( P_{\operatorname{type-1}} ) ,
\label{ineq:main_list:proof}
\end{align}
where
\begin{itemize}
\item
(a) follows by the definition of $\mathcal{R}$ stated in \eqref{def:region_R},
\item
(b) follows by the inclusion $\mathcal{R} \subset \bar{\mathcal{R}}$,
\item
(c) follows from \lemref{lem:max_necessity} and the fact that $\bar{\mathcal{R}}$ has balanced conditional distributions,
\item
(d) follows by the symmetry of both $\phi : \mathcal{P}( \mathcal{X} ) \to [0, \infty]$ and $P_{\mathrm{e}}^{(L)} : \mathcal{P}( \mathcal{X} ) \to [0, 1]$,
\item
(e) follows from \lemref{lem:convex_combination_majorization}, and
\item
(f) follows from \propref{prop:Schur_convex} and \lemref{lem:cond_to_marg}.
\end{itemize}
Inequalities~\eqref{ineq:main_list:proof} are indeed the Fano-type inequality stated in \eqref{eq:main_list} of \thref{th:main_list}.
If $\varepsilon = P_{\mathrm{e}}^{(L)}( Q )$, then it can be verified by the definition of $P_{\operatorname{type-1}}$ stated in \eqref{def:type1} that $P_{\operatorname{type-1}} = Q^{\downarrow}$ (see also \propref{prop:type1}).
In such a case, the supremum in \eqref{def:main_object} can be achieved by a pair $(X, Y)$ satisfying $P_{X} = Q$ and $X \Perp Y$.

Finally, we shall construct a jointly distributed pair $(X, Y)$ satisfying
\begin{align}
\mathsf{H}_{\phi}(X \mid Y)
& =
\phi( P_{\operatorname{type-1}} ) ,
\label{eq:cond-inf-measure_type-1} \\
P_{\mathrm{e}}^{(L)}(X \mid Y)
& =
\varepsilon ,
\label{eq:error-probab_type-1} \\
P_{X}( x )
& =
Q^{\downarrow}( x )
\qquad (\mathrm{for} \ x \in \mathcal{X}) .
\label{eq:fixed_X-marginal}
\end{align}
For the sake of brevity, suppose that $\mathcal{Y}$ is the index set of the set of permutation matrices on $\{ J, J+1, \dots, K_{1} \}$.
Namely, denote by $\Pi^{(y)} = \{ \pi_{i, j}^{(y)} \}_{i, j = J}^{K_{1}}$ a permutation matrix for each index $y \in \mathcal{Y}$.
By the definition of $P_{\operatorname{type-1}}$ stated in \eqref{def:type1} (see also \propref{prop:type1}), we observe that
\begin{align}
\sum_{x = J}^{k} Q^{\downarrow}( x )
& \le
\sum_{x = J}^{k} P_{\operatorname{type-1}}( x )
\qquad \text{for} \ J \le k \le K_{1} ,
\label{eq:majorization_partial1}
\end{align}
and
\begin{align}
\sum_{x = J}^{K_{1}} Q^{\downarrow}( x )
& =
\sum_{x = J}^{K_{1}} P_{\operatorname{type-1}}( x ) .
\label{eq:majorization_partial2}
\end{align}
Noting that $K_{1} < \infty$ if $\varepsilon > 0$ (see \eqref{def:K3}), Equations~\eqref{eq:majorization_partial1} and~\eqref{eq:majorization_partial2} are indeed a majorization relation between two finite-dimensional real vectors; and thus, it follows from the Hardy--Littlewood--P\'{o}lya theorem (see Theorem~8 of \cite{hardy_littlewood_polya_1928} or Theorem~2.B.2 \cite{marshall_olkin_arnold_majorization}) that there exists a $(K_{1} - J + 1) \times (K_{1} - J + 1)$ doubly stochastic matrix $\mathbf{M} = \{ m_{i, j} \}_{i, j = J}^{K_{1}}$ satisfying
\begin{align}
Q^{\downarrow}( i )
=
\sum_{j = J}^{K_{1}} m_{i, j} \, P_{\operatorname{type-1}}( j )
\label{eq:DSM_Past}
\end{align}
for each $J \le i \le K_{1}$.
Moreover, it follows from the finite dimensional version of Birkhoff's theorem \cite{birkhoff_1946} (see also Theorems~2.A.2 and~2.C.2 of \cite{marshall_olkin_arnold_majorization}) that for such a doubly stochastic matrix $\mathbf{M} = \{ m_{i, j} \}_{i, j = J}^{K_{1}}$, there exists a probability vector $\boldsymbol{\lambda} = ( \lambda_{y} )_{y \in \mathcal{Y}}$ satisfying
\begin{align}
m_{i, j}
=
\sum_{y \in \mathcal{Y}} \lambda_{y} \, \pi_{i, j}^{(y)}
\label{eq:DSM_Past_PM}
\end{align}
for every $J \le i, j \le K_{1}$, where a nonnegative vector is called a \emph{probability vector} if the sum of the elements is unity.
Using them, we construct a pair $(X, Y)$ via the following distributions,
\begin{align}
P_{X|Y=y}( x )
& =
\begin{cases}
P_{\operatorname{type-1}}( x )
& \text{if} \ 1 \le x < J \ \text{or} \ K_{1} < x < \infty ,
\\
P_{\operatorname{type-1}}( \tilde{\psi}_{y}( x ) )
& \text{if} \ J \le x \le K_{1} ,
\end{cases}
\label{def:Past_joint1} \\
P_{Y}( y )
& =
\lambda_{y} ,
\label{def:Past_joint2}
\end{align}
where the permutation $\tilde{\psi}_{y}$ on $\{ J, J+1, \dots, K_{1} \}$ is defined by
\begin{align}
{\tilde{\psi}_{y}( i )}
\coloneqq
\sum_{j = J}^{K_{1}} \pi_{i, j}^{(y)} \, j
\label{def:bar_psi}
\end{align}
for each $y \in \mathcal{Y}$.
Then, it follows from \eqref{eq:DSM_Past} and \eqref{eq:DSM_Past_PM} that \eqref{eq:fixed_X-marginal} holds.
Moreover, it is easy to see that $P_{X|Y=y}^{\downarrow} = P_{\operatorname{type-1}}$ for every $y \in \mathcal{Y}$.
Thus, we observe that \eqref{eq:cond-inf-measure_type-1} and \eqref{eq:error-probab_type-1} hold as well.
This implies together with \eqref{ineq:main_list:proof} that the constructed pair $(X, Y)$ achieves the supremum in \eqref{def:main_object}, completes the proof of \thref{th:main_list}.
\end{IEEEproof}

\subsection{Proof of \thref{th:main_zero}}
\label{sect:proof_main_zero}

Even if $\varepsilon = 0$, the inequalities~in \eqref{ineq:main_list:proof} hold as well;
that is, the Fano-type inequality stated in \eqref{eq:main_zero} of \thref{th:main_zero} holds.
In this proof, we shall verify the equality conditions of \eqref{eq:main_zero}.

If $\supp( Q )$ is finite, then it follows by the definition of $K_{1}$ stated in \eqref{def:K3} that $K_{1} < \infty$.
Thus, the same construction of a jointly distributed pair $(X, Y)$ as the last paragraph of \sectref{sect:proof_main_list} proves that \eqref{eq:main_zero} holds with equality if $\supp( Q )$ is finite.

Consider the case where $\supp( Q )$ is infinite and $J = L$.
Since $\varepsilon = 0$, we readily see that $K_{1} = \infty$, $\mathcal{V}( J ) > 0$, and $\mathcal{W}( K_{1} ) = 0$.
Suppose that
\begin{align}
\mathcal{Y}
=
\{ L, L+1, L+2, \dots \} .
\end{align}
We then construct a pair $(X, Y)$ via the following distributions,
\begin{align}
P_{X|Y=y}( x )
& =
\begin{cases}
Q^{\downarrow}( x )
& \mathrm{if} \ 1 \le x < L ,
\\
\mathcal{V}( J )
& \mathrm{if} \ L \le x < \infty \ \mathrm{and} \ x = y ,
\\
0
& \mathrm{if} \ L \le x < \infty \ \mathrm{and} \ x \neq y ,
\end{cases}
\\
P_{Y}( y )
& =
\frac{ Q( y ) }{ \mathcal{V}( J ) } .
\end{align}
We readily see that $P_{X|Y=y}^{\downarrow} = P_{\operatorname{type-1}}$ for every $y \in \mathcal{Y}$; therefore, we have that \eqref{eq:cond-inf-measure_type-1}--\eqref{eq:fixed_X-marginal} hold.
This implies that the constructed pair $(X, Y)$ achieves the supremum in \eqref{def:main_object}.

Finally, suppose that the cardinality of $\mathcal{Y}$ is at least the cardinality of the continuum.
Assume without loss of generality that $\mathcal{Y}$ is the set of $\infty \times \infty$ permutation matrices.
Consider the measurable space $(\mathcal{Y}, \Gamma)$ given in the infinite-dimensional version of Birkhoff's theorem (see \lemref{lem:infinite-Birkhoff}).
In addition, consider a jointly distributed pair $(X, Y)$ satisfying $P_{X|Y}^{\downarrow} = P_{\operatorname{type-1}}$ a.s.
Then, it is easy to see that \eqref{eq:cond-inf-measure_type-1} and \eqref{eq:error-probab_type-1} hold for any induced probability measure $P_{Y}$ on $\mathcal{Y}$.
Similar to the construction of the probability measure $P_{V}$ on $\mathcal{Y}$ below \eqref{eq:HLP}, we can find an induced probability measure $P_{Y}$ satisfying \eqref{eq:fixed_X-marginal}.
Therefore, it follows from \eqref{eq:main_zero} that this pair $(X, Y)$ achieves the supremum in \eqref{def:main_object}.
This completes the proof of \thref{th:main_zero}.
\hfill\IEEEQEDhere

\subsection{Proof of \thref{th:main_finite}}
\label{sect:proof_main_finite}

To prove \thref{th:main_finite}, we need some more preliminary results.
Throughout this subsection, assume that the alphabet $\mathcal{Y}$ is finite and nonempty.
In this case, given a pair $(X, Y)$, one can define
\begin{align}
P_{X|Y = y}( x )
=
\mathbb{P}\{ X = x \mid Y = y \} ,
\end{align}
provided that $P_{Y}( y ) > 0$.

For a subset $\mathcal{Z} \subset \mathcal{X}$, define
\begin{align}
P_{\mathrm{e}}^{(L)}(X \mid Y \ \| \ \mathcal{Z})
\coloneqq
\min_{f : \mathcal{Y} \to \binom{\mathcal{Z}}{L}} \mathbb{P}\{ X \notin f( Y ) \} .
\label{def:PeZ}
\end{align}
Note that the difference between $P_{\mathrm{e}}^{(L)}(X \mid Y)$ and $P_{\mathrm{e}}^{(L)}(X \mid Y \ \| \ \mathcal{Z})$ is the restriction of the decoding range $\mathcal{Z} \subset \mathcal{X}$, and the inequality $P_{\mathrm{e}}^{(L)}(X \mid Y) \le P_{\mathrm{e}}^{(L)}(X \mid Y \ \| \ \mathcal{Z})$ is trivial from these definitions stated in \eqref{def:Pe} and \eqref{def:PeZ}, respectively.
The following propositions are easy consequences of the proofs of Propositions~\ref{prop:listMAP} and~\ref{prop:boundPe}, and so we omit those proofs in this paper.

\begin{proposition}
\label{prop:listMAP_Z}
It holds that
\begin{align}
P_{\mathrm{e}}^{(L)}(X \mid Y \ \| \ \mathcal{Z})
=
1 - \mathbb{E} \Bigg[ \min_{\mathcal{D} \in \binom{\mathcal{Z}}{L}} \sum_{x \in \mathcal{D}} P_{X|Y}( x ) \Bigg] .
\label{eq:listMAPformula_Z}
\end{align}
\end{proposition}

\begin{proposition}
\label{prop:boundPe_Z}
Let $\beta : \{ 1, \dots, |\mathcal{Z}| \} \to \mathcal{Z}$ be a bijection satisfying $P_{X}( \beta( i ) ) \ge P_{X}( \beta( j ) )$ if $i < j$.
It holds that
\begin{align}
1 - \sum_{x \in \mathcal{Z}} P_{X}( x )
\le
P_{\mathrm{e}}^{(L)}(X \mid Y)
\le
1 - \sum_{x = 1}^{L} P_{X}( \beta( x ) ) .
\label{ineq:boundPe_Z}
\end{align}
\end{proposition}

For a finite subset $\mathcal{Z} \subset \mathcal{X}$, denote by $\Psi( \mathcal{Z} )$ the set of $|\mathcal{Z}| \times |\mathcal{Z}|$ permutation matrices in which both rows and columns are indexed by the elements in $\mathcal{Z}$.
The main idea of proving \thref{th:main_finite} is the following lemma.

\begin{lemma}
\label{lem:FiniteY}
For any $\mathcal{X} \times \mathcal{Y}$-valued r.v.\ $(X, Y)$, there exist a subset $\mathcal{Z} \subset \mathcal{X}$ and an $\mathcal{X} \times \Psi(\mathcal{Z})$-valued r.v.\ $(U, W)$ such that
\begin{align}
|\mathcal{Z}|
& =
L \cdot |\mathcal{Y}| ,
\\
P_{U}( x )
& =
P_{X}( x )
\quad \mathrm{for} \ x \in \mathcal{X} ,
\label{eq:FiniteY_property1} \\
\hspace{-0.75em}
P_{\mathrm{e}}^{(L)}(U \mid W)
& \le
P_{\mathrm{e}}^{(L)}(U \mid W \ \| \ \mathcal{Z})
=
P_{\mathrm{e}}^{(L)}(X \mid Y) ,
\label{eq:FiniteY_property2} \\
\mathsf{H}_{\phi}(U \mid W)
& \ge
\mathsf{H}_{\phi}(X \mid Y) ,
\label{eq:FiniteY_property3} \\
P_{U|W = w}( x )
& =
P_{X}( x )
\quad \mathrm{for} \ x \in \mathcal{X} \setminus \mathcal{Z} \ \mathrm{and} \ w \in \Psi( \mathcal{Z} ) .
\label{eq:FiniteY_property4} 
\end{align}
\end{lemma}

\begin{IEEEproof}[Proof of \lemref{lem:FiniteY}]
Suppose without loss of generality that
\begin{align}
\mathcal{Y}
=
\{ 0, 1, \dots, N-1 \}
\end{align}
for some positive integer $N$.
By the definition of cardinality, one can find a subset $\mathcal{Z} \subset \mathcal{X}$ satisfying (i) $|\mathcal{Z}| = L N$, and (ii) for each $x \in \{ 1, 2, \dots, L \}$ and $y \in \mathcal{Y}$, there exists $z \in \mathcal{Z}$ satisfying
\begin{align}
P_{X|Y=y}( z )
=
P_{X|Y=y}^{\downarrow}( x ) .
\label{eq:construction_Z}
\end{align}
For each $\Pi = \{ \pi_{i, j} \}_{i, j \in \mathcal{Z}} \in \Psi( \mathcal{Z} )$, define the permutation $\varphi_{\Pi} : \mathcal{Z} \to \mathcal{Z}$ by
\begin{align}
\varphi_{\Pi}( z )
\coloneqq
\sum_{w \in \mathcal{Z}} \pi_{z, w} \, w ,
\end{align}
as in \eqref{def:psi} and \eqref{def:bar_psi}.
It is clear that for each $y \in \mathcal{Y}$, there exists at least one $\Pi \in \Psi( \mathcal{Z} )$ such that
\begin{align}
P_{X|Y=y}( \varphi_{\Pi}( x_{1} ) )
\ge
P_{X|Y=y}( \varphi_{\Pi}( x_{2} ) )
\end{align}
for every $x_{1}, x_{2} \in \mathcal{Z}$ satisfying $x_{1} \le x_{2}$, which implies that the permutation $\varphi_{\Pi}$ plays the role of a decreasing rearrangement of $P_{X|Y=y}$ on $\mathcal{Z}$.
To denote such a correspondence between $\mathcal{Y}$ and $\Psi( \mathcal{Z} )$, one can choose an injection $\iota : \mathcal{Y} \to \Psi( \mathcal{Z} )$ appropriately.
In other words, one can find an injection $\iota$ so that
\begin{align}
P_{X|Y=y}( \varphi_{\iota(y)}( x_{1} ) )
\ge
P_{X|Y=y}( \varphi_{\iota(y)}( x_{2} ) )
\label{eq:iota}
\end{align}
for every $y \in \mathcal{Y}$ and $x_{1}, x_{2} \in \mathcal{Z}$ satisfying $x_{1} \le x_{2}$.
We now construct an $\mathcal{X} \times \mathcal{Y} \times \Psi(\mathcal{Z})$-valued r.v.\ $(U, V, W)$ as follows:
The conditional distribution $P_{U|V, W}$ is given by
\begin{align}
P_{U|V = v, W = w}( u )
& =
\begin{cases}
P_{X|Y = v}( \varphi_{\iota(v)} \circ \varphi_{w}( u ) )
& \text{if} \ u \in \mathcal{Z} ,
\\
P_{X|Y = v}( u )
& \text{if} \ u \in \mathcal{X} \setminus \mathcal{Z} ,
\end{cases}
\label{eq:U|VW}
\end{align}
where $\sigma_{1} \circ \sigma_{2}$ stands for the composition of two bijections $\sigma_{1}$ and $\sigma_{2}$.
The induced probability distribution $P_{V}$ of $V$ is given by $P_{V} = P_{Y}$.
Suppose that the independence $V \Perp W$ holds.
As
\begin{align}
P_{U, V, W}
=
P_{U|V, W} \, P_{V} \, P_{W} ,
\end{align}
it remains to determine the induced probability distribution $P_{W}$ of $W$, and we defer to determine it until the last paragraph of this proof.
A direct calculation shows
\begin{align}
P_{U|W = w}( u )
& =
\sum_{v \in \mathcal{Y}} P_{V|W = w}( v ) \, P_{U|V = v, W = w}( u )
\notag \\
& \overset{\mathclap{\text{(a)}}}{=}
\sum_{v \in \mathcal{Y}} P_{Y}( v ) \, P_{U|V = v, W = w}( u )
\notag \\
& \overset{\mathclap{\text{(b)}}}{=}
\begin{cases}
\omega( u, w )
& \text{if} \ u \in \mathcal{Z} ,
\\
P_{X}( u )
& \text{if} \ u \in \mathcal{X} \setminus \mathcal{Z} ,
\end{cases}
\label{eq:Q_XW}
\end{align}
where
\begin{itemize}
\item
(a) follows by the independence $V \Perp W$ and $P_{V} = P_{Y}$, and
\item
(b) follows by \eqref{eq:U|VW} and defining $\omega( u, w )$ so that
\begin{align}
\omega( u, w )
& \coloneqq
\sum_{v \in \mathcal{Y}} P_{Y}( v ) \, P_{X | Y = v}( \varphi_{\iota(v)} \circ \varphi_{w}( u ) )
\label{def:omega}
\end{align}
for each $x \in \mathcal{Z}$ and $w \in \Psi( \mathcal{Z} )$.
\end{itemize}
Now, we readily see from \eqref{eq:Q_XW} that \eqref{eq:FiniteY_property4} holds for \emph{any} induced probability distribution $P_{W}$ of $W$.
Therefore, to complete the proof, it suffices to show that $(U, W)$ satisfies \eqref{eq:FiniteY_property2} and \eqref{eq:FiniteY_property3} with an \emph{arbitrary} choice of $P_{W}$, and $(U, W)$ satisfies \eqref{eq:FiniteY_property1} with an \emph{appropriate} choice of $P_{W}$.

Firstly, we shall prove \eqref{eq:FiniteY_property2}.
For each $w \in \Psi( \mathcal{Z} )$, denote by $\mathcal{D}( w ) \in \binom{\mathcal{Z}}{L}$ the set satisfying
\begin{align}
\varphi_{w}( k )
<
\varphi_{w}( x )
\label{eq:Dw}
\end{align}
for every $k \in \mathcal{D}( w )$ and $x \in \mathcal{Z} \setminus \mathcal{D}( w )$, i.e., it stands for the set of first $L$ elements in $\mathcal{Z}$ under the permutation rule $w \in \Psi( \mathcal{Z} )$.
Then, we have
\begin{align}
P_{\mathrm{e}}^{(L)}(U \mid W)
& \overset{\mathclap{\text{(a)}}}{\le}
P_{\mathrm{e}}^{(L)}(U \mid W \ \| \ \mathcal{Z})
\notag \\
& \overset{\mathclap{\text{(b)}}}{=}
1 - \sum_{w \in \Psi(\mathcal{Z})} P_{W}( w ) \min_{\mathcal{D} \in \binom{\mathcal{Z}}{L}} \sum_{u \in \mathcal{D}} P_{U|W = w}( u )
\notag \\
& \overset{\mathclap{\text{(c)}}}{=}
1 - \sum_{w \in \Psi(\mathcal{Z})} P_{W}( w ) \min_{\mathcal{D} \in \binom{\mathcal{Z}}{L}} \sum_{u \in \mathcal{D}} \omega(u, w)
\notag \\
& \overset{\mathclap{\text{(d)}}}{=}
1 - \sum_{w \in \Psi(\mathcal{Z})} P_{W}( w ) \min_{\mathcal{D} \in \binom{\mathcal{Z}}{L}} \sum_{u \in \mathcal{D}} \sum_{v \in \mathcal{Y}} P_{Y}( v ) \, P_{X | Y = v}( \varphi_{\iota(v)} \circ \varphi_{w}( u ) )
\notag \\
& \overset{\mathclap{\text{(e)}}}{=}
1 - \sum_{w \in \Psi(\mathcal{Z})} P_{W}( w ) \sum_{u \in \mathcal{D}( w )} \sum_{v \in \mathcal{Y}} P_{Y}( v ) \, P_{X | Y = v}( \varphi_{\iota(v)} \circ \varphi_{w}( u ) )
\notag \\
& \overset{\mathclap{\text{(f)}}}{=}
1 - \sum_{w \in \Psi(\mathcal{Z})} P_{W}( w ) \sum_{u = 1}^{L} \sum_{v \in \mathcal{Y}} P_{Y}( v ) \, P_{X | Y = v}^{\downarrow}( u )
\notag \\
& =
1 - \sum_{y \in \mathcal{Y}} P_{Y}( y ) \sum_{x = 1}^{L} P_{X | Y = y}^{\downarrow}( x )
\notag \\
& \overset{\mathclap{\text{(g)}}}{=}
P_{\mathrm{e}}^{(L)}(X \mid Y) ,
\end{align}
where
\begin{itemize}
\item
(a) is an obvious inequality (see the definitions stated in \eqref{def:Pe} and \eqref{def:PeZ}),
\item
(b) follows from \propref{prop:listMAP_Z},
\item
(c) follows from \eqref{eq:Q_XW},
\item
(d) follows from the definition of $\omega(u, w)$ stated in \eqref{def:omega},
\item
(e) follows from \eqref{eq:iota} and \eqref{eq:Dw},
\item
(f) follows from \eqref{eq:construction_Z}, \eqref{eq:iota}, and \eqref{eq:Dw}, and
\item
(g) follows from \propref{prop:listMAP}.
\end{itemize}
Therefore, we obtain \eqref{eq:FiniteY_property2}.

Secondly, we shall prove \eqref{eq:FiniteY_property3}.
We get
\begin{align}
\mathsf{H}_{\phi}(X \mid Y)
& =
\sum_{y \in \mathcal{Y}} P_{Y}( y ) \, \phi( P_{X|Y=y} )
\notag \\
& =
\sum_{w \in \Psi(\mathcal{Z})} P_{W}( w ) \sum_{y \in \mathcal{Y}} P_{Y}( y ) \, \phi( P_{X|Y=y} )
\notag \\
& \overset{\mathclap{\text{(a)}}}{=}
\sum_{w \in \Psi(\mathcal{Z})} P_{W}( w ) \sum_{y \in \mathcal{Y}} P_{Y}( y ) \, \phi( P_{U|V=y, W=w} )
\notag \\
& \overset{\mathclap{\text{(b)}}}{=}
\sum_{w \in \Psi(\mathcal{Z})} P_{W}( w ) \sum_{v \in \mathcal{Y}} P_{V}( v ) \, \phi( P_{U|V=v, W=w} )
\notag \\
& \overset{\mathclap{\text{(c)}}}{\le}
\sum_{w \in \Psi(\mathcal{Z})} P_{W}( w ) \, \phi\Bigg( \sum_{v \in \mathcal{Y}} P_{V}( v ) \, P_{U|V=v, W=w} \Bigg)
\notag \\
& \overset{\mathclap{\text{(d)}}}{=}
\sum_{w \in \Psi(\mathcal{Z})} P_{W}( w ) \, \phi( P_{U|W=w} )
\notag \\
& =
\mathsf{H}_{\phi}(U \mid W) ,
\end{align}
where
\begin{itemize}
\item
(a) follows by the symmetry of $\phi$ and \eqref{eq:U|VW},
\item
(b) follows by $P_{V} = P_{Y}$,
\item
(c) follows by Jensen's inequality, and
\item
(d) follows by the independence $U \Perp W$.
\end{itemize}
Therefore, we obtain \eqref{eq:FiniteY_property3}.

Finally, we shall prove that there exists an induced probability distribution $P_{W}$ satisfying \eqref{eq:FiniteY_property1}.
If we denote by $I \in \Psi( \mathcal{Z} )$ the identity matrix, then it follows from \eqref{def:omega} that
\begin{align}
P_{U|W=I}( u )
=
P_{U|W = w}( \varphi_{w}^{-1}( u ) )
\label{eq:permutation_Q_equiv}
\end{align}
for every $(u, w) \in \mathcal{Z} \times \Psi( \mathcal{Z} )$.
It follows from \eqref{eq:Q_XW} that
\begin{align}
\sum_{x \in \mathcal{Z}} P_{X}( x )
=
\sum_{u \in \mathcal{Z}} P_{U|W=I}( u ) .
\label{eq:sum_U|W=I}
\end{align}
Now, denote by $\beta_{1} : \{ 1, 2, \dots, LN \} \to \mathcal{Z}$ and $\beta_{2} : \{ 1, 2, \dots, LN \} \to \mathcal{Z}$ two bijections satisfying $P_{X}( \beta_{1}( i ) ) \ge P_{X}( \beta_{1}( j ) )$ and $\beta_{2}( i ) < \beta_{2}( j )$, respectively, provided that $i < j$.
That is, the bijection $\beta_{1}$ and $\beta_{2}$ play roles of decreasing rearrangements of $P_{X}$ and $P_{U|W=I}$, respectively, on $\mathcal{Z}$.
Using those bijections, one can rewrite \eqref{eq:sum_U|W=I} as
\begin{align}
\sum_{i = 1}^{LN} P_{X}( \beta_{1}( i ) )
=
\sum_{i = 1}^{LN} P_{U|W=I}( \beta_{2}( i ) ) .
\label{eq:finiteY_majorizaiton1}
\end{align}
In the same way as \eqref{ineq:majorization_PXdr_PXYdr}, it can be verified from \eqref{def:omega} by induction that
\begin{align}
\sum_{i = 1}^{k} P_{X}( \beta_{1}( i ) )
\le
\sum_{i = 1}^{k} P_{U|W=I}( \beta_{2}( i ) )
\label{eq:finiteY_majorizaiton2}
\end{align}
for each $k = 1, 2, \dots, LN$.
Equations~\eqref{eq:finiteY_majorizaiton1} and~\eqref{eq:finiteY_majorizaiton2} are indeed a majorization relation between two finite-dimensional real vectors, because $\beta_{1}$ plays a role of a decreasing rearrangement of $P_{X}$ on $\mathcal{Z}$.
Combining \eqref{eq:permutation_Q_equiv} and this majorization relation, it follows from the Hardy--Littlewood--P\'{o}lya theorem derived in Theorem~8 of \cite{hardy_littlewood_polya_1928} (see also Theorem~2.B.2 of \cite{marshall_olkin_arnold_majorization}) and the finite-dimensional version of Birkhoff's theorem \cite{birkhoff_1946} (see also Theorem~2.A.2 of \cite{marshall_olkin_arnold_majorization}) that there exists an induced probability distribution $P_{W}$ satisfying $P_{U} = P_{X}$, i.e., Equation~\eqref{eq:FiniteY_property1} holds, as in \eqref{eq:majorization_partial1}--\eqref{def:Past_joint2}.
This completes the proof of \lemref{lem:FiniteY}.
\end{IEEEproof}

\begin{remark}
\label{rem:FiniteY}
\lemref{lem:FiniteY} can restrict the feasible region of the supremum in \eqref{def:main_object} from a countably infinite alphabet $\mathcal{X}$ to a finite alphabet $\mathcal{Z}$ in the sense of \eqref{eq:FiniteY_property4}.
Specifically, if $\mathcal{Y}$ is finite, it suffices to vary at most $|\mathcal{Z}| = L \cdot |\mathcal{Y}|$ probability masses $\{ P_{X|Y=y}( x ) \}_{x \in \mathcal{Z}}$ for each $y \in \mathcal{Y}$.
\lemref{lem:FiniteY} is useful not only to prove \thref{th:main_finite} but also to prove \propref{prop:infty} of \sectref{sect:infty} (see \appref{app:infty} for the proof).
\end{remark}

As with \eqref{def:region_R}, for a subset $\mathcal{Z} \subset \mathcal{X}$, we define
\begin{align}
\mathcal{R}(Q, L, \varepsilon, \mathcal{Y}, \mathcal{Z})
\coloneqq
\left\{ (X, Y) \ \middle|
\begin{array}{l}
\text{$X$ is $\mathcal{X}$-valued} , \\
\text{$Y$ is $\mathcal{Y}$-valued} , \\
P_{\mathrm{e}}^{(L)}(X \mid Y \, \| \, \mathcal{Z}) \le \varepsilon , \\
P_{X} = Q , \\
P_{X|Y = y}( x )
=
Q( x )
\quad \forall (x, y) \in (\mathcal{X} \setminus \mathcal{Z}) \times \mathcal{Y}
\end{array}
\right\} ,
\label{def:region_R_Z}
\end{align}
provided that $\mathcal{Y}$ is finite.
It is clear that \eqref{def:region_R_Z} coincides with \eqref{def:region_R} if $\mathcal{Z} = \mathcal{X}$, i.e., it holds that
\begin{align}
\mathcal{R}(Q, L, \varepsilon, \mathcal{Y}, \mathcal{X})
=
\mathcal{R}(Q, L, \varepsilon, \mathcal{Y}) .
\end{align}
Note from \lemref{lem:FiniteY} that for each system $(Q, L, \varepsilon, \mathcal{Y})$ satisfying \eqref{eq:range_epsilon_list}, there exists a subset $\mathcal{Z} \subset \mathcal{X}$ such that $|\mathcal{Z}| = L \cdot |\mathcal{Y}|$ and $\mathcal{R}(Q, L, \varepsilon, \mathcal{Y}, \mathcal{Z})$ is nonempty, provided that $\mathcal{Y}$ is finite.

Another important idea of proving \thref{th:main_finite} is to apply \lemref{lem:max_necessity} for this collection of r.v.'s.
The correction $\mathcal{R}(Q, L, \varepsilon, \mathcal{Y}, \mathcal{Z})$ does not, however, have balanced conditional distributions of \eqref{eq:balanced_X|Y} in general, as with \eqref{def:region_R}.
Fortunately, similar to \lemref{lem:balanced_Pe}, the following lemma can avoid this issue by blowing-up the collection $\mathcal{R}(Q, L, \varepsilon, \mathcal{Y}, \mathcal{Z})$ via the \emph{finite-dimensional version of Birkhoff's theorem} \cite{birkhoff_1946}.

\begin{lemma}
\label{lem:Pxy_finite}
Suppose that $\mathcal{Z} \subset \mathcal{X}$ is finite and $\mathcal{R}(Q, L, \varepsilon, \mathcal{Y}, \mathcal{Z})$ is nonempty.
If $|\mathcal{Z}| \le |\mathcal{Y}|! < \infty$, then the collection $\mathcal{R}(Q, L, \varepsilon, \mathcal{Y}, \mathcal{Z})$ has balanced conditional distributions.
\end{lemma}

\begin{IEEEproof}[Proof of \lemref{lem:Pxy_finite}]
\lemref{lem:Pxy_finite} can be proven in a similar fashion to the proof of \lemref{lem:balanced_Pe}.
As this proof is slightly long as with \lemref{lem:balanced_Pe}, we only give a sketch of the proof as follows.

As $|\Psi( \mathcal{Z} )| = |\mathcal{Z}|!$, we may assume without loss of generality that $\mathcal{Y} = \Psi( \mathcal{Z} )$.
For the sake of brevity, we write
\begin{align}
{\tilde{\mathcal{R}}}
=
\mathcal{R}(Q, L, \varepsilon, \mathcal{Y}, \mathcal{Z})
\end{align}
in this proof.
For a pair $(X, Y) \in \tilde{\mathcal{R}}$, construct another $\mathcal{X} \times \mathcal{Y}$-valued r.v.\ $(U, V)$, as in \eqref{eq:cond_Q}, so that $P_{U|V=y}( x ) = Q( x )$ for every $(x, y) \in (\mathcal{X} \setminus \mathcal{Z}) \times \mathcal{Y}$.
By such a construction of \eqref{eq:cond_Q}, the condition stated in \eqref{eq:balanced_X|Y} is obviously satisfied.
In the same way as \eqref{eq:same_Pe_P-Q}, we can verify that
\begin{align}
P_{\mathrm{e}}^{(L)}(U \mid V \, \| \, \mathcal{Z})
=
P_{\mathrm{e}}^{(L)}(X \mid Y \, \| \, \mathcal{Z}) .
\end{align}
Moreover, employing the finite-dimensional version of Birkhoff's theorem \cite{birkhoff_1946} (also known as the Birkhoff--von~Neumann decomposition) instead of \lemref{lem:infinite-Birkhoff}, we can also find an induced probability distribution $P_{V}$ of $V$ so that $P_{U} = Q$ in the same way as \eqref{eq:marginal_PxQx}.
Therefore, for any $(X, Y) \in \tilde{\mathcal{R}}$, one can find $(U, V) \in \tilde{\mathcal{R}}$ satisfying \eqref{eq:balanced_X|Y}.
This completes the proof of \lemref{lem:Pxy_finite}.
\end{IEEEproof}

Let $\mathcal{Z} \subset \mathcal{X}$ be a subset.
Consider a bijection $\beta : \{ 1, 2, \dots, |\mathcal{Z}| \} \to \mathcal{Z}$ satisfying $Q( \beta( i ) ) \ge Q( \beta( j ) )$ whenever $i < j$, i.e., it plays a role of a decreasing rearrangement of $Q$ on $\mathcal{Z}$.
Thereforeforth, suppose that $(Q, L, \varepsilon, \mathcal{Y}, \mathcal{Z})$ satisfies
\begin{align}
1 - \sum_{x \in \mathcal{Z}} Q( x )
\le
\varepsilon
\le
1 - \sum_{x = 1}^{L} Q( \beta( x ) ) .
\label{eq:range_varepsilon_Z}
\end{align}
Define the \emph{extremal distribution of type-3} by the following $\mathcal{X}$-marginal:
\begin{align}
P_{\operatorname{type-3}}( x )
=
P_{\operatorname{type-3}}^{(Q, L, \varepsilon, \mathcal{Y}, \mathcal{Z})}( x )
\coloneqq
\begin{cases}
\mathcal{V}_{3}( J_{3} )
& \mathrm{if} \ x \in \mathcal{Z} \ \mathrm{and} \ J_{3} \le \beta_{1}^{-1}( x ) \le L ,
\\
\mathcal{W}_{3}( K_{3} )
& \mathrm{if} \ x \in \mathcal{Z} \ \mathrm{and} \ L < \beta_{1}^{-1}( x ) \le K_{3} ,
\\
Q( x )
& \mathrm{otherwise} ,
\end{cases}
\label{def:type7}
\end{align}
where the weight $\mathcal{V}_{3}( j )$ is defined by
\begin{align}
\mathcal{V}_{3}( j )
=
\mathcal{V}_{3}^{(Q, L, \varepsilon, \mathcal{Y}, \mathcal{Z})}( j )
\coloneqq
\frac{ (1 - \varepsilon) - \sum_{x = 1}^{j-1} Q( \beta_{1}( x ) ) }{ L - j + 1 }
\label{def:Vbar}
\end{align}
for each integer $1 \le j \le L$, the weight $\mathcal{W}_{3}( k )$ is defined by
\begin{align}
\mathcal{W}_{3}( k )
=
\mathcal{W}_{3}^{(Q, L, \varepsilon, \mathcal{Y}, \mathcal{Z})}( k )
\coloneqq
\begin{dcases}
-1
& \mathrm{if} \ k = L ,
\\
\frac{ \sum_{x = 1}^{k} Q( \beta_{1}( x ) ) - (1 - \varepsilon) }{ k - L }
& \mathrm{if} \ k > L
\end{dcases}
\label{def:Wbar}
\end{align}
for each integer $L \le k \le L \cdot |\mathcal{Y}|$, the integer $J_{3}$ is chosen so that
\begin{align}
J_{3}
=
J_{3}(Q, L, \varepsilon, \mathcal{Y}, \mathcal{Z})
\coloneqq
\min \{ 1 \le j \le L \mid Q( \beta_{1}( j ) ) \le \mathcal{V}_{3}( j ) \} ,
\label{def:Jbar}
\end{align}
and the integer $K_{3}$ is chosen so that
\begin{align}
K_{3}
=
K_{3}(Q, L, \varepsilon, \mathcal{Y}, \mathcal{Z})
\coloneqq
\max \{ L \le k \le L \cdot |\mathcal{Y}| \mid \mathcal{W}_{3}( k ) \le P_{X}( \beta_{1}( k ) ) \} .
\label{def:Kbar}
\end{align}

\begin{remark}
The extremal distribution of type-3 can be specialized to both extremal distribution of type-2 defined in \eqref{def:type2} and Ho--Verd\'{u}'s truncated distribution defined in Equation~(17) of \cite{ho_verdu_2010}, respectively.
\end{remark}

The following lemma shows a relation between the type-2 and the type-3.

\begin{lemma}
\label{lem:prefix}
Suppose that $|\mathcal{Z}| = L \cdot |\mathcal{Y}|$.
Then, it holds that
\begin{align}
P_{\operatorname{type-2}}^{(Q, L, \varepsilon, \mathcal{Y})}
\prec
P_{\operatorname{type-3}}^{(Q, L, \varepsilon, \mathcal{Y}, \mathcal{Z})} .
\end{align}
\end{lemma}

\begin{IEEEproof}[Proof of \lemref{lem:prefix}]
We readily see that
\begin{align}
P_{\operatorname{type-2}}
=
P_{\operatorname{type-3}} ,
\end{align}
provided that $\mathcal{Z} = \{ 1, 2, \dots, L \cdot |\mathcal{Y}| \}$ and $Q = Q^{\downarrow}$, because $\beta : \{ 1, 2, \dots, |\mathcal{Z}| \} \to \mathcal{Z}$ used in \eqref{def:type7} is the identity mapping in this case.
Actually, we may assume without loss of generality that $Q = Q^{\downarrow}$.

Although
\begin{align}
P_{\operatorname{type-2}}
=
P_{\operatorname{type-2}}^{\downarrow}
\end{align}
does not hold in general, we can see from the definition of $P_{\operatorname{type-2}}$ stated in \eqref{def:type2} that
\begin{align}
P_{\operatorname{type-2}}( x )
=
P_{\operatorname{type-2}}^{\downarrow}( x )
\end{align}
for each $x = 1, 2, \dots, L$.
Therefore, as
\begin{align}
P_{\operatorname{type-2}}( x )
=
Q( x )
\le
P_{\operatorname{type-3}}( x )
\end{align}
for each $x = 1, 2, \dots, J-1$, it follows that
\begin{align}
\sum_{x = 1}^{k} P_{\operatorname{type-2}}^{\downarrow}( x )
& \le
\sum_{x = 1}^{k} P_{\operatorname{type-3}}^{\downarrow}( x )
\label{eq:type6_type7}
\end{align}
for each $k = 1, 2, \dots, J-1$.
By the definitions \eqref{def:V}, \eqref{def:J}, \eqref{def:Vbar}, and \eqref{def:Jbar}, it can be verified that
\begin{align}
J
& \ge
J_{3} ,
\\
\mathcal{V}( J )
& \le
\mathcal{V}_{3}( J_{3} ) .
\end{align}
Thus, as
\begin{align}
P_{\operatorname{type-2}}^{\downarrow}( x )
=
\mathcal{V}( J )
\end{align}
for each $x = J, J+1, \dots, L$, it follows that
\begin{align}
P_{\operatorname{type-3}}^{\downarrow}( x )
\ge
\mathcal{V}_{3}( J_{3} )
\end{align}
for each $x = J, J+1, \dots, L$; which implies that \eqref{eq:type6_type7} also holds for each $k = J, J+1, \dots, L$.
Therefore, we observe that $P_{\operatorname{type-3}}$ majorizes $P_{\operatorname{type-2}}$ over the subset $\{ 1, 2, \dots, L \} \subset \mathcal{X}$.

We prove the rest of the majorization relation by contradiction.
Namely, assume that
\begin{align}
\sum_{x = 1}^{l} P_{\operatorname{type-2}}^{\downarrow}( x )
& >
\sum_{x = 1}^{l} P_{\operatorname{type-3}}^{\downarrow}( x )
\label{eq:contradict_type6_type7}
\end{align}
for some integer $l \ge L+1$.
By the definitions stated in \eqref{def:W2}, \eqref{def:K4}, \eqref{def:Wbar}, and \eqref{def:Kbar}, it can be verified that
\begin{align}
K_{2}
& \le
K_{3} ,
\\
\mathcal{W}( K_{2} )
& \ge
\mathcal{W}_{3}( K_{3} ) .
\end{align}
Thus, as
\begin{align}
P_{\operatorname{type-2}}( x )
& =
\mathcal{W}( K_{2} )
\le
Q( x )
&& (\mathrm{for} \ x = L+1, L+2, \dots, K_{2}) ,
\\
P_{\operatorname{type-3}}( x )
& =
\mathcal{W}_{3}( K_{3} )
\le
Q( x )
&& (\mathrm{for} \ x = \beta_{1}( L+1 ), \beta_{1}( L+2 ), \dots, \beta_{1}( K_{3} )) ,
\end{align}
it follows that
\begin{align}
P_{\operatorname{type-2}}( x )
\ge
P_{\operatorname{type-3}}( x )
\end{align}
for every $x = l, l+1, \dots$, which implies together with the hypothesis \eqref{eq:contradict_type6_type7} that
\begin{align}
\sum_{x = l}^{\infty} P_{\operatorname{type-2}}^{\downarrow}( x )
>
\sum_{x = l}^{\infty} P_{\operatorname{type-3}}^{\downarrow}( x ) .
\end{align}
This, however, contradicts to the definition of probability distributions, i.e., the sum of probability masses is strictly larger than one.
This completes the proof of \lemref{lem:prefix}.
\end{IEEEproof}

Similar to \eqref{def:PeZ}, we now define
\begin{align}
P_{\mathrm{e}}^{(L)}( X \, \| \, \mathcal{Z}  )
\coloneqq
\min_{\mathcal{D} \in \binom{\mathcal{Z}}{L}} \mathbb{P}\{ X \in \mathcal{D} \} .
\end{align}
As with \propref{prop:boundPe_Z}, we can verify that
\begin{align}
P_{\mathrm{e}}^{(L)}( X \, \| \, \mathcal{Z}  )
& =
1 - \min_{\mathcal{D} \in \binom{\mathcal{Z}}{L}} \sum_{x \in \mathcal{D}} P_{X}( x )
=
1 - \sum_{x = 1}^{L} P_{X}( \beta( x ) ) .
\label{eq:listMAPmarg_Z}
\end{align}
Therefore, the restriction stated in \eqref{eq:range_varepsilon_Z} comes from the same observation as \eqref{eq:range_epsilon_list} (see Propositions~\ref{prop:boundPe} and~\ref{prop:boundPe_Z}).
In view of \eqref{eq:listMAPmarg_Z}, we write $P_{\mathrm{e}}^{(L)}( Q \, \| \, \mathcal{Z}  ) = P_{\mathrm{e}}^{(L)}( X \, \| \, \mathcal{Z}  )$ if $P_{X} = Q$.
As in \lemref{lem:cond_to_marg}, the following lemma holds.

\begin{lemma}
\label{lem:majorization_Z}
Suppose that an $\mathcal{X}$-marginal $R$ satisfies that (i) $R$ majorizes $Q$, (ii) $P_{\mathrm{e}}^{(L)}(R \, \| \, \mathcal{Z}) \le \varepsilon$, and (iii) $R( k ) = Q( k )$ for each $k \in \mathcal{X} \setminus \mathcal{Z}$.
Then, it holds that $R$ majorizes $P_{\operatorname{type-3}}$ as well.
\end{lemma}

\begin{IEEEproof}[Proof of \lemref{lem:majorization_Z}]
Since
\begin{align}
R( x )
=
P_{\operatorname{type-3}}( x )
=
Q( x )
\end{align}
for every $x \in \mathcal{X} \setminus \mathcal{Z}$, it suffices to verify the majorization relation over $\mathcal{Z}$.
Denote by $\beta_{1} : \{ 1, 2, \dots, L \cdot |\mathcal{Y}| \} \to \mathcal{Z}$ and $\beta_{2} : \{ 1, 2, \dots, L \cdot |\mathcal{Y}| \} \to \mathcal{Z}$ two bijection satisfying $R( \beta_{1}( i ) ) \ge R( \beta_{1}( j ) )$ and $\beta_{2}( i ) \le \beta_{2}( j )$, respectively, whenever $i < j$.
In other words, two bijections $\beta_{1}$ and $\beta_{2}$ play roles of decreasing rearrangements of $R$ and $P_{3}$, respectively, on $\mathcal{Z}$.
That is, we shall prove that
\begin{align}
\sum_{x = 1}^{k} P_{\operatorname{type-3}}( \beta_{2}( x ) )
\le
\sum_{x = 1}^{k} R( \beta_{1}( x ) )
\label{eq:majorization_Z}
\end{align}
for every $k = 1, 2, \dots, |\mathcal{Z}|$.

As $R$ majorizes $Q$, it follows from \eqref{def:type7} that \eqref{eq:majorization_Z} holds for each $k = 1, 2, \dots, J_{3}-1$.
Moreover, we readily see from \eqref{def:type7} that
\begin{align}
\sum_{x = 1}^{L} P_{\operatorname{type-3}}( \beta_{2}( x ) )
& =
1 - \varepsilon .
\end{align}
Therefore, it follows from \lemref{lem:majorizes_uniform} and the hypothesis $P_{\mathrm{e}}^{(L)}(R \, \| \, \mathcal{Z}) \le \varepsilon$ that \eqref{eq:majorization_Z} holds for each $k = J_{3}, J_{3} + 1, \dots, L$.
Similarly, since \eqref{eq:majorization_Z} holds with equality if $k = |\mathcal{Z}|$, it also follows from \lemref{lem:majorizes_uniform} that \eqref{eq:majorization_Z} holds for each $k = L+1, L+2, \dots |\mathcal{Z}|$.
Therefore, we observe that $R$ majorizes $P_{\operatorname{type-3}}$.
This completes the proof of \lemref{lem:majorization_Z}.
\end{IEEEproof}

Finally, we can prove \thref{th:main_finite} by using the above lemmas.

\begin{IEEEproof}[Proof of \thref{th:main_finite}]
For the sake of brevity, we define
\begin{align}
\mathcal{R}_{1}
& \coloneqq
\mathcal{R}(Q, L, \varepsilon, \mathcal{Y}) ,
\\
\mathcal{R}_{2}
& \coloneqq
\bigcup_{\mathcal{Z} \subset \mathcal{X} : |\mathcal{Z}| = L \cdot |\mathcal{Y}|} \mathcal{R}(Q, L, \varepsilon, \mathcal{Y}, \mathcal{Z}) ,
\\
\mathcal{R}_{3}
& \coloneqq
\bigcup_{\mathcal{Z} \subset \mathcal{X} : |\mathcal{Z}| = L \cdot |\mathcal{Y}|} \mathcal{R}(Q, L, \varepsilon, \mathcal{Y} \cup \Psi( \mathcal{Z} ), \mathcal{Z}) ,
\\
\mathcal{P}_{4}
& \coloneqq
\left\{ R \in \mathcal{P}( \mathcal{X} ) \ \middle|
\begin{array}{r}
\exists \mathcal{Z} \subset \mathcal{X} \ \mathrm{s.t.} \ |\mathcal{Z}| = L \cdot |\mathcal{Y}| , \\ P_{\mathrm{e}}^{(L)}(R \, \| \, \mathcal{Z}) \le \varepsilon , \\
R(x) = Q(x) \ \mathrm{for} \ x \in \mathcal{X} \setminus \mathcal{Z}
\end{array}
\right\} .
\end{align}
Then, we have
\begin{align}
\mathbb{H}_{\phi}(Q, L, \varepsilon, \mathcal{Y})
& \overset{\mathclap{\text{(a)}}}{=}
\sup_{(X, Y) \in \mathcal{R}_{1}} \mathsf{H}_{\phi}(X \mid Y)
\notag \\
& \overset{\mathclap{\text{(b)}}}{=}
\sup_{(X, Y) \in \mathcal{R}_{2}} \mathsf{H}_{\phi}(X \mid Y)
\notag \\
& \overset{\mathclap{\text{(c)}}}{\le}
\sup_{(X, Y) \in \mathcal{R}_{3}} \mathsf{H}_{\phi}(X \mid Y)
\notag \\
& \overset{\mathclap{\text{(d)}}}{=}
\sup_{\substack{(X, Y) \in \mathcal{R}_{3} : \\ \text{$(X, Y)$ is connected uniform-dispersively} }} \mathsf{H}_{\phi}(X \mid Y)
\notag \\
& \overset{\mathclap{\text{(e)}}}{\le}
\sup_{R \in \mathcal{P}_{4}} \phi( R )
\notag \\
& \overset{\mathclap{\text{(f)}}}{\le}
\sup_{\mathcal{Z} \subset \mathcal{X} : |\mathcal{Z}| = L \cdot |\mathcal{Y}|} \phi( P_{\operatorname{type-3}} )
\notag \\
& \overset{\mathclap{\text{(g)}}}{\le}
\phi( P_{\operatorname{type-2}} ) ,
\label{eq:chain_main_finite}
\end{align}
where
\begin{itemize}
\item
(a) follows from the definition of $\mathcal{R}(Q, L, \varepsilon, \mathcal{Y})$ stated in \eqref{def:region_R},
\item
(b) follows from \lemref{lem:FiniteY} and the definition of $\mathcal{R}(Q, L, \varepsilon, \mathcal{Y}, \mathcal{Z})$ stated in \eqref{def:region_R_Z},
\item
(c) follows from the inclusion relation
\begin{align}
\mathcal{R}(Q, L, \varepsilon, \mathcal{Y}, \mathcal{Z}) \:
\subset
\mathcal{R}(Q, L, \varepsilon, \mathcal{Y} \cup \Psi( \mathcal{Z} ), \mathcal{Z}) ,
\end{align}
\item
(d) follows from Lemmas~\ref{lem:max_necessity} and~\ref{lem:Pxy_finite},
\item
(e) follows from \lemref{lem:convex_combination_majorization},
\item
(f) follows from \lemref{lem:majorization_Z}, and
\item
(g) follows from \propref{prop:Schur_convex} and \lemref{lem:prefix}.
\end{itemize}
Inequalities~\eqref{eq:chain_main_finite} are indeed the Fano-type inequality stated in \eqref{eq:main_finite} of \thref{th:main_finite}.

Finally, supposing that $|\mathcal{Y}| \ge (K_{2} - J)^{2} + 1$, we shall construct a jointly distributed pair $(X, Y)$ satisfying
\begin{align}
\mathsf{H}_{\phi}(X \mid Y)
& =
\phi( P_{\operatorname{type-2}} ) ,
\label{eq:cond-inf-measure_type-2} \\
P_{\mathrm{e}}^{(L)}(X \mid Y)
& =
\varepsilon ,
\label{eq:error-probab_type-2} \\
P_{X}( x )
& =
Q^{\downarrow}( x )
\qquad (\mathrm{for} \ x \in \mathcal{X}) .
\label{eq:fixed_X-marginal_type-2}
\end{align}
Similar to \eqref{eq:majorization_partial1} and~\eqref{eq:majorization_partial2}, we see that
\begin{align}
\sum_{x = J}^{k} Q^{\downarrow}( x )
& \le
\sum_{x = J}^{k} P_{\operatorname{type-2}}( x )
\qquad \text{for} \ J \le k \le K_{2} ,
\end{align}
and
\begin{align}
\sum_{x = J}^{K_{2}} Q^{\downarrow}( x )
& =
\sum_{x = J}^{K_{2}} P_{\operatorname{type-2}}( x ) .
\end{align}
This is a majorization relation between two $(K_{2} - J + 1)$-dimensional real vectors; and thus, it follows from the Hardy--Littlewood--P\'{o}lya theorem \cite[Theorem~8]{hardy_littlewood_polya_1928} (see also \cite{marshall_olkin_arnold_majorization}, Theorem~2.B.2) that there exists a $(K_{2} - J + 1) \times (K_{2} - J + 1)$ doubly stochastic matrix $\mathbf{M} = \{ m_{i, j} \}_{i, j = J}^{K_{2}}$ satisfying
\begin{align}
Q^{\downarrow}( i )
=
\sum_{j = J}^{K_{2}} m_{i, j} \, P_{\operatorname{type-2}}( j )
\label{eq:DSM_Past_type-2}
\end{align}
for each $J \le i \le K_{2}$.
Moreover, it follows from Marcus--Ree's or Farahat--Mirsky's refinement of the finite-dimensional version of Birkhoff's theorem derived in \cite{marcus_ree_1959} or Theorem~3 of \cite{farahat_mirsky_1960}, respectively (see also Theorem~2.F.2 of \cite{marshall_olkin_arnold_majorization}), that there exists a pair of a probability vector $\bvec{\lambda} = ( \lambda_{y} )_{y \in \mathcal{Y}}$ and a collection $\{ \{ \pi_{i, j}^{(y)} \}_{i, j = J}^{K_{2}} \}_{y \in \mathcal{Y}}$ of $(K_{2} - J + 1) \times (K_{2} - J + 1)$ permutation matrices such that
\begin{align}
m_{i, j}
=
\sum_{y \in \mathcal{Y}} \lambda_{y} \, \pi_{i, j}^{(y)}
\end{align}
for every $J \le i, j \le K_{2}$.
Using them, construct a pair $(X, Y)$ via the following distributions,
\begin{align}
P_{X|Y=y}( x )
& =
\begin{cases}
P_{\operatorname{type-2}}( x )
& \mathrm{if} \ 1 \le x < J \ \mathrm{or} \ K_{2} < x < \infty ,
\\
P_{\operatorname{type-2}}( \tilde{\psi}_{y}( x ) )
& \mathrm{if} \ J \le x \le K_{2} ,
\end{cases}
\\
P_{Y}( y )
& =
\lambda_{y} ,
\end{align}
where $\tilde{\psi}_{y}$ is defined as in \eqref{def:bar_psi}.
Similar to \sectref{sect:proof_main_list}, we now observe that \eqref{eq:cond-inf-measure_type-2}--\eqref{eq:fixed_X-marginal_type-2} hold.
This implies together with \eqref{eq:chain_main_finite} that the constructed pair $(X, Y)$ achieves the supremum in \eqref{def:main_object}.
Furthermore, since $P_{\operatorname{type-2}}$ and $Q^{\downarrow}$ differ at most $\binom{ K_{2} - J + 1 }{ L - J + 1 }$ probability masses, it follows that the collection $\{ P_{X|Y=y} \}_{y \in \mathcal{Y}}$ consists of at most $\binom{ K_{2} - J + 1 }{ L - J + 1 }$ distinct distributions.
Namely, the condition that $|\mathcal{Y}| \ge \binom{ K_{2} - J + 1 }{ L - J + 1 }$ is also sufficient to construct a jointly distributed pair $(X, Y)$ satisfying \eqref{eq:cond-inf-measure_type-2}--\eqref{eq:fixed_X-marginal_type-2}.
This completes the proof of \thref{th:main_finite}.
\end{IEEEproof}

\begin{remark}
Step~(b) in \eqref{eq:chain_main_finite} is a key of proving \thref{th:main_finite}; it is the reduction step from infinite to finite-dimensional settings via \lemref{lem:FiniteY} (see also \remref{rem:FiniteY}).
Note that this proof technique is not applicable when $\mathcal{Y}$ is infinite, while the proof of \thref{th:main_list} works well for infinite $\mathcal{Y}$.
\end{remark}

\subsection{Proof of \thref{th:main_M}}
\label{sect:main_M}

It is known that every discrete probability distribution on $\{ 1, \dots, M \}$ majorizes the uniform distribution on $\{ 1, \dots, M \}$.
Thus, since
\begin{align}
P_{\operatorname{type-0}}^{(M, L, \varepsilon)}
=
P_{\operatorname{type-1}}^{(\mathrm{Unif}_{M}, L, \varepsilon)}
\end{align}
with the uniform distribution $\mathrm{Unif}_{M}$ on $\{ 1, \dots, M \}$, it follows from \lemref{lem:cond_to_marg} that
\begin{align}
P_{\operatorname{type-0}}^{(M, L, \varepsilon)}
\prec
P_{\operatorname{type-1}}^{(Q, L, \varepsilon)}
\end{align}
if $\supp( Q ) \subset \{ 1, \dots, M \}$.
Therefore, it follows from \propref{prop:Schur_convex} and Theorems~\ref{th:main_list} and~\ref{th:main_zero} that
\begin{align}
\mathbb{H}_{\phi}(M, L, \varepsilon, \mathcal{Y})
\le
\phi( P_{\operatorname{type-0}} ) .
\label{eq:UB_type0}
\end{align}

Finally, it is easy to see that
\begin{align}
\mathsf{H}_{\phi}(X \mid Y)
& =
\phi( P_{\operatorname{type-0}} ) ,
\\
P_{\mathrm{e}}^{(L)}(X \mid Y)
& =
\varepsilon ,
\end{align}
provided that
\begin{align}
P_{X|Y}( x )
=
P_{\operatorname{type-0}}( x )
\qquad (\mathrm{a.s.})
\end{align}
for every $1 \le x \le M$.
This implies the existence of a pair $(X, Y)$ achieving the maximum in \eqref{def:H_M}; and therefore, the equality \eqref{eq:UB_type0} holds.
This completes the proof of \thref{th:main_M}.
\hfill\IEEEQEDhere

\section{Proofs of Asymptotic Behaviors on Equivocations}
\label{sect:proof_equivocation}

In this section, we prove Theorems~\ref{th:Fano_meets_AEP}--\ref{th:VanishingAMsymbolerror}.

\subsection{Proof of \thref{th:Fano_meets_AEP}}
\label{sect:Fano_meets_AEP}

Defining the \emph{variational distance} between two $\mathcal{X}$-marginals $P$ and $Q$ by
\begin{align}
d(P, Q)
\coloneqq
\frac{ 1 }{ 2 } \sum_{x \in \mathcal{X}} \big| P( x ) - Q( x ) \big| ,
\end{align}
we now introduce the following lemma, which is useful to prove \thref{th:Fano_meets_AEP}.

\begin{lemma}[{\cite{ho_yeung_2010_TV}, Theorem~3}]
\label{lem:TV_continuity}
Let $Q$ be an $\mathcal{X}$-marginal, and $0 \le \delta \le 1 - Q^{\downarrow}( 1 )$ a real number.
Then, it holds that
\begin{align}
\min_{ R \in \mathcal{P}(\mathcal{X}) : d(Q, R) \le \delta } H( R )
=
H( S^{(Q, \delta)} ) ,
\end{align}
where the $\mathcal{X}$-marginal $S^{(Q, \delta)}$ is defined by
\begin{align}
S^{(Q, \delta)}( x )
\coloneqq
\begin{dcases}
Q^{\downarrow}( x ) + \delta
& \mathrm{if} \ x = 1 ,
\\
Q^{\downarrow}( x )
& \mathrm{if} \ 1 < x < B ,
\\
\sum_{k = B}^{\infty} Q^{\downarrow}( k ) - \delta
& \mathrm{if} \ x = B ,
\\
0
& \mathrm{if} \ x > B ,
\end{dcases}
\end{align}
and the integer $B$ is chosen so that
\begin{align}
B
\coloneqq
\sup\bigg\{ b \ge 1 \ \bigg| \ \sum_{k = b}^{\infty} Q^{\downarrow}( k ) \ge \delta \bigg\} .
\end{align}
\end{lemma}

For the sake of brevity, in this proof, we write
\begin{align}
\varepsilon_{n}
& \coloneqq
P_{\mathrm{e}}^{(L_{n})}(X_{n} \mid Y_{n}) ,
\\
P_{n}
& \coloneqq
P_{X_{n}}^{\downarrow} ,
\\
P_{1, n}
& \coloneqq
P_{\operatorname{type-1}}^{(P_{n}, L_{n}, \varepsilon_{n})}
\end{align}
for each $n \ge 1$.
Suppose that $\varepsilon_{n} = \mathrm{o}( 1 )$ as $n \to \infty$.
By \corref{cor:Shannon_list}, instead of \eqref{eq:vanishing_normalizedRenyi_list1}, it suffices to verify that
\begin{align}
\big| H( P_{1, n} ) - \log L_{n} \big|^{+} = \mathrm{o}\big( H(X_{n}) \big) .
\label{eq:continuity_Shannon_type5_normalized}
\end{align}
As $\supp( P_{1, n} ) = \{ 1, \dots, L_{n} \}$ if $\varepsilon_{n} = 0$, we may assume without loss of generality that $0 < \varepsilon_{n} < 1$.

Define two $\mathcal{X}$-marginals $Q_{n}^{(1)}$ and $Q_{n}^{(2)}$ by
\begin{align}
Q_{n}^{(1)}( x )
& =
\begin{dcases}
\frac{ P_{1, n}( x ) }{ 1 - \varepsilon_{n} }
& \mathrm{if} \ 1 \le x \le L_{n} ,
\\
0
& \mathrm{if} \ x \ge L_{n} + 1 ,
\end{dcases}
\label{def:Q1} \\
Q_{n}^{(2)}( x )
& =
\begin{dcases}
0
& \mathrm{if} \ 1 \le x \le L_{n} ,
\\
\frac{ P_{1, n}( x ) }{ \varepsilon_{n} }
& \mathrm{if} \ x \ge L_{n} + 1
\end{dcases}
\label{def:Q2}
\end{align}
for each $n \ge 1$.
As $Q_{n}^{(1)}$ majorizes the uniform distribution on $\{ 1, 2, \dots, L_{n} \}$, it is clear from the Schur-concavity property of the Shannon entropy that
\begin{align}
H( Q_{n}^{(1)} )
\le
\log L_{n} .
\end{align}
Thus, since
\begin{align}
P_{1, n}
=
(1 - \varepsilon_{n}) \, Q_{n}^{(1)} + \varepsilon_{n} \, Q_{n}^{(2)} ,
\end{align}
it follows by strong additivity of the Shannon entropy (cf.\ Property~(1.2.6) of \cite{aczel_daroczy_1975}) that
\begin{align}
H( P_{1, n} )
& =
h_{2}( \varepsilon_{n} ) + (1 - \varepsilon_{n}) \, H( Q_{n}^{(1)} ) + \varepsilon_{n} \, H( Q_{n}^{(2)} )
\notag \\
& \le
h_{2}( \varepsilon_{n} ) + (1 - \varepsilon_{n}) \log L_{n} + \varepsilon_{n} \, H( Q_{n}^{(2)} ) .
\label{eq:H_Ptype5}
\end{align}
Thus, since $h_{2}( \varepsilon_{n} ) = \mathrm{o}( 1 )$, it suffices to verify the asymptotic behavior of the third term in the right-hand side of \eqref{eq:H_Ptype5}, i.e., whether
\begin{align}
\varepsilon_{n} \, H( Q_{n}^{(2)} )
=
\mathrm{o}\big( H(X_{n}) \big)
\label{eq:vanishing_H(Q)}
\end{align}
holds or not.

Consider the $\mathcal{X}$-marginal $Q_{n}^{(3)}$ given by
\begin{align}
Q_{n}^{(3)}( x )
=
\frac{ P_{n}( x ) - \varepsilon_{n} \, Q_{n}^{(2)}( x ) }{ 1 - \varepsilon_{n} }
\label{def:Q3}
\end{align}
for each $n \ge 1$.
As
\begin{align}
P_{n}
=
\varepsilon_{n} \, Q_{n}^{(2)} + (1 - \varepsilon_{n}) \, Q_{n}^{(3)} ,
\end{align}
it follows by the concavity of the Shannon entropy that
\begin{align}
H( X_{n} )
\ge
\varepsilon_{n} \, H( Q_{n}^{(2)} ) + (1 - \varepsilon_{n}) \, H( Q_{n}^{(3)} )
\label{eq:concavity_Shannon_Q_tildeQ}
\end{align}
for each $n \ge 1$.
A direct calculations shows
\begin{align}
d( P_{n}, Q_{n}^{(3)} )
& =
\frac{ 1 }{ 2 } \sum_{x = 1}^{\infty} \Big| P_{n}( x ) - Q_{n}^{(3)}( x ) \Big|
\notag \\
& =
\frac{ 1 }{ 2 } \sum_{x = 1}^{\infty} \bigg| P_{n}( x ) - \frac{ P_{n}( x ) - \varepsilon_{n} \, Q_{n}^{(3)}( x ) }{ 1 - \varepsilon_{n} } \bigg|
\notag \\
& =
\frac{ 1 }{ 2 } \frac{ \varepsilon_{n} }{ 1 - \varepsilon_{n} } \sum_{x = 1}^{\infty} \Big| P_{n}( x ) - Q_{n}^{(2)}( x ) \Big|
\notag \\
& =
\frac{ \varepsilon_{n} }{ 1 - \varepsilon_{n} } d( P_{n}, Q_{n}^{(2)} )
\notag \\
& \le
\frac{ \varepsilon_{n} }{ 1 - \varepsilon_{n} }
\notag \\
& \eqqcolon
\delta_{n}
\end{align}
for each $n \ge 1$, where note that $\varepsilon_{n} = \mathrm{o}( 1 )$ implies $\delta_{n} = \mathrm{o}( 1 )$ as well.
Thus, it follows from \lemref{lem:TV_continuity} that
\begin{align}
H( Q_{n}^{(3)} )
& \ge
H( S^{(P_{n}, \delta_{n})} )
\notag \\
& \overset{\mathclap{\text{(a)}}}{=}
(P_{n}( 1 ) + \delta_{n}) \log \frac{ 1 }{ P_{n}( 1 ) + \delta_{n} } + \sum_{x = 2}^{B_{n} - 1} P_{n}( x ) \log \frac{ 1 }{ P_{n}( x ) } - \left( \sum_{k = B_{n}}^{\infty} P_{n}( k ) - \delta_{n} \right) \log \left( \sum_{k = B_{n}}^{\infty} P_{n}( k ) - \delta_{n} \right)
\notag \\
& \overset{\mathclap{\text{(b)}}}{\ge}
\sum_{x = 1}^{B_{n}} P_{n}( x ) \log \frac{ 1 }{ P_{n}( x ) } - 2 \, \gamma_{n}
\notag \\
& \overset{\mathclap{\text{(c)}}}{=}
\sum_{x \in \mathcal{B}^{(n)}} P_{X_{n}}( x ) \log \frac{ 1 }{ P_{X_{n}}( x ) } - 2 \, \gamma_{n}
\notag \\
& \overset{\mathclap{\text{(d)}}}{\ge}
\sum_{x \in \mathcal{A}_{\epsilon}^{(n)} \cap \mathcal{B}^{(n)}} P_{X_{n}}( x ) \log \frac{ 1 }{ P_{X_{n}}( x ) } - 2 \, \gamma_{n}
\notag \\
& \overset{\mathclap{\text{(e)}}}{\ge}
\sum_{x \in \mathcal{A}_{\epsilon}^{(n)} \cap \mathcal{B}^{(n)}} P_{X_{n}}( x ) \, (1 - \epsilon) \, H( X_{n} ) - 2 \, \gamma_{n}
\notag \\
& =
\mathbb{P}\{ X_{n} \in \mathcal{A}_{\epsilon}^{(n)} \cap \mathcal{B}^{(n)} \} \, (1 - \epsilon) \, H( X_{n} ) - 2 \, \gamma_{n}
\label{eq:tildeQ_vs_PX}
\end{align}
for every $\epsilon > 0$ and each $n \ge 1$,
where
\begin{itemize}
\item
(a) follows by the definition
\begin{align}
B_{n}
\coloneqq
\sup \bigg\{ b \ge 1 \ \bigg| \ \sum_{k = b}^{\infty} P_{n}( k ) \ge \delta_{n} \bigg\}
\end{align}
for each $n \ge 1$,
\item
(b) follows by the continuity of the map $u \mapsto - u \log u$ and the fact that $\delta_{n} = \mathrm{o}( 1 )$ as $n \to \infty$, i.e., there exists a sequence $\{ \gamma_{n} \}_{n = 1}^{\infty}$ of positive reals satisfying $\gamma_{n} = \mathrm{o}( 1 )$ as $n \to \infty$ and
\begin{align}
\left| P_{n}( 1 ) \log \frac{ 1 }{ P_{n}( 1 ) } - (P_{n}( 1 ) + \delta_{n}) \log \frac{ 1 }{ P_{n}( 1 ) + \delta_{n} } \right|
& \le
\gamma_{n} ,
\\
\left| P_{n}( B_{n} ) \log \frac{ 1 }{ P_{n}( B_{n} ) } + \left( \sum_{k = B_{n}}^{\infty} P_{n}( k ) - \delta_{n} \right) \log \left( \sum_{k = B_{n}}^{\infty} P_{n}( k ) - \delta_{n} \right) \right|
& \le
\gamma_{n}
\end{align}
for each $n \ge 1$,
\item
(c) follows by constructing the subset $\mathcal{B}^{(n)} \subset \mathcal{X}$ so that
\begin{align}
|\mathcal{B}^{(n)}|
=
\min_{\substack{ \mathcal{B} \subset \mathcal{X} : \\ \mathbb{P}\{ X_{n} \in \mathcal{B} \} \ge 1 - \delta_{n} }} |\mathcal{B}|
\end{align}
for each $n \ge 1$,
\item
(d) follows by defining the typical set $\mathcal{A}_{\epsilon}^{(n)} \subset \mathcal{X}$ so that
\begin{align}
\mathcal{A}_{\epsilon}^{(n)}
\coloneqq
\bigg\{ x \in \mathcal{X} \ \bigg| \ \log \frac{ 1 }{ P_{X_{n}}( x ) } \le (1 - \epsilon) \, H( X_{n} ) \bigg\}
\end{align}
with some $\epsilon > 0$ for each $n \ge 1$, and
\item
(e) follows by the definition of $\mathcal{A}_{\epsilon}^{(n)}$.
\end{itemize}
As $\{ X_{n} \}_{n = 1}^{\infty}$ satisfies the AEP and
\begin{align}
\mathbb{P}\{ X_{n} \in \mathcal{B}^{(n)} \}
& \ge
1 - \delta_{n} ,
\\
\lim_{n \to \infty} \delta_{n}
& = 0 ,
\end{align}
it is clear that
\begin{align}
\lim_{n \to \infty} \mathbb{P}\{ X_{n} \notin \mathcal{A}_{\epsilon}^{(n)} \cap \mathcal{B}^{(n)} \}
=
0
\end{align}
(see, e.g., Problem~3.11 of \cite{cover_thomas_ElementsofInformationTheory}).
Thus, since $\epsilon > 0$ can be arbitrarily small and $\varepsilon_{n} = \mathrm{o}( 1 )$ as $n \to \infty$, it follows from \eqref{eq:tildeQ_vs_PX} that
there exists a sequence $\{ \lambda_{n} \}_{n = 1}^{\infty}$ of positive real numbers satisfying $\lambda_{n} = \mathrm{o}( 1 )$ as $n \to \infty$ and
\begin{align}
(1 - \varepsilon_{n}) \, H( Q_{n}^{(3)} )
\ge
(1 - \lambda_{n}) \, H( X_{n} ) - \frac{ 2 \, \gamma_{n} }{ 1 - \varepsilon_{n} }
\label{eq:LB_tildeQ}
\end{align}
for each $n \ge 1$.
Combining \eqref{eq:concavity_Shannon_Q_tildeQ} and \eqref{eq:LB_tildeQ}, we observe that
\begin{align}
\lambda_{n} \, H( X_{n} ) + \frac{ 2 \, \gamma_{n} }{ 1 - \varepsilon_{n} }
\ge
\varepsilon_{n} \, H( Q_{n}^{(2)} )
\end{align}
for each $n \ge 1$.
Therefore, Equation~\eqref{eq:vanishing_H(Q)} is indeed valid, which proves \eqref{eq:continuity_Shannon_type5_normalized} together with \eqref{eq:H_Ptype5}.
This completes the proof of \thref{th:Fano_meets_AEP}.
\hfill\IEEEQEDhere

\begin{remark}
The construction of $Q_{n}^{(3)}$ defined in \eqref{def:Q3} is a special case of the \emph{splitting technique;}
it was used to derive limit theorems of Markov processes by Nummelin \cite{nummelin_1978} and Athreya--Ney \cite{athreya_ney_1978}.
This technique has many applications in information theory \cite{ho_verdu_2010, kumar_li_elgamal_isit2014, vellambi_kliewer_2016, vellambi_kliewer_isit2018, yu_tan_2018_common, yu_tan_2018_channel} and to the Markov chain Monte Carlo (MCMC) algorithm \cite{roberts_rosenthal_2004}.
\end{remark}

\subsection{Proof of \thref{th:vanishing_conditionalRenyi_list}}
\label{sect:vanishing_conditionalRenyi_list}

Condition (b) is a direct consequence of \thref{th:Fano_meets_AEP}; and we shall verify Conditions (a), (c), and (d) in the proof.
For the sake of brevity, in the proof, we write
\begin{align}
\varepsilon_{n}
& \coloneqq
P_{\mathrm{e}}^{(L_{n})}(X_{n} \mid Y_{n}) ,
\\
P_{n}
& \coloneqq
P_{X_{n}}^{\downarrow} ,
\\
P
& \coloneqq
P_{X}^{\downarrow} ,
\\
P_{1, n}
& \coloneqq
P_{\operatorname{type-1}}^{(P_{n}, L_{n}, \varepsilon_{n})}
\end{align}
for each $n \ge 1$.
By \corref{cor:Fano_Renyi_5}, instead on \eqref{eq:vanishing_conditionalRenyi_list1}, it suffices to verify that
\begin{align}
\lim_{n \to \infty} \Big| H_{\alpha}( P_{1, n} ) - \log L_{n} \Big|^{+} = 0
\label{eq:continuity_Renyi_type5}
\end{align}
under any one of Conditions (a), (b), and (c).
Similar to the proof of \thref{th:Fano_meets_AEP}, we may assume without loss of generality that $0 < \varepsilon_{n} < 1$.

Firstly, we shall verify Condition (a).
Let $Q_{n}$ be an $\mathcal{X}$-marginal given by
\begin{align}
Q_{n}( x )
=
\begin{dcases}
\frac{ 1 - \varepsilon_{n} }{ L_{n} }
& \mathrm{if} \ 1 \le x \le L_{n} ,
\\
P_{\mathrm{type5, n}}( x )
& \mathrm{if} \ x \ge L_{n} + 1
\end{dcases}
\end{align}
for each $n \ge 1$.
As $P_{1, n}$ majorizes $Q_{n}$, it follows by the Schur-concavity property of the R\'{e}nyi entropy that
\begin{align}
H_{\alpha}( P_{1, n} )
& \le
H_{\alpha}( Q_{n} )
\notag \\
& =
\frac{ 1 }{ 1 - \alpha } \log \bigg( (1 - \varepsilon_{n})^{\alpha} \, L_{n}^{1-\alpha}  + \sum_{x = L_{n}}^{\infty} P_{1, n}( x )^{\alpha} \bigg)
\notag \\
& \le
\frac{ 1 }{ 1 - \alpha } \log \Big( (1 - \varepsilon_{n})^{\alpha} \, L_{n}^{1-\alpha} \Big)
\notag \\
& =
\log L_{n} + \frac{ \alpha }{ 1 - \alpha } \log( 1 - \varepsilon_{n} ) ,
\end{align}
where the second inequality follows by the hypothesis that $\alpha > 1$, i.e., by Condition (a).
These inequalities immediately ensure \eqref{eq:continuity_Renyi_type5} under Condition (a).

Second, we shall verify Condition (d) of \thref{th:vanishing_conditionalRenyi_list}.
As $X$ and $\{ X_{n} \}_{n}$ are discrete r.v.'s, note that the convergence in distribution $X_{n} \overset{d}{\to} X$ is equivalent to $P_{n}( x ) \to P( x )$ as $n \to \infty$ for each $x \in \mathcal{X}$, i.e., the pointwise convergence $P_{n} \to P$ as $n \to \infty$.
It is well-known that the R\'{e}nyi entropy $\alpha \mapsto H_{\alpha}( P )$ is nonincreasing for $\alpha \ge 0$;
hence, it suffices to verify \eqref{eq:continuity_Renyi_type5} with $\alpha = 1$, i.e.,
\begin{align}
\lim_{n \to \infty} \Big| H( P_{1, n} ) - \log L_{n} \Big|^{+} = 0 .
\label{eq:continuity_Shannon_type5}
\end{align}
We now define two $\mathcal{X}$-marginals $Q_{n}^{(1)}$ and $Q_{n}^{(2)}$ in the same ways as \eqref{def:Q1} and \eqref{def:Q2}, respectively, for each $n \ge 1$.
By \eqref{eq:H_Ptype5}, it suffices to verify whether the third term in the right-hand side of \eqref{eq:H_Ptype5} approaches to zero, i.e., 
\begin{align}
\lim_{n \to \infty} \varepsilon_{n} \, H( Q_{n}^{(2)} ) = 0 .
\label{eq:limit_Q2_Shannon}
\end{align}
This can be verified in a similar fashion to the proof of Lemma~3 of \cite{ho_verdu_2010} as follows:
Consider the $\mathcal{X}$-marginal $Q_{n}^{(3)}$ defined in \eqref{def:Q3} for each $n \ge 1$.
Since $Q_{n}^{(2)}( 1 ) = 0$ and $\varepsilon_{n} \, Q_{n}^{(2)}( x ) \le \varepsilon_{n}$ for each $x \ge 2$, we observe that
\begin{align}
\lim_{n \to \infty} \varepsilon_{n} \, Q_{n}^{(2)}( x )
=
0
\end{align}
for every $x \ge 1$; therefore,
\begin{align}
\lim_{n \to \infty} Q_{n}^{(3)}( x ) = \lim_{n \to \infty} P_{X_{n}}^{\downarrow}( x )
\end{align}
for every $x \ge 1$.
Therefore, since $P_{n}$ converges pointwise to $P$ as $n \to \infty$, we see that $Q_{n}^{(3)}$ also converges pointwise to $P_{X}^{\downarrow}$ as $\varepsilon_{n}$ vanishes.
Therefore, by the lower semicontinuity property of the Shannon entropy, we observe that
\begin{align}
\liminf_{n \to \infty} H( Q_{n}^{(3)} )
\ge
H( X ) ,
\end{align}
and we then have
\begin{align}
H( X )
& =
\lim_{n \to \infty} H( X_{n} )
\notag \\
& \overset{\mathclap{\text{(a)}}}{\ge}
\limsup_{n \to \infty} \Big( \varepsilon_{n} \, H( Q_{n}^{(2)} ) + (1 - \varepsilon_{n}) \, H( Q_{n}^{(3)} ) \Big)
\notag \\
& \ge
\limsup_{n \to \infty} \Big( \varepsilon_{n} \, H( Q_{n}^{(2)} ) \Big) + \liminf_{n \to \infty} \Big( (1 - \varepsilon_{n}) \, H( Q_{n}^{(3)} ) \Big)
\notag \\
& =
\limsup_{n \to \infty} \Big( \varepsilon_{n} \, H( Q_{n}^{(2)} ) \Big) + \liminf_{n \to \infty} H( Q_{n}^{(3)} )
\notag \\
& \ge
\limsup_{n \to \infty} \Big( \varepsilon_{n} \, H( Q_{n}^{(2)} ) \Big) + H( X ) ,
\label{ineq:limsup_Shannon}
\end{align}
where (a) follows from \eqref{eq:concavity_Shannon_Q_tildeQ}.
Thus, it follows from \eqref{ineq:limsup_Shannon}, the hypothesis $H(X) < \infty$, and the nonnegativity of the Shannon entropy that \eqref{eq:limit_Q2_Shannon} is valid, which proves \eqref{eq:continuity_Shannon_type5}  together with \eqref{eq:H_Ptype5}.

Finally, we shall verify Condition (c) of \thref{th:vanishing_conditionalRenyi_list}.
Define the $\mathcal{X}$-marginal $\tilde{Q}_{n}^{(2)}$ by
\begin{align}
{\tilde{Q}_{n}^{(2)}( x )}
=
\begin{dcases}
0
& \mathrm{if} \ 1 \le x \le L_{n} ,
\\
\frac{ \tilde{P}_{1, n}( x ) }{ \varepsilon_{n} }
& \mathrm{if} \ x \ge L_{n} + 1 ,
\end{dcases}
\end{align}
for each $n \ge 1$, where $\tilde{P}_{1, n} = P_{\operatorname{type-1}}^{(P, L_{n}, \varepsilon_{n})}$.
Note that the difference between $Q_{n}^{(2)}$ and $\tilde{Q}_{n}^{(2)}$ is the difference between $P_{n}$ and $P$.
It can be verified by the same way as \eqref{ineq:limsup_Shannon} that
\begin{align}
\lim_{n \to \infty} \Big( \varepsilon_{n} \, H( \tilde{Q}_{n}^{(2)} ) \Big) = 0 .
\label{eq:limit_tildeQ2_Shannon}
\end{align}
It follows by the same manner as Lemma~1 of \cite{ho_verdu_2010} that if $P_{n}$ majorizes $P$, then $Q_{n}^{(2)}$ majorizes $\tilde{Q}_{n}^{(2)}$ as well.
Therefore, it follows from the Schur-concavity property of the Shannon entropy that if $P_{n}$ majorizes $P$ for sufficiently large $n$, then
\begin{align}
H( Q_{n}^{(2)} )
\le
H( \tilde{Q}_{n}^{(2)} )
\label{ineq:Shannon_SchurConcavity}
\end{align}
for sufficiently large $n$.
Combining \eqref{eq:limit_tildeQ2_Shannon} and \eqref{ineq:Shannon_SchurConcavity}, Equation~\eqref{eq:limit_Q2_Shannon} also holds under Condition~(c).
This completes the proof of \thref{th:vanishing_conditionalRenyi_list}.
\hfill\IEEEQEDhere

\subsection{Proof of \thref{th:VanishingAMsymbolerror}}
\label{sect:VanishingAMsymbolerror}

To prove \thref{th:VanishingAMsymbolerror}, we now give the following lemma.

\begin{lemma}
\label{lem:concave_RD}
If $H(Q) < \infty$, then the map $\varepsilon \mapsto H( P_{\operatorname{type-1}}^{(Q, L, \varepsilon)} )$ is concave in the interval \eqref{eq:range_epsilon_list} with $|\mathcal{Y}| = \infty$.
\end{lemma}

\begin{IEEEproof}[Proof of \lemref{lem:concave_RD}]
It is well-known that for a fixed $P_{X}$, the conditional Shannon entropy $H(X \mid Y)$ is concave in $P_{Y|X}$ (cf.\ \cite{cover_thomas_ElementsofInformationTheory}, Theorem~2.7.4).
Defining the distortion measure $d : \mathcal{X} \times \binom{\mathcal{X}}{L} \to \{ 0, 1 \}$ by
\begin{align}
d( x, \hat{x} )
=
\begin{cases}
1
& \mathrm{if} \ x \notin \hat{x} ,
\\
0
& \mathrm{if} \ x \in \hat{x} ,
\end{cases}
\end{align}
the average probability of list decoding error is equal to the average distortion, i.e., 
\begin{align}
\mathbb{P}\{ X \notin f(Y) \}
=
\mathbb{E}[ d(X, f(Y)) ]
\end{align}
for any list decoder $f : \mathcal{Y} \to \binom{\mathcal{X}}{L}$.
Therefore, by following \thref{th:main_list}, the concavity property of \lemref{lem:concave_RD} can be proved by the same argument as the proof of the convexity of the rate-distortion function (cf.\ Lemma~10.4.1 of \cite{cover_thomas_ElementsofInformationTheory}).
\end{IEEEproof}

For the sake of brevity, we write
\begin{align}
P
& =
P_{X} ,
\\
P_{n}
& =
P_{X_{n}} ,
\\
\varepsilon_{n}
& =
P_{\mathrm{e}}^{(L_{n})}(X_{n} \mid Y_{n}) ,
\\
P_{1, n}
& =
P_{\operatorname{type-1}}^{(P_{n}, L_{n}, \varepsilon_{n})} ,
\\
{\bar{P}_{1, n}}
& =
P_{\operatorname{type-1}}^{(P_{n}, \bar{L}_{n}, \varepsilon_{n})}
\end{align}
in this proof.
Define
\begin{align}
{\bar{L}}
\coloneqq
\limsup_{n \to \infty} L_{n} .
\end{align}
If $\bar{L} = \infty$, then \eqref{eq:VanishingAMsymbolerror} is a trivial inequality.
Therefore, it suffices to consider the case where $\bar{L} < \infty$.

It is clear that there exists an integer $n_{0} \ge 1$ such that $L_{n} \le \bar{L}$ for every $n \ge n_{0}$.
Then, we can verify that $P_{1, n}$ majorizes $\bar{P}_{1, n}$ for every $n \ge n_{0}$ as follows.
Let $J_{n}$ and $J_{3}$ be given by \eqref{def:J} with $(Q, L, \varepsilon) = (P_{n}, L_{n}, \varepsilon_{n})$ and $(Q, L, \varepsilon) = (P_{n}, \bar{L}, \varepsilon_{n})$, respectively.
Similarly, let $K_{n}$ and $K_{3}$ be given by \eqref{def:K3} with $(Q, L, \varepsilon) = (P_{n}, L_{n}, \varepsilon_{n})$ and $(Q, L, \varepsilon) = (P_{n}, \bar{L}, \varepsilon_{n})$, respectively.
As $L_{n} \le \bar{L}$ implies that $J_{n} \le J_{3}$ and $K_{n} \le K_{3}$, it can be seen from \eqref{def:type1} that
\begin{align}
P_{1, n}( x )
& =
\bar{P}_{1, n}( x )
\qquad
\mathrm{for} \ 1 \le x < J_{n} \ \mathrm{or} \ x \ge K_{3} ,
\\
P_{1, n}( x )
& \ge
\bar{P}_{1, n}( x )
\qquad
\mathrm{for} \ J_{n} \le x \le L_{n} \ \mathrm{or} \ \bar{L} < x \le K_{3} ,
\\
P_{1, n}( x )
& \le
\bar{P}_{1, n}( x )
\qquad
\mathrm{for} \ L_{n} < x \le \bar{L} .
\end{align}
Therefore, noting that
\begin{align}
\sum_{x = 1}^{L_{n}} P_{1, n}( x )
=
\sum_{x = 1}^{\bar{L}} \bar{P}_{1, n}( x )
=
1 - \varepsilon_{n} ,
\end{align}
we obtain the majorization relation $P_{1, n} \succ \bar{P}_{1, n}$ for every $n \ge n_{0}$.

By hypothesis, there exists an integer $n_{1} \ge 1$ such that $P_{n}$ majorizes $P$ for every $n \ge n_{1}$.
Letting $n_{2} = \max\{ n_{0}, n_{1} \}$, we observe that 
\begin{align}
\frac{ 1 }{ n } H(X^{n} \mid Y^{n})
& \le
\frac{ 1 }{ n } \sum_{i = 1}^{n} H(X_{i} \mid Y_{i})
\notag \\
& \le
\frac{ 1 }{ n } \sum_{i = 1}^{n_{2}-1} H(X_{i} \mid Y_{i}) + \frac{ 1 }{ n } \sum_{j = n_{2}}^{n} H(X_{i} \mid Y_{i})
\notag \\
& \overset{\mathrm{(a)}}{\le}
\frac{ n_{2} - 1 }{ n } \Big( \max_{1 \le i < n_{2}} H( X_{i} ) \Big) + \frac{ 1 }{ n } \sum_{j = n_{2}}^{n} H\Big( \bar{P}_{1, j} \Big)
\notag \\
& \overset{\mathrm{(b)}}{\le}
\frac{ n_{2} - 1 }{ n } \Big( \max_{1 \le i < n_{2}} H( X_{i} ) \Big) + \frac{ 1 }{ n } \sum_{j = n_{2}}^{n} H\Big( P_{\operatorname{type-1}}^{(P, \bar{L}, \varepsilon_{j})} \Big)
\notag \\
& \overset{\mathrm{(c)}}{\le}
\frac{ n_{2} - 1 }{ n } \Big( \max_{1 \le i < n_{2}} H( X_{i} ) \Big) + \frac{ n - n_{2} + 1 }{ n } H\Big( P_{\operatorname{type-1}}^{(P, \bar{L}, \bar{\varepsilon}_{n})} \Big)
\label{eq:InProof_VanishingAMsymbolerror_1}
\end{align}
for every $n \ge n_{2}$, where
\begin{itemize}
\item
(a) follows by \corref{cor:Fano_Renyi_5} and $P_{1, n} \succ \bar{P}_{1, n}$,
\item
(b) follows by Condition (b) of \thref{th:vanishing_conditionalRenyi_list} and the same manner as (\cite{ho_verdu_2010}, Lemma~1), and
\item
(c) follows by \lemref{lem:concave_RD} together with the following definition
\begin{align}
{\bar{\varepsilon}_{n}}
\coloneqq
\frac{ 1 }{ n - n_{2} + 1 } \sum_{j = n_{2}}^{n} \varepsilon_{j}
=
\frac{ 1 }{ n - n_{2} + 1 } \sum_{j = n_{2}}^{n} P_{\mathrm{e}}^{(L_{j})}(X_{j} \mid Y_{j}) .
\end{align}
\end{itemize}
Note that the Schur-concavity property of the Shannon entropy is used in both (b) and (c) of \eqref{eq:InProof_VanishingAMsymbolerror_1}.
As
\begin{align}
\lim_{n \to \infty} P_{\mathrm{e, sym.}}^{(\mathbf{L})}(X^{n} \mid Y^{n}) = 0
\iff
\lim_{n \to \infty} \bar{\varepsilon}_{n} =  0,
\end{align}
it follows from \eqref{eq:continuity_Renyi_type5} that there exists an integer $n_{3} \ge 1$ such that
\begin{align}
H\Big( P_{\operatorname{type-1}}^{(P, \bar{L}, \bar{\varepsilon}_{n})} \Big)
\le
\log \bar{L}
\end{align}
for every $n \ge n_{3}$.
Therefore, it follows from \eqref{eq:InProof_VanishingAMsymbolerror_1} that
\begin{align}
\frac{ 1 }{ n } H(X^{n} \mid Y^{n})
& \le
\frac{ n_{2} - 1 }{ n } \Big( \max_{1 \le i < n_{2}} H( X_{i} ) \Big) + \frac{ n - n_{2} + 1 }{ n } \log \bar{L}
\label{eq:InProof_VanishingAMsymbolerror_2}
\end{align}
for every $n \ge \max\{ n_{2}, n_{3} \}$.
Therefore, letting $n \to \infty$ in \eqref{eq:InProof_VanishingAMsymbolerror_2}, we have \eqref{eq:VanishingAMsymbolerror}.
This completes the proof of \thref{th:VanishingAMsymbolerror}.
\hfill\IEEEQEDhere

\section{Concluding Remarks}
\label{sect:conclusion}

\subsection{Impossibility of Establishing Fano-Type Inequality}
\label{sect:infty}

In \sectref{sect:Fano_h}, we explored the principal maximization problem $\mathbb{H}_{\phi}(Q, L, \varepsilon, \mathcal{Y})$ defined in \eqref{def:main_object} without any explicit form of $\phi$ under the three postulates: $\phi$ is symmetric, concave, and lower semicontinuous.
If $\varepsilon > 0$ and we impose another postulate on $\phi$, then we can also avoid the (degenerate) case in which $\phi(Q) = \infty$.
The following proposition shows this fact.

\begin{proposition}
\label{prop:infty}
Let $g_{1} : [0, 1] \to [0, \infty)$ be a function satisfying $g_{1}( 0 ) = 0$, and $g_{2} : [0, \infty] \to [0, \infty]$ a function satisfying $g_{2}( u ) = \infty$ only if $u = \infty$.
Suppose that $\varepsilon > 0$ and $\phi : \mathcal{P}( \mathcal{X} ) \to [0, \infty]$ is of the form
\begin{align}
\phi( Q )
& =
g_{2} \bigg( \sum_{x \in \mathcal{X}} g_{1}\big( Q( x ) \big) \bigg) .
\label{eq:phi_sum}
\end{align}
Then, it holds that
\begin{align}
\mathbb{H}_{\phi}(Q, L, \varepsilon, \mathcal{Y})
<
\infty
\quad \iff \quad
\phi( Q )
<
\infty .
\end{align}
\end{proposition}

\begin{IEEEproof}[Proof of \propref{prop:infty}]
See \appref{app:infty}.
\end{IEEEproof}

As seen in \sectref{sect:Fano_Renyi}, the conditional Shannon and R\'{e}nyi entropies can be expressed by $\mathsf{H}_{\phi}(X \mid Y)$; and then $\phi$ must satisfy \eqref{eq:phi_sum}.
\propref{prop:infty} shows that we cannot establish an effective Fano-type inequality based on the conditional information measure $\mathsf{H}_{\phi}(X \mid Y)$ subject to our original postulates in \sectref{sect:majorization}, provided that (i) $\phi$ satisfies the additional postulate of \eqref{eq:phi_sum}, (ii) $\varepsilon > 0$, and (iii) $\phi( Q ) = \infty$.
This generalizes a pathological example given in Example~2.49 of \cite{yeung_2008}, which states issues of the interplay between conditional information measures and error probabilities over countably infinite alphabets $\mathcal{X}$; see \sectref{sect:countably-infinite}.

\subsection{Postulational Characterization of Conditional Information Measures}
\label{sec:philo}

Our Fano-type inequalities were stated in terms of the general conditional information $\mathsf{H}_{\phi}(X \mid Y)$ defined in \sectref{sect:majorization}.
As shown in \sectref{sect:Fano_Renyi}, the quantity $\mathsf{H}_{\phi}(X \mid Y)$ can be specialized to Shannon's and R\'{e}nyi's information measures.
Moreover, the quantity $\mathsf{H}_{\phi}(X \mid Y)$ can be further specialized to the following quantities:

\begin{enumerate}
\item
If $\phi = \| \cdot \|_{1/2}$, then $\mathsf{H}_{\phi}(X \mid Y)$ coincides with the (unnormalized) \emph{Bhattacharyya parameter} (cf.\ Definition~17 of \cite{mori_tanaka_2014} and Section~4.2.1 of \cite{sasoglu_2012}) defined by
\begin{align}
B(X \mid Y)
\coloneqq
\mathbb{E}\Bigg[ \sum_{ x, x^{\prime} \in \mathcal{X} } \sqrt{ P_{X|Y}( x ) \, P_{X|Y}( x^{\prime} ) } \Bigg] .
\label{def:Bhattacharyya}
\end{align}
Note that the Bhattacharyya parameter is often defined so that $Z(X \mid Y) \coloneqq (B(X \mid Y)-1) / (M-1)$ to normalize as $0 \le Z(X \mid Y) \le 1$, provided that $X$ is $\{ 0, 1, \dots, M-1 \}$-valued.
When $X$ takes values in a finite alphabet with a certain algebraic structure, the Bhattacharyya parameter $B(X \mid Y)$ is useful in analyzing the speed of polarization for non-binary polar codes (cf.\ \cite{mori_tanaka_2014, sasoglu_2012}).
Note that $B(X \mid Y)$ is a monotone function of Arimoto's conditional R\'{e}nyi entropy \eqref{def:arimoto} of order $\alpha = 1/2$.
\item
If $\phi = 1 - \| \cdot \|_{2}^{2}$, then $\mathsf{H}_{\phi}(X \mid Y)$ coincides with the \emph{conditional quadratic entropy} \cite{cover_hart_1967} defined by
\begin{align}
H_{\mathrm{o}}(X \mid Y)
& \coloneqq
\mathbb{E} \bigg[ \sum_{x \in \mathcal{X}} P_{X|Y}( x ) \, \Big( 1 - P_{X|Y}( x ) \Big) \bigg] ,
\label{def:quadratic}
\end{align}
which is used in the analysis of stochastic decoding (see, e.g., \cite{muramatsu_miyake_isit2017}).
Note that $H_{\mathrm{o}}(X \mid Y)$ is a monotone function of Hayashi's conditional R\'{e}nyi entropy \eqref{def:hayashi} of order $\alpha = 2$.
\item
If $X$ is $\{ 1, 2, \dots, M \}$-valued, then one can define the following (variational distance-like) conditional quantity:
\begin{align}
K(X \mid Y)
& \coloneqq
\mathbb{E} \bigg[ \frac{ 1 }{ 2 (M - 1) } \sum_{x = 1}^{M} \sum_{x^{\prime} = 1}^{M} \Big| P_{X|Y}( x ) - P_{X|Y}( x^{\prime} ) \Big| \bigg] .
\label{def:K_TV}
\end{align}
Note that $0 \le K(X \mid Y) \le 1$.
This quantity $K(X \mid Y)$ was introduced by Shuval--Tal \cite{shuval_tal_2018} to analyze the speed of polarization of non-binary polar codes for sources with memory.
When we define the function $\bar{d} : \mathcal{P}( \{ 1, 2, \dots, M \} ) \to [0, 1]$ by
\begin{align}
{\bar{d}( P )}
& \coloneqq
\frac{ 1 }{ 2(M - 1) } \sum_{x = 1}^{M} \sum_{x^{\prime} = 1}^{M} \Big| P( x ) - P( x^{\prime} ) \Big| ,
\end{align}
it holds that $K(X \mid Y) = \mathsf{H}_{\bar{d}}(X \mid Y)$.
Clearly, the function $\bar{d}$ is symmetric, convex, and continuous.
\end{enumerate}

On the other hand, the quantity $\mathsf{H}_{\phi}(X \mid Y)$ has the following properties that are appealing in information theory:

\begin{enumerate}
\item
As $\phi$ is concave, lower bounded, and lower semicontinuous, it follows from Jensen's inequality for an extended real-valued function on a closed, convex, and  bounded subset of a Banach space (\cite{shirokov_2010}, Proposition~A-2) that
\begin{align}
\mathsf{H}_{\phi}(X \mid Y)
\le
\phi( P_{X} ) .
\label{eq:obviousUB}
\end{align}
This bound is analogous  to the property that conditioning reduces entropy (cf.\ \cite{cover_thomas_ElementsofInformationTheory}, Theorem~2.6.5).
\item
It is easy to check that for any (deterministic) mapping $g : \mathcal{X} \to \mathcal{A}$ with $\mathcal{A} \subset \mathcal{X}$, the conditional distribution $P_{g(X)|Y}$ majorizes $P_{X|Y}$ a.s.
Thus, it follows from \propref{prop:Schur_convex} that for any mapping $g : \mathcal{X} \to \mathcal{A}$,
\begin{align}
\mathsf{H}_{\phi}(g( X ) \mid Y)
\le
\mathsf{H}_{\phi}(X \mid Y) ,
\label{eq:DPI}
\end{align}
which is a counterpart of the \emph{data processing inequality} (cf.\ Equations~(26)--(28) of \cite{hayashi_tan_2017}).
\item
As shown in \sectref{sect:Fano_h}, the quantity $\mathsf{H}_{\phi}(X \mid Y)$ also satisfies appropriate generalizations of \emph{Fano's inequality.}
\end{enumerate}

Therefore, similar to the family of $f$-divergences~\cite{ali_silvery_1966, csiszar_1963}, the quantity $\mathsf{H}_{\phi}(X \mid Y)$ is a generalization of various information-theoretic conditional quantities that also admit certain desirable  properties. In addition, we can establish Fano-type inequalities based on $\mathsf{H}_{\phi}(X \mid Y)$; this characterization provides insights on {\em how to measure conditional information} axiomatically.

\subsection{When Does Vanishing Error Probabilities Imply Vanishing Equivocations?}

In the list decoding setting, the rate of a block code with codeword length $n$, message size $M_{n}$, and list size $L_{n}$ can be defined as $(1/n) \log(M_{n} / L_{n})$  (cf.\  \cite{elias_1957}).
Motivated by this, we established asymptotic behaviors of this quantity in Theorems~\ref{th:Fano_meets_AEP} and~\ref{th:vanishing_conditionalRenyi_list}.
We would like to emphasize that \exref{ex:list} shows that Ahlswede--G\'{a}cs--K\"{o}rner's proof technique described in Chapter~5 of \cite{ahlswede_gacs_korner_1976} (see also Section~3.6.2 of \cite{raginsky_sason_2014}) works for an i.i.d.\ source on a {\em countably infinite alphabet}, provided that the alphabets $\{ \mathcal{Y}_{n} \}_{n = 1}^{\infty}$ are finite.

\thref{th:Fano_meets_AEP} states that the asymptotic growth of $H(X_{n} \mid Y_{n}) - \log L_{n}$ is \emph{strictly slower} than $H( X_{n} )$, provided that the general source $\mathbf{X} = \{ X_{n} \}_{n = 1}^{\infty}$ satisfies the AEP and the error probabilities vanish (i.e., $P_{\mathrm{e}}^{(L_{n})}(X_{n} \mid Y_{n}) = \mathrm{o}( 1 )$ as $n \to \infty$).
This is a novel characterization of the AEP via Fano's inequality.
An instance of this characterization using the Poisson source (cf.\ Example~4 of \cite{verdu_han_1997}) was provided in \exref{ex:poisson}.

\subsection{Future Works}

\begin{enumerate}
\item
While there are various studies of the \emph{reverse} Fano inequalities \cite{kovalevsky_1968, chu_chueh_1966, tebbe_dwyer_1968, feder_merhav_1994, sakai_iwata_isit2017, sason_verdu_2017}, this study has focused only on the \emph{forward} Fano inequality.
Generalizing the reverse Fano inequality in the same spirit as was done in this study would be of interest.
\item
Important technical tools used in our analysis include the finite-  and infinite-dimensional versions of Birkhoff's theorem;
they were employed to satisfy the constraint that $P_{X} = Q$.
As a similar constraint is imposed in many information-theoretic problems, e.g., coupling problems (cf.\ \cite{yu_tan_2018_coupling, sason_2013, thorisson_2000}), finding further applications of the infinite-dimensional version of Birkhoff's theorems would refine technical tools, and potentially results, when we are dealing with communication systems on countably infinite alphabets.
\item
We have described a novel connection between  the AEP and Fano's inequality in \thref{th:Fano_meets_AEP}; its role in the classifications of sources and channels and its applications to other coding problems are of interest.  
\end{enumerate}

\section*{Acknowledgements}

The author would like to thank Prof.\ Ken-ichi~Iwata for his valuable comments on an earlier version of this paper.
Prof.\ Vincent~Y.~F.~Tan gave insightful comments and suggestions that greatly improved this paper.
The author also would like to express my gratitude to an anonymous reviewer in IEEE Transactions on Information Theory and three anonymous reviewers in Entropy for carefully following the technical parts and giving a lot of his/her valuable comments.
Finally, the author would like to thank the Guest Editor, Prof.\ Amos~Lapidoth, of the special issue Information Measures with Applications in Entropy for inviting the author to this special issue and supporting this paper.

\appendices

\section{Proof of \propref{prop:listMAP}}
\label{app:listMAP}

The proposition is quite obvious; it is similar to (\cite{merhav_2014}, Equation~(1)).
Here, we prove it to make this paper self-contained.
For a given list decoder $f : \mathcal{Y} \to \binom{\mathcal{X}}{L}$ with list size $1 \le L < \infty$, it follows that
\begin{align}
\mathbb{P}\{ X \notin f(Y) \}
& =
\mathbb{E} [ \mathbb{E}[ \boldsymbol{1}_{\{ X \notin f(Y) \}} \mid Y] ]
\notag \\
& =
\mathbb{E} \left[ \sum_{x \notin f(Y)} P_{X|Y}( x ) \right]
\notag \\
& \overset{\mathclap{\text{(a)}}}{\ge}
\mathbb{E} \Bigg[ \sum_{x = L+1}^{\infty} P_{X|Y}^{\downarrow}( x ) \Bigg] ,
\end{align}
where the equality of (a) can be achieved by an optimal list decoder $f^{\ast}$ satisfying that $X \notin f^{\ast}( Y )$ only if $P_{X|Y}( X ) = P_{X|Y}^{\downarrow}( k )$ for some $k \ge L+1$.
This completes the proof of \propref{prop:listMAP}.
\hfill\IEEEQEDhere

\section{Proof of \propref{prop:boundPe}}
\label{app:boundPe}

The second inequality in \eqref{ineq:boundPe} is indeed a direct consequence of \propref{prop:listMAP} and \eqref{ineq:majorization_PXdr_PXYdr}.
The sharpness of the second bound can be easily verified by setting that $X$ and $Y$ are statistically independent.

We next prove the first inequality in \eqref{ineq:boundPe}.
When $\mathcal{Y}$ is infinite, the first inequality is an obvious one $P_{\mathrm{e}}^{(L)}(X \mid Y) \ge 0$, and its equality holds by setting $\mathcal{X} \subset \mathcal{Y}$ and $X = Y$ a.s.
Therefore, it suffices to consider the case where $\mathcal{Y}$ is finite.
Assume without loss of generality that
\begin{align}
\mathcal{Y}
=
\{ 0, 1, \dots, N-1 \}
\end{align}
for some positive integer $N$.
By the definition of cardinality, there exists a subset $\mathcal{Z} \subset \mathcal{X}$ satisfying (i) $|\mathcal{Z}| = L N$ and (ii) for each $x \in \{ 1, 2, \dots, L \}$ and $y \in \{ 0, 1, \dots, N-1 \}$, there exists an element $z \in \mathcal{Z}$ satisfying $P_{X|Y=y}( z ) = P_{X|Y=y}^{\downarrow}( x )$.
Then,
\begin{align}
P_{\mathrm{e}}(X \mid Y)
& \overset{\mathclap{\text{(a)}}}{=}
1 - \sum_{y \in \mathcal{Y}} P_{Y}( y ) \, \sum_{x = 1}^{L} P_{X|Y=y}^{\downarrow}( x )
\notag \\
& \overset{\mathclap{\text{(b)}}}{\ge}
1 - \sum_{y \in \mathcal{Y}} P_{Y}( y ) \sum_{x \in \mathcal{Z}} P_{X|Y=y}( x )
\notag \\
& = \,
1 - \sum_{x \in \mathcal{Z}} P_{X}( x )
\notag \\
& \overset{\mathclap{\text{(c)}}}{\ge}
1 - \sum_{x = 1}^{L N} Q^{\downarrow}( x ) ,
\end{align}
where
\begin{itemize}
\item
(a) follows from \propref{prop:listMAP},
\item
(b) follows from by the construction of $\mathcal{Z}$, and
\item
(c) follows from the facts that $|\mathcal{Z}| = L N$ and $P_{X} = Q$.
\end{itemize}
This is indeed the first inequality in \eqref{ineq:boundPe}.
Finally, the sharpness of the first inequality can be verified by the $\mathcal{X} \times \mathcal{Y}$-valued r.v.\ $(U, V)$ determined by
\begin{align}
P_{U|V = v}( u )
& =
\begin{dcases}
\frac{ \omega_{2}(Q, L, \varepsilon) }{ \omega_{1}(Q, v, L) } \, Q^{\downarrow}( u )
& \mathrm{if} \ v L < u \le (1 + v) \, L ,
\\
Q^{\downarrow}( u )
& \mathrm{if} \ LN < u < \infty ,
\\
0
& \mathrm{otherwise} ,
\end{dcases}
\\
P_{V}( v )
& =
\frac{ \omega_{1}(Q, v, L) }{ \omega_{2}(Q, L, \varepsilon) } ,
\end{align}
where $\omega_{1}(Q, v, L)$ and $\omega_{2}(Q, L, \varepsilon)$ are defined by
\begin{align}
\omega_{1}(Q, v, L)
& \coloneqq
\sum_{u = 1 + v L}^{(1 + v) L} Q^{\downarrow}( u ) ,
\\
\omega_{2}(Q, L, \varepsilon)
& \coloneqq
\sum_{v = 0}^{LN - 1} \omega_{1}(Q, v, L) .
\end{align}
A direct calculation shows that $P_{U} = Q^{\downarrow}$ and
\begin{align}
P_{\mathrm{e}}^{(L)}(U \mid V)
=
1 - \sum_{x = 1}^{L N} Q^{\downarrow}( x ) ,
\end{align}
which implies the sharpness of the first inequality.
This completes the proof of \propref{prop:boundPe}.
\hfill\IEEEQEDhere

\section{Proof of \propref{prop:type1}}
\label{app:type1}

Equations~\eqref{eq:sort_type1}, \eqref{eq:type1_first-L-sum}, and \eqref{eq:type1_below-J}--\eqref{eq:type1_above-K} directly follow from the definitions stated in \eqref{def:type1}--\eqref{def:K3}.
Equation~\eqref{eq:Pe_type1} follows from \eqref{def:Pe_Q} and \eqref{eq:type1_first-L-sum}.

Finally, we shall verify that $P_{\operatorname{type-1}}$ majorizes $Q$; in other words, we prove that
\begin{align}
\sum_{x = 1}^{k} P_{\operatorname{type-1}}( x )
\ge
\sum_{x = 1}^{k} Q^{\downarrow}( x )
\label{eq:majorization_type1-Q}
\end{align}
for every $k \ge 1$.
Equation~\eqref{eq:type1_below-J} implies that \eqref{eq:majorization_type1-Q} holds with equality for every $1 \le k < J$.
Moreover, it follows from \eqref{def:J} and \eqref{eq:type1_below-J} that \eqref{eq:majorization_type1-Q} holds for every $J \le k \le L$.
On the other hand, Equation~\eqref{eq:type1_above-K} implies that
\begin{align}
\sum_{x = k}^{\infty} P_{\operatorname{type-1}}( x )
=
\sum_{x = k}^{\infty} Q^{\downarrow}( x )
\label{eq:type1_tail}
\end{align}
for every $k > K_{1}$.
Combining \eqref{eq:type1-W}, \eqref{eq:majorization_type1-Q} with $k = L$, and \eqref{eq:type1_tail}, we observe that \eqref{eq:majorization_type1-Q} holds for every $k > L$.
Therefore, we have that $P_{\operatorname{type-1}}$ majorizes $Q$, as desired.
\hfill\IEEEQEDhere

\section{Proof of \propref{prop:infty}}
\label{app:infty}

The ``if'' part $\Leftarrow$ of \propref{prop:infty} is quite obvious from Jensen's inequality even if $\phi : \mathcal{P}( \mathcal{X} ) \to [0, \infty]$ is not of the form \eqref{eq:phi_sum}.
Therefore, it suffices to prove the ``only if'' part $\Rightarrow$.
In other words, we shall prove the following contraposition
\begin{align}
\phi( Q ) = \infty
\quad \Longrightarrow \quad
\sup_{(X, Y) : P_{\mathrm{e}}^{(L)}(X \mid Y) \le \varepsilon, P_{X} = Q} \mathsf{H}_{\phi}(X \mid Y)
=
\infty .
\label{eq:assertion_infty}
\end{align}
In the following, we show \eqref{eq:assertion_infty} by employing \lemref{lem:FiniteY} of \sectref{sect:proof_main_finite}.

Since $g_{2}( u ) = \infty$ only if $u = \infty$, it is immediate from \eqref{eq:phi_sum} that
\begin{align}
\phi( Q )
=
\infty
\quad
\Longrightarrow
\quad
\sum_{x \in \mathcal{X}} g_{1}\big( Q( x ) \big)
=
\infty ,
\end{align}
where note that $\phi(Q) = \infty$ implies that $g_{2}( \infty ) = \infty$ as well.
Moreover, since $g_{1}( 0 ) = 0$, we get
\begin{align}
\sum_{x \in \mathcal{X}} g_{1}\big( Q( x ) \big)
=
\infty
\quad \Longrightarrow \quad
|\supp( Q )| = \infty .
\end{align}
Due to \eqref{eq:range_epsilon_list}, we can find a finite subset $\mathcal{S} \subset \mathcal{Y}$ satisfying
\begin{align}
1 - \sum_{x = 1}^{L \cdot |\mathcal{S}|} Q^{\downarrow}( x )
\le
\varepsilon
\label{eq:subset_S}
\end{align}
by taking a finite but sufficiently large cardinality $|\mathcal{S}| < \infty$.
This implies that the new system $(Q, L, \varepsilon, \mathcal{S})$ still satisfies \eqref{eq:range_epsilon_list}; thus, it follows from \propref{prop:boundPe} that there exists an $\mathcal{X} \times \mathcal{S}$-valued r.v.\ $(X, Y)$ satisfying $P_{\mathrm{e}}^{(L)}(X \mid Y) \le \varepsilon$ and $P_{X} = Q$.
Therefore, the feasible region
\begin{align}
\mathcal{R}_{2}
=
\mathcal{R}(Q, L, \varepsilon, \mathcal{S})
\end{align}
defined in \eqref{def:region_R} is nonempty by this choice of $\mathcal{S}$.
As $\mathcal{S} \subset \mathcal{Y}$, it is clear that $\mathcal{R}_{2} \subset \mathcal{R}_{1}$, where
\begin{align}
\mathcal{R}_{1}
=
\mathcal{R}(Q, L, \varepsilon, \mathcal{Y}) .
\end{align}

By \lemref{lem:FiniteY}, one can find $\mathcal{Z} \subset \mathcal{X}$ so that $|\mathcal{Z}| = L \cdot |\mathcal{Y}|$ and
\begin{align}
\mathcal{R}_{3}
=
\mathcal{R}(Q, L, \varepsilon, \mathcal{S}, \mathcal{Z})
\end{align}
defined in \eqref{def:region_R_Z} is nonempty as well.
Moreover, since $P_{\mathrm{e}}^{(L)}(X \mid Y) \le P_{\mathrm{e}}^{(L)}(X \mid Y \, \| \, \mathcal{Z})$, if follows that $\mathcal{R}_{3} \subset \mathcal{R}_{2}$.
Then, we have
\begin{align}
\mathbb{H}_{\phi}(Q, L, \varepsilon, \mathcal{Y})
& \overset{\mathclap{\text{(a)}}}{=}
\sup_{(X, Y) \in \mathcal{R}_{1}} \mathsf{H}_{\phi}(X \mid Y)
\notag \\
& \overset{\mathclap{\text{(b)}}}{\ge}
\sup_{(X, Y) \in \mathcal{R}_{3}} \mathsf{H}_{\phi}(X \mid Y)
\notag \\
& \overset{\mathclap{\text{(c)}}}{\ge}
\inf_{\substack{ R \in \mathcal{P}( \mathcal{X} ) : \\ \forall x \in \mathcal{X} \setminus \mathcal{Z}, R(x) = Q(x) }} g_{2} \Bigg( \sum_{x \in \mathcal{X}} g_{1}\big( R( x ) \big) \Bigg)
\notag \\
& \overset{\mathclap{\text{(d)}}}{=}
\infty ,
\label{ineq:onlyif}
\end{align}
where
\begin{itemize}
\item
(a) follows by the definition of $\mathcal{R}_{1}$ stated in \eqref{def:region_R},
\item
(b) follows by the inclusions
\begin{align}
\emptyset
\neq
\mathcal{R}_{3}
\subset
\mathcal{R}_{2}
\subset
\mathcal{R}_{1} ,
\end{align}
\item
(c) follows from the fact that $(X, Y) \in \mathcal{R}_{3}$ implies that
\begin{align}
P_{X|Y=y}( x )
=
Q( x )
\end{align}
for $x \in \mathcal{X} \setminus \mathcal{Z}$ and $y \in \mathcal{S}$, and
\item
(d) follows from the facts that
\begin{align}
|\supp(Q) \setminus \mathcal{Z}| 
& =
\infty ,
\\
g_{1}( u )
& \ge
0
\qquad (\mathrm{for} \ 0 \le u \le 1),
\\
g_{2}( \infty )
& =
\infty .
\end{align}
\end{itemize}
Inequalities~\eqref{ineq:onlyif} imply \eqref{eq:assertion_infty}, completing the proof of \propref{prop:infty}.
\hfill\IEEEQEDhere

\section{Proof of \lemref{lem:majorizes_uniform}}
\label{app:majorizes_uniform}

This lemma is quite trivial, but we prove it to make the paper self-contained.
Actually, this can be directly proved by contradiction.
Suppose that \eqref{eq:majorizes_uniform1} and \eqref{eq:majorizes_uniform2} hold, but \eqref{eq:weak_majorization} does not hold.
Then, there must exist an $l \in \{ k, k+1, \dots, n-1 \}$ satisfying
\begin{align}
\sum_{i = 1}^{l} p_{i}
<
\sum_{i = 1}^{l} q_{i} .
\label{eq:contradiction}
\end{align}
As $q_{j}$ is constant for each $j = k, k+1, \dots, n$, it follows from \eqref{eq:majorizes_uniform1} and \eqref{eq:contradiction} that $p_{j} < q_{j}$ for every $j = l, l+1, \dots, n$.
Then, we observe that
\begin{align}
\sum_{i = 1}^{n} p_{i}
<
\sum_{i = 1}^{n} q_{i} ,
\end{align}
which contradicts to the hypothesis of \eqref{eq:majorizes_uniform2}, and therefore \lemref{lem:majorizes_uniform} must hold.
\hfill\IEEEQEDhere

\bibliographystyle{IEEEtran}
\bibliography{IEEEabrv,mybib}

\end{document}